\documentstyle[12pt,color,epsfig]{article}

\newlength{\dinwidth}                       
\newlength{\dinmargin}                      
\newcommand{\hdick}{\noalign{\hrule height1.4pt}}

\newcommand{\GeV} {\mathrm{GeV}}
\newcommand{\TeV} {\mathrm{TeV}}

\def\ti    {\tilde}

\def\sq    {{\ti q}}
\def\st    {{\ti t}}
\def\sb    {{\ti b}}

\def\stau  {{\ti\tau}^-}
\def\staup {{\ti\tau}^+}

\def\cp    {\ti \chi^+}
\def\cm    {\ti \chi^-}
\def\ch    {\ti \chi^\pm}

\def\nt    {\ti \chi^0}

\def\sg    {\ti g}

\def\sq    {\ti q}

\def\ur    {\ti u_R}

\def\dr    {\ti d_R}

\def\smul  {{\ti\mu}^-_L}
\def\smur  {{\ti\mu}^-_R}
\def\smulp {{\ti\mu}^+_L}
\def\smurp {{\ti\mu}^+_R}
\def\snm   {{\ti\nu}_\mu}

\def\sel   {{\ti e}^-_L}
\def\ser   {{\ti e}^-_R}
\def\selp  {{\ti e}^+_L}
\def\serp  {{\ti e}^+_R}
\def\sne  {{\ti\nu}_e}

\def\snt      {{\ti\nu}_\tau}

\def\anue {{\bar\nu}_e}
\def\anum {{\bar\nu}_\mu}
\def\anut {{\bar\nu}_\tau}

\def\isajet  {{\sc Isajet}}
\def\susygen {{\sc Susygen}}
\def\pythia  {{\sc Pythia}}
\def\suspect {{\tt SuSpect}}

\begin{document}
 
\pagenumbering{roman}
\begin{titlepage} 
\begin{flushright}
  {\tt LC-TH-2001-079} \\
  {\tt hep-ph/0201233} \\
  January 18, 2002
\end{flushright}

\bigskip
\begin{center}
  {\Large\bf Compilation of SUSY particle spectra from \\[.5ex]
    Snowmass~2001 benchmark models} \\*[3ex]
    {\large Nabil Ghodbane$^\dagger$,  Hans-Ulrich Martyn$^\star$} \\*[2ex] 
    {\em $^\dagger$~DESY, Hamburg, Germany,    
         $^\star$~I. Physikalisches Institut der RWTH, Aachen, Germany}    
\end{center}

\begin{abstract}\noindent
  A comparative study of supersymmetric particle spectra 
  calculated by the programs \isajet, \susygen \ and \pythia \ is presented
  for various SUSY scenarios defined at the Snowmass 2001 workshop.
\end{abstract}

\noindent
At the Snowmass~2001 
{\em 'Summer Study on the Future of Particle Physics' } 
a concensus was reached to define a list of
SUSY models as benchmarks to be investigated in future collider studies.
Various scenarios, so-called 'Snowmass Points and Slopes' (SPS),
were proposed\footnote{
  {\em 'SUSY benchmark discussion'} at Snowmass 2001,
  http://lotus.phys.nwu.edu/~schmittm/snowmass 
  }$^{,\,}$\footnote{
  M Battaglia et al, {\em 'The Snowmass points and slopes:
    benchmarks for SUSY searches'}, Snowmass pro\-ceedings,
  in preparation }  
in terms of a few parameters describing 
`typical' to `extreme' $R_p$ conserving supersymmetry breaking mechanisms of
mSUGRA, GMSB and AMSB. 
All benchmark points respect currently existing experimental constraints.

\begin{center}
\begin{tabular}{ll ccccc}
  \hdick & & & & & &  \\[-2.ex]
  & mSUGRA scenario & $m_0$ & $m_{1/2}$ & $A_0$ & $\tan\beta$ & sign $\mu$ 
  \\[.5ex] \hdick \\[-2.ex]
  SPS 1 & typical point      &  100 & 250 & $-100$ & 10 & + \\                 
  SPS 2 & focus point region & 1450 & 300 &    0 & 10 & + \\                 
  SPS 3 &\hspace*{-2mm}\begin{tabular}{l}
           model line into \\
           coannihilation region  \end{tabular}
        & 90 & 400 & 0 & 10 & + \\                 
  SPS 4 & large $\tan\beta$  &  400 & 300 &    0 & 50 & + \\                 
  SPS 5 & light stop         &  150 & 300 &$-1000$ &  5 & + \\                 
  SPS 6 & \hspace*{-2mm}\begin{tabular}{l} 
           non-unified gaugino masses \\ 
           {\small $M_1 = 480, \ M_2 = M_3 = 300$} \end{tabular}
        &  150 & 300 & 0 & 10 & + \\[.5ex]
  \hdick & & & & & &  \\[-2.ex]
  & GMSB scenario & $\Lambda$ & $M_{mess}$ & $N_{mess}$ 
                  & $\tan\beta$ & sign $\mu$ 
  \\[.5ex] \hdick \\[-2.ex]
  SPS 7 &  NLSP = $\tilde\tau_1$     &  40,000 &  80,000 & 3 & 15 & + \\ 
  SPS 8 &  NLSP = $\tilde\chi^0_1$ & 100,000 & 200,000 & 1 & 15 & + \\[.5ex]
  \hdick & & & & & &  \\[-2.ex]
  & AMSB scenario & $m_0$ & $m_{3/2}$ & 
                  & $\tan\beta$ & sign $\mu$ 
  \\[.5ex] \hdick \\[-2.ex]
  SPS 9 & small $\Delta m(\tilde\chi^+_1 - \tilde\chi^0_1)$ 
        & 400 & 60,000 &   & 10 & + \\[.5ex]
  \hdick & & & & & &  \\[-2.ex]  
  \multicolumn{5}{l}{ {\footnotesize masses and scales in GeV}}
\end{tabular} 
\end{center}

However, at Snowmass it was recommended 
to take the SUSY particle spectrum as generated by the program 
\isajet\footnote{
  H Baer et al, hep-ph/0001086, \isajet, 
  http://paige.home.cern.ch/paige}
as the reference for benchmark models,
instead of the few high energy parameters.
This arbitrary choice certainly does not imply that this program is superior 
to any other modern program.
In fact, the purpose of this note is to compare different codes and thus
to get a feeling on the reliability of presently available calculations.
We present SPS model spectra, i.e. using the same input parameters,
as generated by \isajet~7.58 in comparison with results from the programs
\susygen~3.00/27\footnote{ 
  N Ghodbane et al, hep-ph/9909499, \susygen, 
  http://lyoinfo.in2p3.fr/susygen/susygen3.html}
and \pythia~6.2/00\footnote{ 
  T Sj\"ostrand et al, hep-ph/0108264, \pythia, 
  http://www.thep.lu.se/$\sim$torbjorn//Pythia.html}.
These event generators are most popular and frequently used for 
SUSY studies and simulations.

\isajet \ uses the package {\sc Isasugra} to solve numerically the two-loop
RGE's in the MSSM couplings in the $3^{rd}$ generation approximation.
Theta-function thresholds are included in the gauge and Yukawa one-loop 
running couplings.
Decay modes are calculated with {\sc Isasusy} applying
decay matrix elements.
Besides generic MSSM models, specific scenarios can be selected with the 
mSUGRA, GMSB or AMSB options.

In \susygen \ the mSUGRA scenarios are calculated with 
the program \suspect~2.0\footnote{
  A Djouadi et al, hep-ph/9901246, {\tt SuSpect},
  http://www.lpm.univ-montp2.fr:6714/$\sim$kneur/ suspect.html}.
All masses and couplings (except scalar masses which are run at one-loop)
are evaluated numerically using two-loop RGE's 
in the $3^{rd}$ generation approximation
and including smooth thresholds.
The Higgs branching ratios are calculated using {\sc Hdecay}\footnote{
  A Djouadi et al, Comp. Phys. Comm. 108 (1998) 56}.
General models with a limited number of parameters can be chosen; 
GMSB and AMSB scenarios can be generated by setting 
the appropriate MSSM parameters.

\pythia \ is a fast and robust program and
uses semi-analytical formulae to solve the RGE's with one-loop
beta functions. Thresholds are not included.
Several SUSY models such as mSUGRA or fixed gaugino masses
can be chosen.
GMSB and AMSB scenarios can be realised via generic MSSM parameters.

In general all three programs provide mass spectra which agree at the
level of about 10\,\% for not too extreme choices of parameters.
However, huge mass differences of up to factors of 2 or even more
may occur in models with very large parameters, 
e.g high $m_0  \gg m_{1/2}$ (SPS~2) or high $\tan\beta$ (SPS~4). 
Such extreme situations are very sensitive to higher order corrections,
for which the programs may not be prepared.
In a study of SPS model lines it was observed\footnote{
  B C Allanach, 
  {\em `Theoretical uncertainties in sparticle mass predictions'},
  hep-ph/0110227}
that \isajet\ exhibits mass instabilities or wiggles as a function of
gaugino masses at the level of a few percent, 
but which may increase up to $\sim 30\,\%$ for the light chargino in SPS~3
depending on the value of $m_{1/2}$.

Another concern are the decay modes 
and branching ratios which are treated in the three programs
with different sophistication. Some of them seem to be incomplete, e.g. 
missing sfermion decays into gauge and Higgs bosons
and missing chargino decays into Higgs (\susygen) 
or missing Higgs decays into supersymmetric particles (\pythia).
Obvious problems arise due to mass differences, which may suppress 
or open certain decay channels. 
The user has to carefully examine the decay modes and possibly consult the
original literature.
For the Higgs sector dedicated programs like {\sc Hdecay}$^7$
or {\sc FeynHiggs}\footnote{S Heinemeyer et al, 
  Comp. Phys. Comm. 124 (2000) 76}
may be used as cross checks.

We adopt a user's point of view and feel unable to judge which is the 
optimal code or event generator to produce SUSY particle spectra.
However, if benchmark studies should be useful and should allow 
to do comparisons, we strongly advise the authors 
of future analyses to give the complete list of sparticle masses and
decay branching ratios which have been used.
In order to reproduce the \isajet\ spectra with \susygen\ and \pythia\ 
all necessary information is given in the SPS compilations below.
Besides the physical masses and branching ratios, 
further relevant MSSM parameters are: 
\\[2ex] \phantom{xxxx}
\begin{tabular}{lll}
  MSSMA                  
  & $m_{\tilde g}$, $\mu$, $m_A$, $\tan\beta$ 
  & gluino mass, $\mu$, $A$ mass, $\tan\beta$ \\[1ex]
  MSSMB                 
  & $m_{\sq_{1\,L}}$, $m_{\dr}$, $m_{\ur}$, $m_{\ti e_L}$, $m_{\ti e_R}$    
  & $1^{st}$ generation squark and  slepton masses \\[1ex]
  MSSMC                 
  & $m_{\sq_{3\,L}}$, $m_{\sb_R}$, $m_{\st_R}$, 
    $m_{\ti\tau_L}$, $m_{\ti\tau_R}$,  
  & $3^{rd}$ generation squark and  slepton masses, \\[1ex]
  & $A_t$, $A_b$, $A_\tau$   
  & squark and slepton mixings \\[1ex]
  MSSMD                
  & $m_{\sq_{2\,L}}$, $m_{\ti s_R}$, $m_{\ti c_R}$, 
    $m_{\ti\mu_L}$, $m_{\ti\mu_R}$  
  & $2^{nd}$ generation squark and  slepton masses \\[1ex]
  MSSME                
  & $M_1$, $M_2$  & gaugino masses  \\[1ex]
\end{tabular}

Keeping in mind the previous remarks, we briefly comment the main
characteristics of the superparticle spectra obtained with the program
codes \isajet, \susygen\ and \pythia.
Note that there exist variants to SPS~1 and SPS~8 with $\tan\beta=30$,
leading to very similar spectra$^2$.

\paragraph{SPS 1} 
The spectra of this `typical' mSUGRA scenario provided by the
three programs are in remarkable agreement with each other
to within a few percent.
Also the decay modes are reasonably similar.

\paragraph{SPS 2} 
Most striking in this `focus point' scenario with large $m_0 \gg m_{1/2}$
are the completely different neutralino and chargino
mass spectra of all three programs.
Factors of 2 with respect to some central value may easily occur.
This is related to the derived values of $\mu$ which exhibit discrepancies 
of the same order, casting doubt on the calculations.
The other sparticle masses are in reasonable agreement with each other.
However, large controversies exist in the treatment of the decays.
The heavy sleptons and squarks can decay via all charginos and neutralinos,
which apparently do have completely different gaugino and
higgsino admixtures, thus leading to different $\chi$'s in the final states.
This model as a whole needs certainly more care and a better 
theoretical understanding.

\paragraph{SPS 3} 
All spectra of this scenario which lies on a model line into
the `coannihilation region' are in good agreement with each other.
The masses differ within a few up to ten percent, some 
branching ratios may need to be adjusted.

\paragraph{SPS 4} 
In general the mass spectra are consistent to within 10\%. But due to the
high $\tan\beta$ value larger discrepancies are observed for the third
generation of sleptons and squarks and for the Higgs sector of \pythia.
The original purpose of this parameter choice was to provide a model with
relatively light Higgs bosons. 
At the time of the Snowmass meeting$^1$ a value of $m_A \simeq 310~\GeV$
was supported by \isajet~7.51 and \suspect~2.0 (point d'Aix~3).
Meanwhile the new release \isajet~7.58 gives a considerably
higher mass $m_A = 404~\GeV$, 
probably a consequence of a new treatment of the Yukawa bottom
coupling at two-loops in the RG evolution 
Similarly, if \suspect\ takes into account additional radiative corrections 
to all squarks and gauginos, one also gets a larger value of 
$m_A = 354~\GeV$.
Thus, in this parameter space with large $\tan\beta$
the Higgs spectrum depends sensitively on the 
details of higher order corrections. 
This also explains the lowish Higgs masses of \pythia, which uses a one-loop
approximation.

\paragraph{SPS 5} 
The sparticle spectra of all three programs are in general agreement.
But the scalar top $\st_1$, which should be light in this scenario,
is about 20\% heavier in \pythia.
This leads to quite different decay modes.

\paragraph{SPS 6} 
Very good agreement among the spectra is observed, since 
in this mSUGRA like scenario with non-unified gaugino masses
the MSSM parameters
calculated by \isajet\ serve as input to \susygen\ and \pythia.
Slight discrepancies occur in the branching ratios.

\paragraph{SPS 7 \& 8}
Only \isajet\ is able to generate genuine GMSB scenarios.
Since the gravitino is incorporated,
the MSSM parameters of \susygen\ and \pythia\ 
are adjusted accordingly.
\pythia\ does not offer to treat $\ti\tau_1$ as NLSP (SPS~7).
All the masses are in agreement by construction, except for
discrepancies of 5 - 20\% in the heavy Higgs sector of \susygen;
for the time being a more precise tuning appears problematic.
Again there are some differences in the decay channels, 
most important are those of $\ser$ and $\smur$ decays in model SPS~7.
The specified parameters do not fix the lifetime of the NLSP, 
which depends on the fundamental scale $\sqrt{F}$ of SUSY breaking.
\isajet\ allows to set this scale, such that finite decay length
distributions $c \tau \sim F^2 \cdot m^{-5}_{NLSP}$ of the NLSP can be studied.

\paragraph{SPS 9} 
An elementary treatment of AMSB scenarios is only possible in \isajet.
Using the corresponding MSSM parameters the appropriate spectra can be produced
with \susygen\ and \pythia .
Again there are some inconsistent decay modes, e.g. $\ser$ and $\smur$ decays
and, more important, the $\ch_1$ decays.
The characteristic feature is the near degeneracy of the lightest chargino 
and neutralino masses, which critically determines the $\ch_1$ lifetime and
decay modes.
The present parameters give $\tau = 165$~ps ($c \tau = 50$~mm).
The lifetime and proper branching ratios 
determine the $\ch_1 / \nt_1$ search strategy 
and it is advisable to consult the literature\footnote{
  J F Gunion, S Mrenna, Phys. Rev. D 64 (2001) 075002}.

\paragraph{Conclusions}
The three event generators \isajet, \susygen\ and \pythia\ produce consistent
supersymmetric particle spectra for a wide range of commonly accepted SUSY
parameters. However, severe discrepancies exist for scenarios with extreme
values of parameters, {\em e.g.} large $m_0 \gg m_{1/2}$ in the focus point
scenario SPS~2 (chargino and neutralino sector) and high $\tan\beta$ of SPS~4
(Higgs sector, $3^{rd}$ generation sleptons and squarks), 
where some particle characteristics depend sensitively on 
higher order corrections.
Although the programs are flexible enough to be adjusted to any mass spectrum
(including decay modes), 
it is {\em a priori} not obvious which program to prefer.
It is the aim of the present compilation to provide detailed information
and to help those who are interested to study the properties of the Snowmass
benchmark models.

\bigskip\bigskip\noindent 
{\em 
  It's a pleasure to thank S~Kraml, E~Perez, W~Porod and G~Weiglein  
  for useful discussions on the SPS spectra and on the manuscript. }  

\end{titlepage}                         

\clearpage
\tableofcontents
\clearpage
                                                                 
\pagestyle{headings}
\pagenumbering{arabic}
\setcounter{page}{1}

\section{SPS 1 -- mSUGRA scenario}
\setcounter{figure}{0}
\setcounter{table}{0}

\large\boldmath
\hspace{20mm}
\begin{tabular}{|l c|}
  \hline
  $m_0$       & \ \, $ 100~\GeV$ \\ 
  $m_{1/2}$   & \ \, $ 250~\GeV$ \\
  $A_0$       & $-100~\GeV$ \\
  $\tan\beta$ & $10$       \\
  ${\rm sign}~\mu$ & $+$ \\
  \hline
\end{tabular} \hspace{10mm}
\begin{tabular}{l}
   {\bf `typical' scenario} \\
    $m_0 = 0.4\,m_{1/2} = -A_0 $ \\
\end{tabular}
\unboldmath\normalsize
\bigskip

\subsection{Spectrum \& parameters of ISAJET 7.58}

\begin{figure}[h] \centering
  \epsfig{file=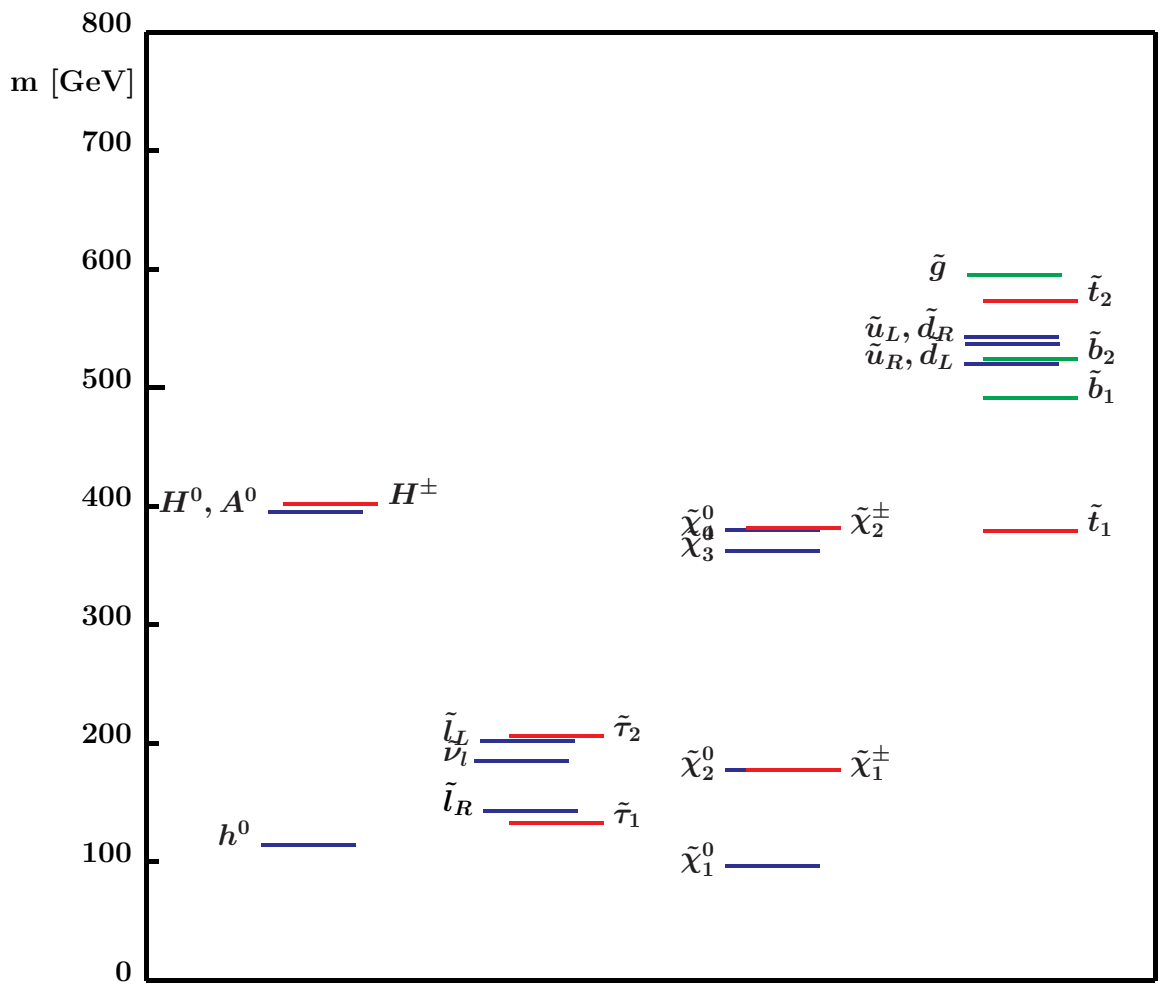,width=.9\textwidth}
  \caption{SPS 1 mass spectrum of \isajet}
\end{figure}

\clearpage
\normalsize
\noindent
{\bf \isajet\ parameters} 
\begin{small}
\begin{verbatim}
 Minimal supergravity (mSUGRA) model:

 M_0,  M_(1/2),  A_0,  tan(beta),  sgn(mu),  M_t =
   100.000   250.000  -100.000    10.000     1.0   175.000

 ISASUGRA unification:
 M_GUT      = 0.218E+17   g_GUT          =0.714      alpha_GUT =0.041
 FT_GUT     = 0.481       FB_GUT         = 0.046     FL_GUT = 0.069

 1/alpha_em =  127.70     sin**2(thetaw) =0.2309     alpha_s   =0.119
 M_1        =   99.13     M_2            =  192.74   M_3       =  580.51
 mu(Q)      =  352.39     B(Q)           =   44.54   Q         =  454.65
 M_H1^2     = 0.333E+05   M_H2^2         =-0.124E+06

 ISAJET masses (with signs):
 M(GL)  =   595.19
 M(UL)  =   537.25   M(UR)  =   520.45   M(DL)  =   543.04   M(DR) =   520.14
 M(B1)  =   491.91   M(B2)  =   524.59   M(T1)  =   379.11   M(T2) =   574.71
 M(SN)  =   186.00   M(EL)  =   202.14   M(ER)  =   142.97
 M(NTAU)=   185.06   M(TAU1)=   133.22   M(TAU2)=   206.13
 M(Z1)  =   -96.05   M(Z2)  =  -176.82   M(Z3)  =   358.81   M(Z4) =  -377.81
 M(W1)  =  -176.38   M(W2)  =  -378.23
 M(HL)  =   113.97   M(HH)  =   394.15   M(HA)  =   393.63   M(H+) =   401.77

 theta_t=   0.9603   theta_b=   0.4916   theta_l=   1.2876   alpha_h=   0.1107

 NEUTRALINO MASSES (SIGNED) =   -96.051  -176.820   358.815  -377.811
 EIGENVECTOR 1       =   0.05441  -0.15001  -0.05711   0.98553
 EIGENVECTOR 2       =   0.16023  -0.27963  -0.94070  -0.10592
 EIGENVECTOR 3       =  -0.71046  -0.69495   0.09245  -0.06120
 EIGENVECTOR 4       =   0.68309  -0.64525   0.32137  -0.11730

 CHARGINO MASSES (SIGNED)  =  -176.383  -378.229
 GAMMAL, GAMMAR             =   1.99234   1.80868


 ISAJET equivalent input:
 MSSMA:   595.19  352.39  393.63   10.00
 MSSMB:   539.86  519.53  521.66  196.64  136.23
 MSSMC:   495.91  516.86  424.83  195.75  133.55 -510.01 -772.66 -254.20
 MSSMD: SAME AS MSSMB (DEFAULT)
 MSSME:    99.13  192.74

\end{verbatim}
\end{small}

\clearpage
\subsection{Spectrum \& parameters of SUSYGEN 3.00/27}
\begin{figure}[h] \centering
  \epsfig{file=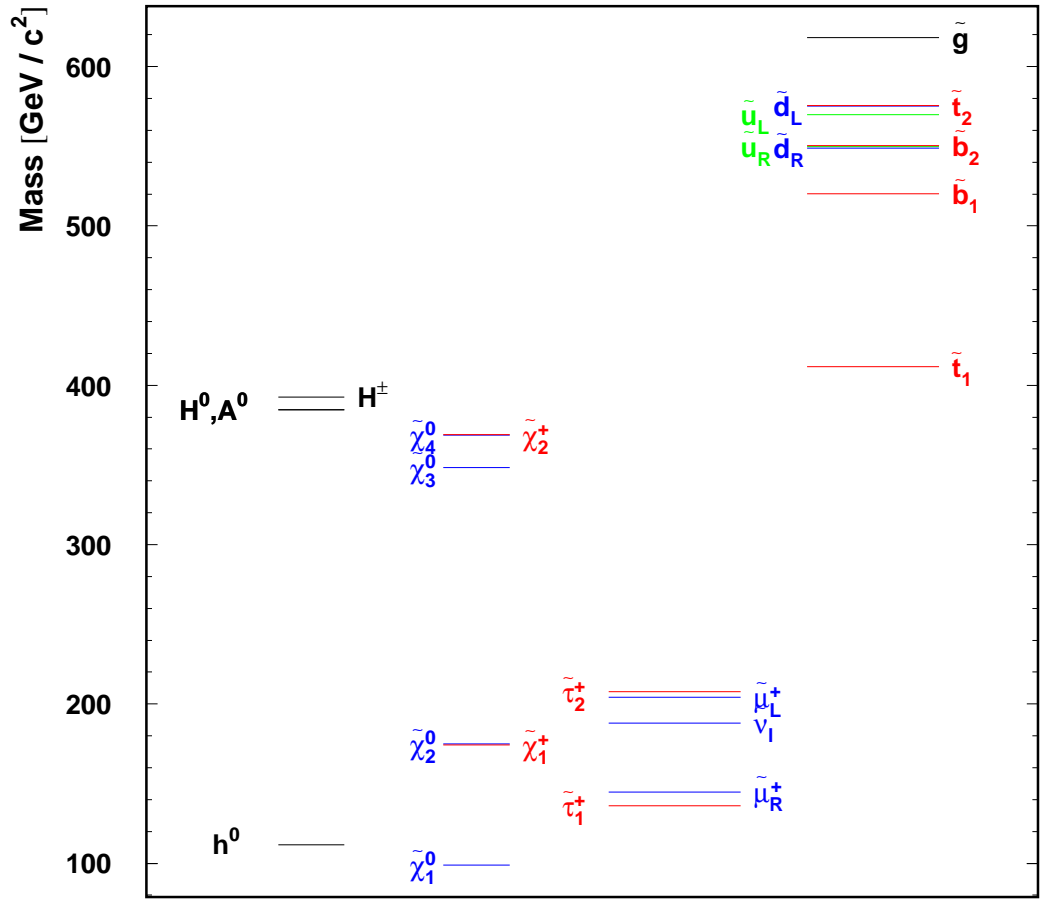,width=.9\textwidth}
  \caption{SPS 1 mass spectrum of \susygen}
\end{figure}
\newpage
\normalsize
\noindent
{\bf \susygen\  parameters} 
\begin{small}
\begin{verbatim}
 Susygen Inputs:
 --------------

 m0      =   100.000   TANB     =    10.000
 m1/2    =   250.000    mu/|mu| =     1
 A0      =  -100.000

 Sparticle masses: 
 ----------------
 SUPR      550.  SUPL       570.
 SDNR      549.  SDNL       575.
 SELR      145.  SELL       204.
 SNU       188.
 STP1      412.  STP2       576. cosmix =  0.534
 SBT1      520.  SBT2       550. cosmix =  0.913
 STA1      136.  STA2       208. cosmix=   0.271
 SGLU      618.


 Gaugino masses:
 --------------

 M1 =    102.191 M2 =    191.812 M3 =    588.293

 NEUTRALINO m, CP, ph/zi/ha/hb 1 =   98.8  1.  0.831 -0.530  0.075  0.152
 NEUTRALINO m, CP, ph/zi/ha/hb 2 =  174.9  1. -0.554 -0.762  0.196  0.271
 NEUTRALINO m, CP, ph/zi/ha/hb 3 =  348.4 -1. -0.009  0.113 -0.638  0.762
 NEUTRALINO m, CP, ph/zi/ha/hb 4 =  368.7  1. -0.054 -0.354 -0.741 -0.569

 CHARGINO MASSES    =   174.211   369.022
 CHARGINO ETA       =    -1.000     1.000

 U matrix      WINO      HIGGSINO   V matrix      WINO     HIGGSINO 
 W1SS+        -0.905     0.426      W1SS-         0.968    -0.249
 W2SS+         0.426     0.905      W2SS-         0.249     0.968


 Higgses masses: 
 --------------

 Light CP-even Higgs =   111.794
 Heavy CP-even Higgs =   384.532
       CP-odd  Higgs =   384.598
       Charged Higgs =   392.561
       sin(a-b)      =    -0.111
       cos(a-b)      =     0.994

\end{verbatim}
\end{small}

\clearpage

\subsection{Spectrum \& parameters of  PYTHIA 6.2/00}

\begin{figure}[h] \centering
  \epsfig{file=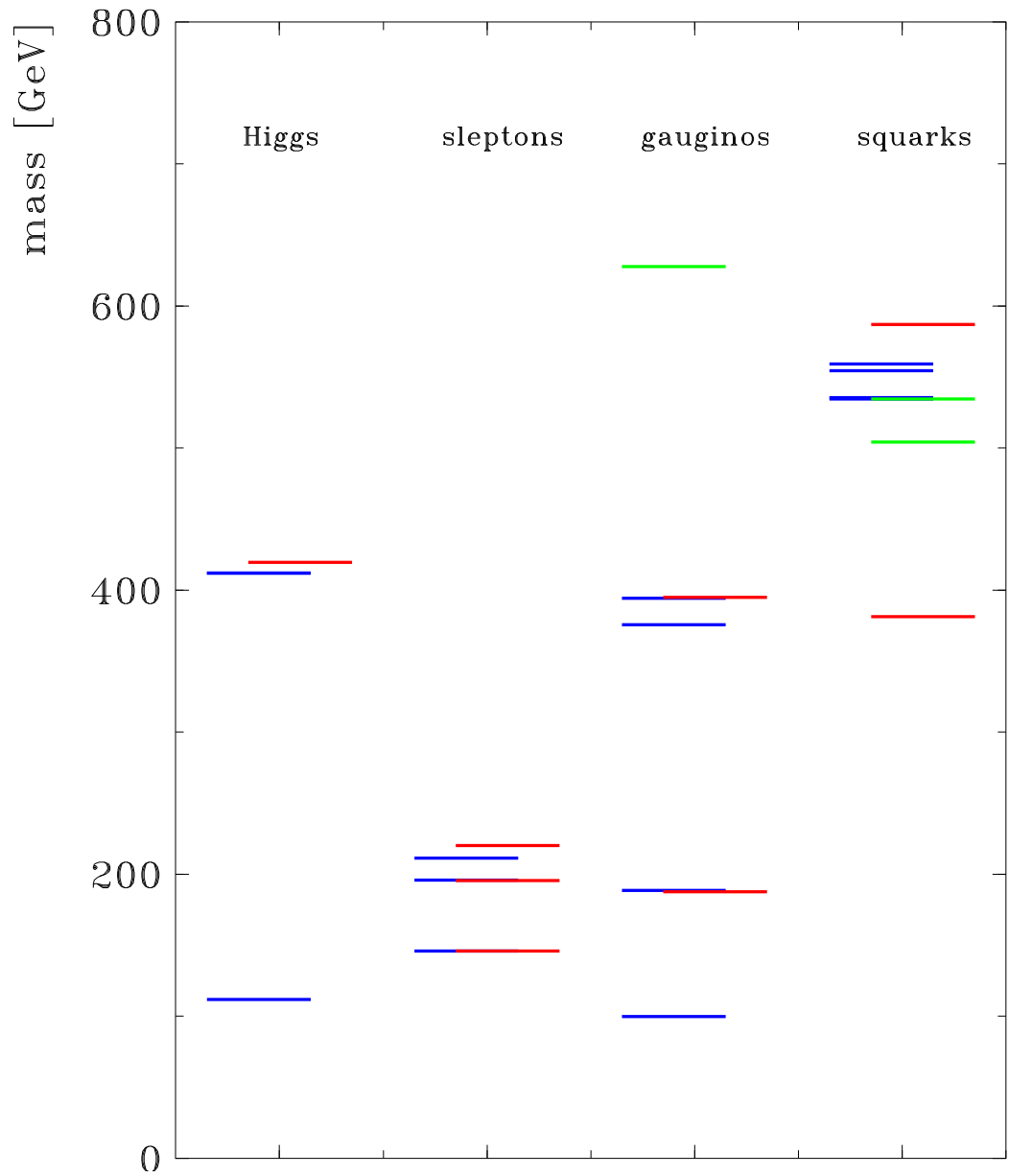,width=.9\textwidth}
  \caption{SPS 1 mass spectrum of \pythia}
\end{figure}
\newpage
\normalsize
\noindent
{\bf \pythia\  parameters} 
\begin{small}
\begin{verbatim}
   SUGRA input parameters
   ----------------------
   m_0         RMSS( 8) =   100.0    
   m_1/2       RMSS( 1) =   250.0    
   A_0         RMSS(16) =  -100.0    
   tan_beta    RMSS( 5) =   10.00    
   sign mu     RMSS( 4) =   1.000    

   sparticle masses & widths
   -------------------------
   M_se_R   145.8 ( 0.204)   M_se_L   211.4 ( 0.235)   M_sne_L  195.8 ( 0.173)
   M_sm_R   145.8 ( 0.204)   M_sm_L   211.4 ( 0.235)   M_snm_L  195.8 ( 0.173)
   M_st_1   146.0 ( 0.198)   M_st_2   220.2 ( 0.335)   M_snt_L  195.6 ( 0.171)

   M_ch0_1   99.9 ( 0.000)   M_ch0_2  188.4 ( 0.015)   M_ch0_3  375.7 ( 2.209)
   M_ch0_4  394.2 ( 2.946)
   M_ch+_1  187.7 ( 0.012)   M_ch+_2  394.8 ( 2.888)

   M_h0     111.7 ( 0.004)   M_H0     412.1 ( 0.910)   M_A0     411.9 ( 0.980)
   M_H+     419.5 ( 0.855)

   M_g~     627.8 (11.779)
   M_uL     554.4 ( 5.376)   M_uR     535.3 ( 1.127)   M_dL     559.3 ( 5.159)
   M_dR     534.4 ( 0.281)
   M_b1     504.2 ( 3.590)   M_b2     534.6 ( 1.060)   M_t1     381.3 ( 1.768)
   M_t2     587.1 ( 7.620)

   parameter settings IMSS, RMSS
   -----------------------------
   IMSS( 1) = 2   IMSS( 4) = 1   IMSS( 7) = 0   IMSS(10) = 0
   IMSS( 2) = 0   IMSS( 5) = 0   IMSS( 8) = 0   IMSS(11) = 0
   IMSS( 3) = 0   IMSS( 6) = 0   IMSS( 9) = 0   IMSS(12) = 0

   RMSS( 1) =   250.0       RMSS( 9) =   700.0       RMSS(17) =   100.0    
   RMSS( 2) =   204.0       RMSS(10) =   504.9       RMSS(18) = -0.1098    
   RMSS( 3) =   600.2       RMSS(11) =   530.1       RMSS(19) =   412.1    
   RMSS( 4) =   369.6       RMSS(12) =   430.3       RMSS(20) =  0.4100E-01
   RMSS( 5) =   10.00       RMSS(13) =   211.2       RMSS(21) =   1.000    
   RMSS( 6) =   211.4       RMSS(14) =   145.3       RMSS(22) =   800.0    
   RMSS( 7) =   145.8       RMSS(15) =   100.0       RMSS(23) =  0.1000E+05
   RMSS( 8) =   100.0       RMSS(16) =  -532.9       RMSS(24) =  0.1000E+05
\end{verbatim}
\end{small}
\clearpage

\subsection{Decay modes}
\normalsize
 \begin{table}[h] \centering
   \begin{tabular}{|l|c c c|c|c c c|}
   \hdick & & & & & & & \\[-2.ex]
   particle & $m_I$ & $m_S$ & $m_P$ & \ decay \ & ${\cal B}_I$ & ${\cal B}_S$ & ${\cal B}_P$ 
   \\[.5ex] \hdick
   $\ser    $ &  143.0 & 144.9 & 145.8 & $\nt_1    e^-               $ & 1.000 & 1.000 & 1.000  \\
   \hline
   $\sel    $ &  202.1 & 204.1 & 211.4 & $\nt_1    e^-               $ & 0.490 & 0.408 & 0.556  \\
              &        &       &       & $\nt_2    e^-               $ & 0.187 & 0.218 & 0.159  \\
              &        &       &       & $\cm_1    \nu_e             $ & 0.323 & 0.374 & 0.285  \\
   \hline
   $\sne    $ &  186.0 & 188.2 & 195.8 & $\nt_1    \nu_e             $ & 0.885 & 0.786 & 0.920  \\
              &        &       &       & $\nt_2    \nu_e             $ & 0.031 & 0.057 &        \\
              &        &       &       & $\cp_1    e^-               $ & 0.083 & 0.157 & 0.059  \\
   \hline
   $\smur   $ &  143.0 & 144.9 & 145.8 & $\nt_1    \mu^-             $ & 1.000 & 1.000 & 1.000  \\
   \hline
   $\smul   $ &  202.1 & 204.1 & 211.4 & $\nt_1    \mu^-             $ & 0.490 & 0.408 & 0.556  \\
              &        &       &       & $\nt_2    \mu^-             $ & 0.187 & 0.218 & 0.159  \\
              &        &       &       & $\cm_1    \nu_\mu           $ & 0.323 & 0.374 & 0.285  \\
   \hline
   $\snm    $ &  186.0 & 188.2 & 195.8 & $\nt_1    \nu_\mu           $ & 0.885 & 0.786 & 0.920  \\
              &        &       &       & $\nt_2    \nu_\mu           $ & 0.031 & 0.057 &        \\
              &        &       &       & $\cp_1    \mu^-             $ & 0.083 & 0.157 & 0.059  \\
   \hline
   $\stau_1 $ &  133.2 & 136.0 & 146.0 & $\nt_1    \tau^-            $ & 1.000 & 1.000 & 1.000  \\
   \hline
   $\stau_2 $ &  206.1 & 207.7 & 220.2 & $\nt_1    \tau^-            $ & 0.526 & 0.453 & 0.504  \\
              &        &       &       & $\nt_2    \tau^-            $ & 0.174 & 0.203 & 0.179  \\
              &        &       &       & $\cm_1    \nu_\tau          $ & 0.300 & 0.344 & 0.316  \\
   \hline
   $\snt    $ &  185.1 & 187.2 & 195.6 & $\nt_1    \nu_\tau          $ & 0.906 & 0.806 & 0.926  \\
              &        &       &       & $\nt_2    \nu_\tau          $ &       & 0.052 &        \\
              &        &       &       & $\cp_1    \tau^-            $ & 0.067 & 0.142 & 0.054  \\
   \hline
   \end{tabular}
   \caption{Slepton masses (GeV) and significant branching ratios ($>3\%$) 
            from \isajet~(I), \susygen~(S) and \pythia~(P)}
 \end{table}
  
 \begin{table}[h] \centering
   \begin{tabular}{|l|c c c|c|c c c|}
   \hdick & & & & & & & \\[-2.ex]
   particle & $m_I$ & $m_S$ & $m_P$ & \ decay \ & ${\cal B}_I$ & ${\cal B}_S$ & ${\cal B}_P$ 
   \\[.5ex] \hdick
   $\nt_1   $ &   96.1 &  98.8  & 99.9 & $                           $ & 1.000 & 1.000 & 1.000  \\
   \hline
   $\nt_2   $ &  176.8 & 174.9 & 188.4 & $\ser     e^+               $ & 0.031 & 0.037 & 0.044  \\
              &        &       &       & $\serp    e^-               $ & 0.031 & 0.037 & 0.044  \\
              &        &       &       & $\smur    \mu^+             $ & 0.031 & 0.037 & 0.044  \\
              &        &       &       & $\smurp   \mu^-             $ & 0.031 & 0.037 & 0.044  \\
              &        &       &       & $\stau_1  \tau^+            $ & 0.437 & 0.424 & 0.407  \\
              &        &       &       & $\staup_1 \tau^-            $ & 0.437 & 0.424 & 0.407  \\
   \hline
   $\nt_3   $ &  358.8 & 348.4 & 375.7 & $\cp_1    W^-               $ & 0.298 & 0.349 & 0.299  \\
              &        &       &       & $\cm_1    W^+               $ & 0.298 & 0.349 & 0.299  \\
              &        &       &       & $\nt_1    Z^0               $ & 0.108 & 0.087 & 0.106  \\
              &        &       &       & $\nt_2    Z^0               $ & 0.215 & 0.155 & 0.218  \\
   \hline
   $\nt_4   $ &  377.8 & 368.7 & 394.2 & $\cp_1    W^-               $ & 0.263 & 0.302 & 0.263  \\
              &        &       &       & $\cm_1    W^+               $ & 0.263 & 0.302 & 0.263  \\
              &        &       &       & $\nt_1    h^0               $ & 0.064 & 0.054 & 0.062  \\
              &        &       &       & $\nt_2    h^0               $ & 0.134 & 0.103 & 0.144  \\
   \hline
   \end{tabular}
   \caption{Neutralino masses (GeV) and significant branching ratios ($>3\%$) 
            from \isajet~(I), \susygen~(S) and \pythia~(P)}
 \end{table}
  
 \begin{table}[h] \centering
   \begin{tabular}{|l|c c c|c|c c c|}
   \hdick & & & & & & & \\[-2.ex]
   particle & $m_I$ & $m_S$ & $m_P$ & \ decay \ & ${\cal B}_I$ & ${\cal B}_S$ & ${\cal B}_P$ 
   \\[.5ex] \hdick
   $\cp_1   $ &  176.4 & 174.2 & 187.7 & $\staup_1 \nu_\tau          $ & 0.979 & 0.864 & 0.847 \\
              &        &       &       & $\nt_1    W^+               $ &       &       & 0.153  \\
   \hline
   $\cp_2   $ &  378.2 & 369.0 & 394.8 & $\nt_1    W^+               $ & 0.064 & 0.062 & 0.065  \\
              &        &       &       & $\nt_2    W^+               $ & 0.296 & 0.303 & 0.065  \\
              &        &       &       & $\selp    \nu_e             $ & 0.052 & 0.056 & 0.049  \\
              &        &       &       & $\smulp   \nu_\mu           $ & 0.052 & 0.056 & 0.049  \\
              &        &       &       & $\staup_2 \nu_\tau          $ & 0.056 & 0.055 & 0.051  \\
              &        &       &       & $\cp_1    Z^0               $ & 0.244 & 0.393 & 0.243  \\
              &        &       &       & $\cp_1    h^0               $ & 0.170 &       & 0.173  \\
   \hline
   \end{tabular}
   \caption{Chargino masses (GeV) and significant branching ratios ($>3\%$) 
            from \isajet~(I), \susygen~(S) and \pythia~(P)}
 \end{table}
  
 \begin{table}[h] \centering
   \begin{tabular}{|l|c c c|c|c c c|}
   \hdick & & & & & & & \\[-2.ex]
   particle & $m_I$ & $m_S$ & $m_P$ & \ decay \ & ${\cal B}_I$ & ${\cal B}_S$ & ${\cal B}_P$ 
   \\[.5ex] \hdick
   $h^0     $ &  114.0 & 111.8 & 111.7 & $\tau^-   \tau^+            $ & 0.051 & 0.080 & 0.068  \\
              &        &       &       & $b        \bar b            $ & 0.847 & 0.792 & 0.809  \\
              &        &       &       & $c        \bar c            $ & 0.035 &       & 0.043  \\
              &        &       &       & $g         g                $ &       & 0.060 &        \\
              &        &       &       & $W^+       W^-              $ &       & 0.040 &        \\
   \hline
   $H^0     $ &  394.1 & 384.5 & 412.1 & $\tau^-   \tau^+            $ & 0.059 & 0.091 & 0.094  \\
              &        &       &       & $b        \bar b            $ & 0.807 & 0.703 & 0.848  \\
              &        &       &       & $t        \bar t            $ & 0.031 & 0.038 & 0.043  \\
              &        &       &       & $\nt_1    \nt_2             $ & 0.034 &       &        \\
              &        &       &       & $\cp_1    \cm_1             $ &       & 0.042 &        \\
   \hline
   $A^0     $ &  393.6 & 384.6 & 411.9 & $\tau^-   \tau^+            $ & 0.049 & 0.060 & 0.087  \\
              &        &       &       & $b        \bar b            $ & 0.681 & 0.465 & 0.791  \\
              &        &       &       & $t        \bar t            $ & 0.092 & 0.099 & 0.119  \\
              &        &       &       & $\nt_1    \nt_2             $ & 0.065 & 0.082 &        \\
              &        &       &       & $\nt_2    \nt_2             $ & 0.058 & 0.075 &        \\
              &        &       &       & $\cm_1    \cp_1             $ &       & 0.194 &        \\
   \hline
   $H^+     $ &  401.8 & 392.6 & 419.5 & $\nu_\tau \tau^+            $ & 0.077 & 0.093 & 0.102  \\
              &        &       &       & $t        \bar b            $ & 0.770 & 0.727 & 0.895  \\
              &        &       &       & $\cp_1    \nt_1             $ & 0.130 & 0.165 &        \\
   \hline
   \end{tabular}
   \caption{Higgs masses (GeV) and significant branching ratios ($>3\%$) 
            from \isajet~(I), \susygen~(S) and \pythia~(P)}
 \end{table}
  
 \begin{table}[h] \centering
   \begin{tabular}{|l|c c c|c|c c c|}
   \hdick & & & & & & & \\[-2.ex]
   particle & $m_I$ & $m_S$ & $m_P$ & \ decay \ & ${\cal B}_I$ & ${\cal B}_S$ & ${\cal B}_P$ 
   \\[.5ex] \hdick
   $\st_1   $ &  379.1 & 411.6 & 381.3 & $\nt_1    t                 $ & 0.179 & 0.146 & 0.189  \\
              &        &       &       & $\nt_2    t                 $ & 0.095 & 0.115 & 0.078  \\
              &        &       &       & $\cp_1    b                 $ & 0.726 & 0.664 & 0.732  \\
              &        &       &       & $\cp_2    b                 $ &       & 0.075 &        \\
   \hline
   $\st_2   $ &  574.7 & 575.7 & 587.1 & $\cp_1    b                 $ & 0.206 & 0.241 & 0.193  \\
              &        &       &       & $\cp_2    b                 $ & 0.216 & 0.361 & 0.198  \\
              &        &       &       & $Z^0      \st_1             $ & 0.225 &       & 0.270  \\
              &        &       &       & $h^0      \st_1             $ & 0.042 &       & 0.034  \\
              &        &       &       & $\nt_1    t                 $ & 0.030 &       & 0.032  \\
              &        &       &       & $\nt_2    t                 $ & 0.080 & 0.056 & 0.078  \\
              &        &       &       & $\nt_3    t                 $ & 0.033 & 0.163 & 0.036  \\
              &        &       &       & $\nt_4    t                 $ & 0.166 & 0.164 & 0.158  \\
   \hline
   $\sb_1   $ &  491.9 & 520.0 & 504.2 & $\nt_1    b                 $ & 0.062 & 0.061 & 0.035  \\
              &        &       &       & $\nt_2    b                 $ & 0.362 & 0.357 & 0.330  \\
              &        &       &       & $\cm_1    t                 $ & 0.428 & 0.542 & 0.401  \\
              &        &       &       & $W^-      \st_1             $ & 0.133 &       & 0.224  \\
   \hline
   $\sb_2   $ &  524.6 & 550.4 & 534.6 & $\nt_1    b                 $ & 0.148 & 0.196 & 0.242  \\
              &        &       &       & $\nt_2    b                 $ & 0.171 & 0.118 & 0.192  \\
              &        &       &       & $\nt_3    b                 $ & 0.053 & 0.181 &        \\
              &        &       &       & $\nt_4    b                 $ & 0.072 & 0.205 &        \\
              &        &       &       & $\cm_1    t                 $ & 0.213 & 0.153 & 0.255  \\
              &        &       &       & $\cm_2    t                 $ &       & 0.147 &        \\
              &        &       &       & $W^-      \st_1             $ & 0.344 &       & 0.304  \\
   \hline
   \end{tabular}
   \caption{Light squark masses (GeV) and significant branching ratios ($>3\%$) 
            from \isajet~(I), \susygen~(S) and \pythia~(P)}
 \end{table}

\clearpage

\section{SPS 2 -- mSUGRA scenario}
\setcounter{figure}{0}
\setcounter{table}{0}

\large\boldmath
\hspace{20mm}
\begin{tabular}{|l c|}
  \hline
  $m_0$       & $ 1450~\GeV$ \\ 
  $m_{1/2}$   & $\ 300~\GeV$ \\
  $A_0$       & $\ \ \ \ 0~\GeV$ \\
  $\tan\beta$ & $10$       \\
  ${\rm sign}~\mu$ & $+$ \\
  \hline
\end{tabular} \hspace{10mm}
\begin{tabular}{l}
   {\bf `focus point' scenario} \\
    $m_0 = 2\,m_{1/2} + 800~\GeV$ \\
\end{tabular}
\unboldmath\normalsize
\bigskip

\subsection{Spectrum \& parameters of ISAJET 7.58}

\begin{figure}[h] \centering
  \epsfig{file=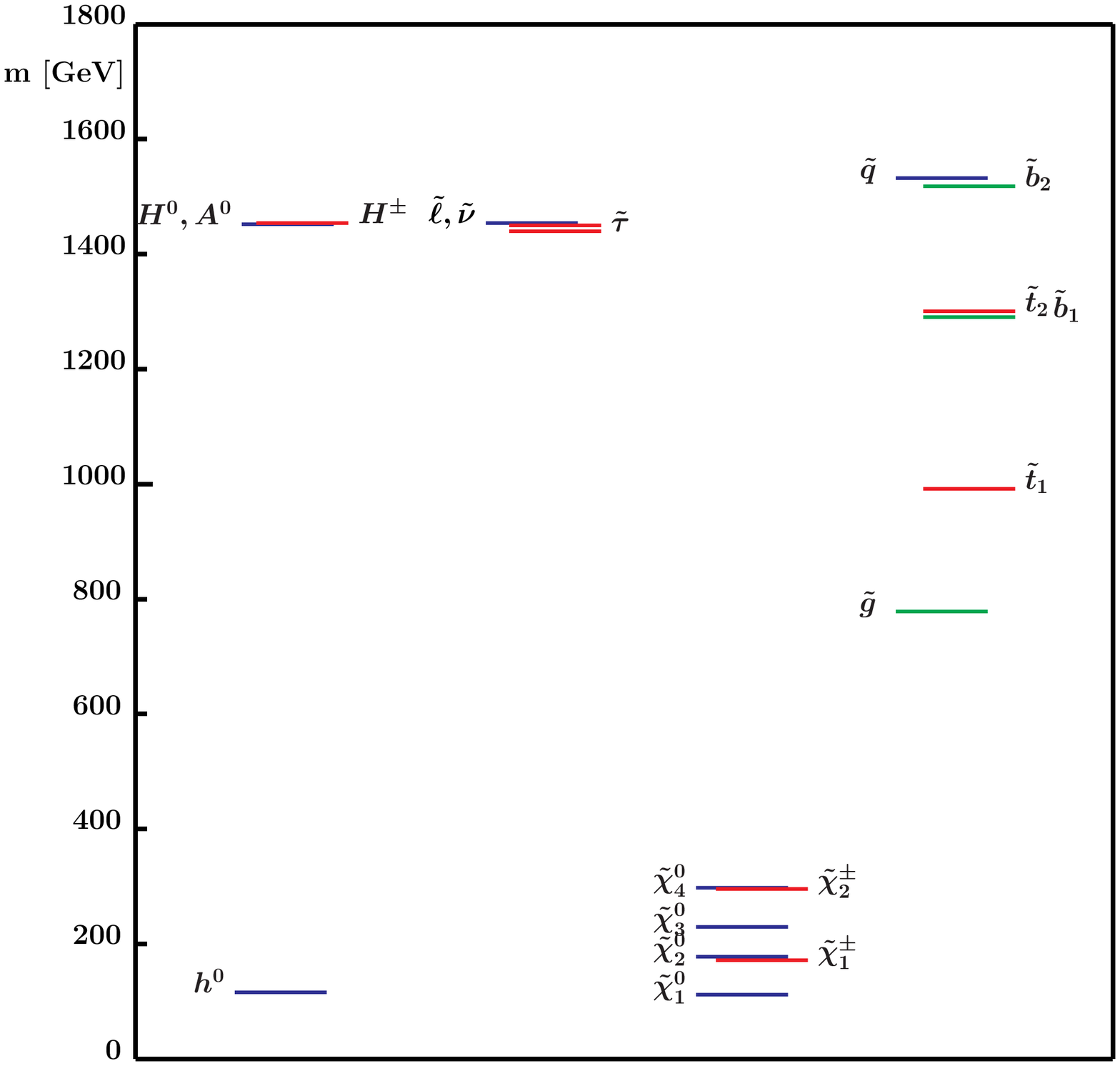,width=.9\textwidth}
  \caption{SPS 2 mass spectrum of \isajet}
\end{figure}

\clearpage
\normalsize
\noindent
{\bf \isajet\ parameters} 
\begin{small}
\begin{verbatim}
 Minimal supergravity (mSUGRA) model:

 M_0,  M_(1/2),  A_0,  tan(beta),  sgn(mu),  M_t =
  1450.000   300.000     0.000    10.000     1.0   175.000

 ISASUGRA unification:
 M_GUT      = 0.270E+17   g_GUT          =0.701      alpha_GUT =0.039
 FT_GUT     = 0.460       FB_GUT         = 0.048     FL_GUT = 0.068

 1/alpha_em =  127.71     sin**2(thetaw) =0.2310     alpha_s   =0.119
 M_1        =  120.36     M_2            =  234.12   M_3       =  696.46
 mu(Q)      =  124.77     B(Q)           = 1653.71   Q         = 1077.13
 M_H1^2     = 0.205E+07   M_H2^2         = 0.319E+05

 ISAJET masses (with signs):
 M(GL)  =   784.37
 M(UL)  =  1532.70   M(UR)  =  1530.08   M(DL)  =  1534.74   M(DR) =  1530.49
 M(B1)  =  1296.56   M(B2)  =  1520.09   M(T1)  =  1003.88   M(T2) =  1307.41
 M(SN)  =  1454.17   M(EL)  =  1456.33   M(ER)  =  1451.69
 M(NTAU)=  1448.15   M(TAU1)=  1439.46   M(TAU2)=  1450.38
 M(Z1)  =   -79.54   M(Z2)  =   135.34   M(Z3)  =  -140.84   M(Z4) =  -269.45
 M(W1)  =  -104.03   M(W2)  =  -269.03
 M(HL)  =   115.71   M(HH)  =  1444.10   M(HA)  =  1442.95   M(H+) =  1446.18

 theta_t=   1.4446   theta_b=   0.0094   theta_l=   1.4909   alpha_h=   0.1007

 NEUTRALINO MASSES (SIGNED) =   -79.537   135.343  -140.839  -269.450
 EIGENVECTOR 1       =   0.45040  -0.64968  -0.26524   0.55200
 EIGENVECTOR 2       =   0.72890   0.65991  -0.14274   0.11336
 EIGENVECTOR 3       =   0.35117  -0.31805  -0.32663  -0.81782
 EIGENVECTOR 4       =  -0.37752   0.20316  -0.89587   0.11668

 CHARGINO MASSES (SIGNED)  =  -104.031  -269.026
 GAMMAL, GAMMAR             =   2.85552   2.58052


 ISAJET equivalent input:
 MSSMA:   784.37  124.77 1442.95   10.00
 MSSMB:  1533.62 1530.29 1530.49 1455.57 1451.04
 MSSMC:  1295.25 1519.86  998.47 1449.56 1438.88 -563.70 -797.21 -187.83
 MSSMD: SAME AS MSSMB (DEFAULT)
 MSSME:   120.36  234.12

\end{verbatim}
\end{small}

\clearpage
\subsection{Spectrum \& parameters of SUSYGEN 3.00/27}

\begin{figure}[h!] \centering
  \epsfig{file=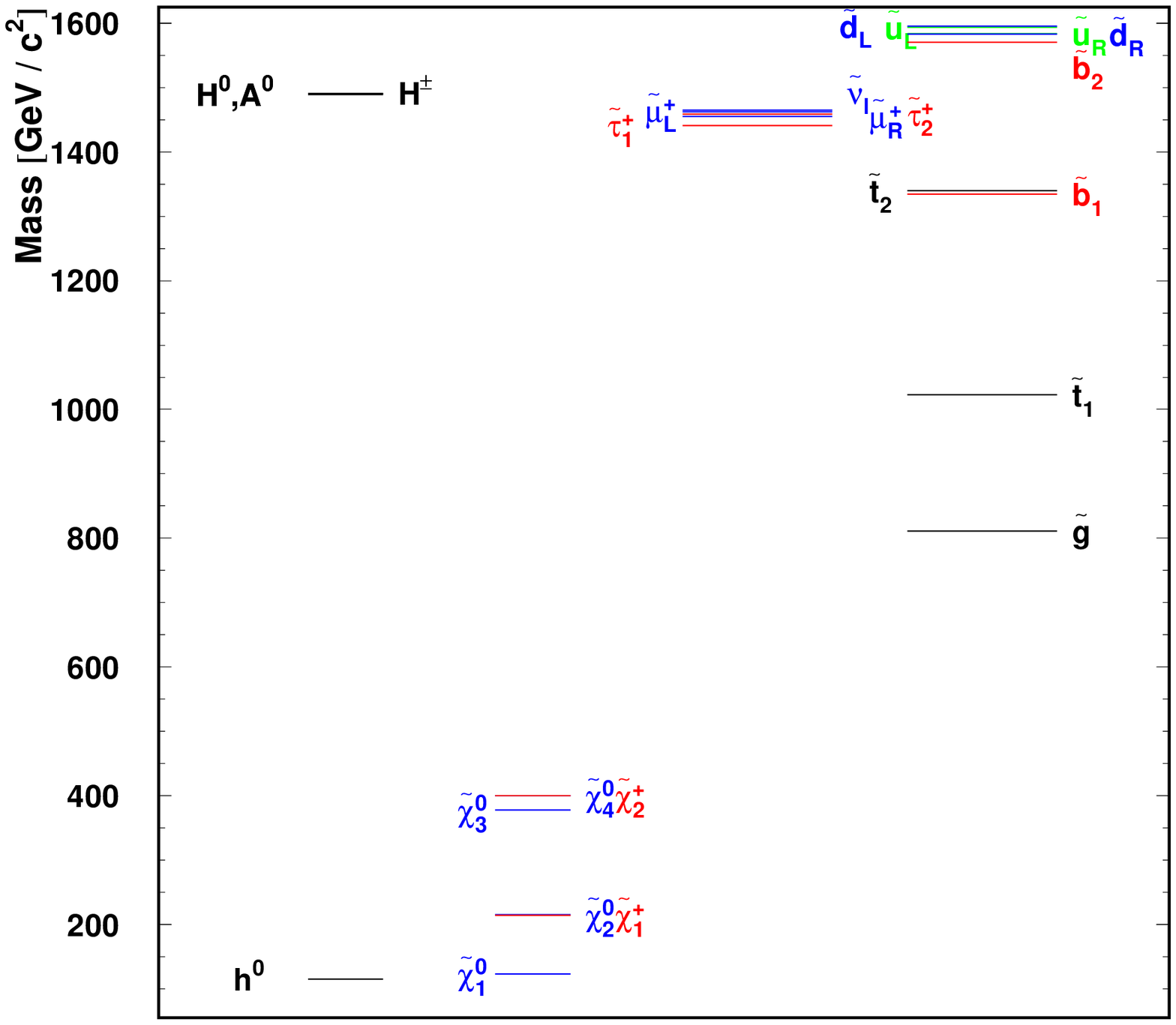,width=.9\textwidth}
  \caption{SPS 2 mass spectrum of \susygen}
\end{figure}

\newpage
\normalsize
\noindent
{\bf \susygen\  parameters} 
\begin{small}
\begin{verbatim}
 Susygen inputs:
 --------------

 m0      =  1450.000   TANB     =    10.000
 m1/2    =   300.000    mu/|mu| =     1
 A0      =     0.000

 Sparticle masses:
 ----------------

 SUPR     1584.  SUPL      1593.
 SDNR     1583.  SDNL      1595.
 SELR     1455.  SELL      1465.
 SNU      1463.
 STP1     1022.  STP2      1340. cosmix =  0.103
 SBT1     1335.  SBT2      1571. cosmix =  1.000
 STA1     1441.  STA2      1459. cosmix=   0.140
 SGLU      811.

 Gaugino masses:
 --------------

 NEUTRALINO m, CP, ph/zi/ha/hb 1 =  123.1  1.  0.837 -0.523  0.077  0.142
 NEUTRALINO m, CP, ph/zi/ha/hb 2 =  214.6  1. -0.543 -0.763  0.218  0.276
 NEUTRALINO m, CP, ph/zi/ha/hb 3 =  377.6 -1. -0.009  0.100 -0.637  0.764
 NEUTRALINO m, CP, ph/zi/ha/hb 4 =  399.9  1. -0.067 -0.368 -0.735 -0.566

 CHARGINO MASSES    =   213.915   399.841
 CHARGINO ETA      =    -1.000     1.000

 U matrix      WINO      HIGGSINO    V matrix      WINO      HIGGSINO 
 W1SS+        -0.901     0.434       W1SS-         0.960    -0.279
 W2SS+         0.434     0.901       W2SS-         0.279     0.960

 Higgses masses: 
 --------------

 Light CP-even Higgs =   114.873
 Heavy CP-even Higgs =  1490.196
       CP-odd  Higgs =  1489.165
       Charged Higgs =  1491.171
       sin(a-b)      =    -0.100
       cos(a-b)      =     0.995

\end{verbatim}
\end{small}

\clearpage

\subsection{Spectrum \& parameters of  PYTHIA 6.2/00}

\begin{figure}[h] \centering
  \epsfig{file=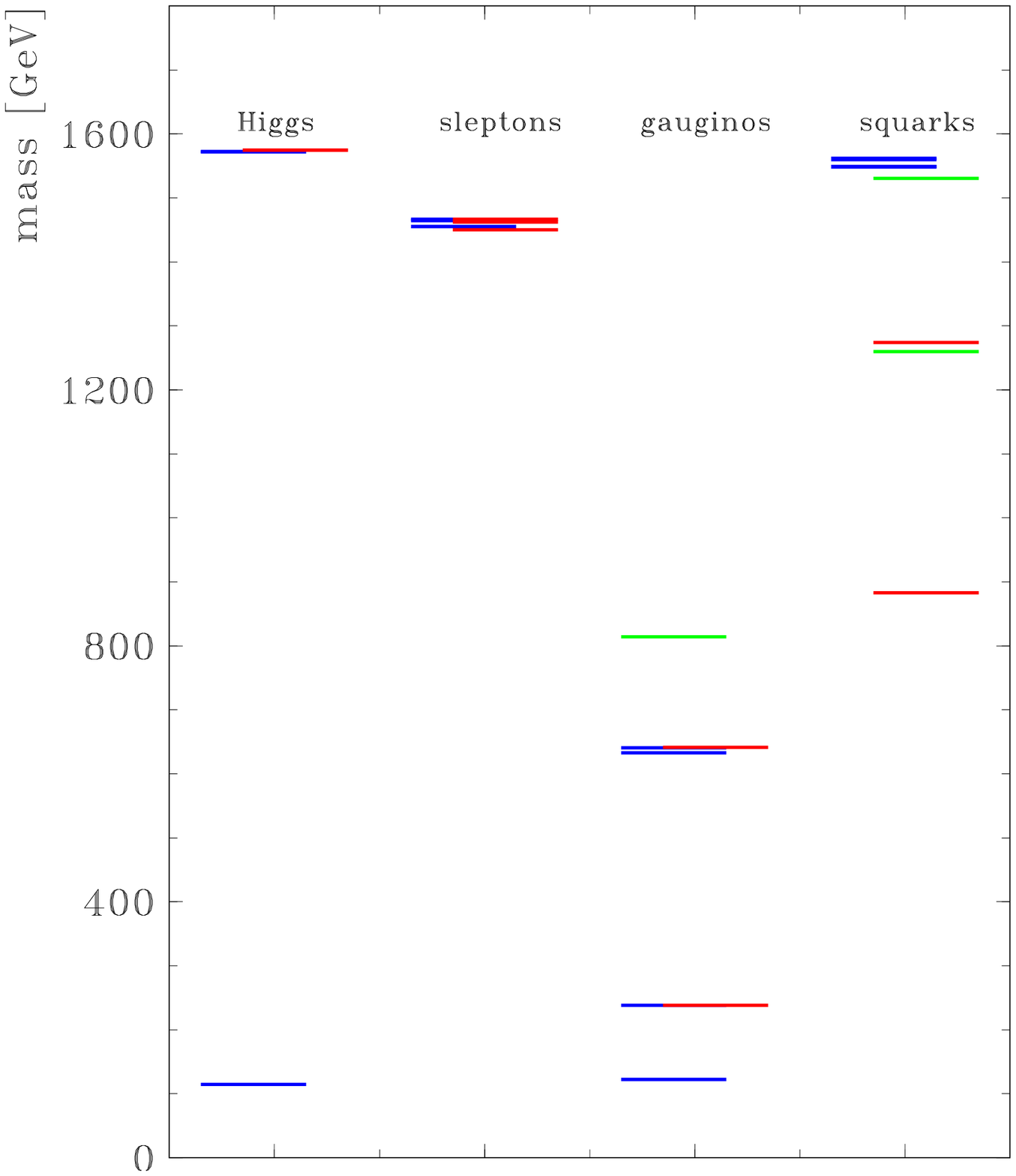,width=.9\textwidth}
  \caption{SPS 2 mass spectrum of \pythia}
\end{figure}
\newpage
\normalsize
\noindent
{\bf \pythia\  parameters} 
\begin{small}
\begin{verbatim}
   SUGRA input parameters
   ----------------------
   m_0         RMSS( 8) =   1450.    
   m_1/2       RMSS( 1) =   300.0    
   A_0         RMSS(16) =   0.000    
   tan_beta    RMSS( 5) =   10.00    
   sign mu     RMSS( 4) =   1.000    

   sparticle masses & widths
   -------------------------
   M_se_R  1455.3 ( 7.470)   M_se_L  1466.8 (19.692)   M_sne_L 1464.6 (19.753)
   M_sm_R  1455.3 ( 7.471)   M_sm_L  1466.8 (19.693)   M_snm_L 1464.6 (19.754)
   M_st_1  1450.0 ( 8.659)   M_st_2  1466.1 (19.071)   M_snt_L 1462.2 (19.923)

   M_ch0_1  122.0 ( 0.000)   M_ch0_2  238.1 ( 0.001)   M_ch0_3  632.3 ( 4.785)
   M_ch0_4  640.3 ( 4.759)
   M_ch+_1  238.0 ( 0.002)   M_ch+_2  641.2 ( 4.815)

   M_h0     114.1 ( 0.004)   M_H0    1572.3 ( 3.711)   M_A0    1571.6 ( 3.733)
   M_H+    1574.6 ( 3.802)

   M_g~     814.1 ( 0.004)
   M_uL    1559.7 (68.025)   M_uR    1549.1 (51.354)   M_dL    1561.7 (68.063)
   M_dR    1548.3 (48.633)
   M_b1    1259.8 (55.517)   M_b2    1530.2 (47.344)   M_t1     882.7 ( 8.311)
   M_t2    1274.0 (52.448)

   parameter settings IMSS, RMSS
   -----------------------------
   IMSS( 1) = 2   IMSS( 4) = 1   IMSS( 7) = 0   IMSS(10) = 0
   IMSS( 2) = 0   IMSS( 5) = 0   IMSS( 8) = 0   IMSS(11) = 0
   IMSS( 3) = 0   IMSS( 6) = 0   IMSS( 9) = 0   IMSS(12) = 0

   RMSS( 1) =   300.0       RMSS( 9) =   700.0       RMSS(17) =   0.000    
   RMSS( 2) =   244.9       RMSS(10) =   1259.       RMSS(18) = -0.1004    
   RMSS( 3) =   706.3       RMSS(11) =   1530.       RMSS(19) =   1573.    
   RMSS( 4) =   628.3       RMSS(12) =   876.4       RMSS(20) =  0.4100E-01
   RMSS( 5) =   10.00       RMSS(13) =   1464.       RMSS(21) =   1.000    
   RMSS( 6) =   1467.       RMSS(14) =   1450.       RMSS(22) =   800.0    
   RMSS( 7) =   1455.       RMSS(15) =   0.000       RMSS(23) =  0.1000E+05
   RMSS( 8) =   1450.       RMSS(16) =  -610.2       RMSS(24) =  0.1000E+05

\end{verbatim}
\end{small}

\clearpage

\subsection{Decay modes}
\normalsize
 \begin{table}[h] \centering
   \vskip -0.3cm
   \begin{tabular}{|l|c c c|c|c c c|}
   \hdick & & & & & & & \\[-2.ex]
   particle & $m_I$ & $m_S$ & $m_P$ & \ decay \ & ${\cal B}_I$ & ${\cal B}_S$ & ${\cal B}_P$ 
   \\[.5ex] \hdick
   $\ser    $ & 1451.7 & 1455.1 & 1455.3 & $\nt_1    e^-               $ & 0.308 & 0.973 & 0.995  \\
              &        &        &        & $\nt_3    e^-               $ & 0.667 &       &        \\
   \hline
   $\sel    $ & 1456.3 & 1464.9 & 1466.8 & $\nt_1    e^-               $ &       & 0.074 &        \\
              &        &        &        & $\nt_2    e^-               $ &       & 0.300 & 0.307  \\
              &        &        &        & $\nt_3    e^-               $ & 0.190 &       &        \\
              &        &        &        & $\nt_4    e^-               $ & 0.207 &       &        \\
              &        &        &        & $\cm_1    \nu_e             $ & 0.051 & 0.498 & 0.582  \\
              &        &        &        & $\cm_2    \nu_e             $ & 0.551 & 0.104 &        \\
   \hline
   $\sne    $ & 1454.2 & 1462.7 & 1464.6 & $\nt_1    \nu_e             $ & 0.103 & 0.112 & 0.101 \\
              &        &        &        & $\nt_2    \nu_e             $ &       & 0.231 & 0.285 \\
              &        &        &        & $\nt_4    \nu_e             $ & 0.274 & 0.048 &       \\
              &        &        &        & $\cp_1    e^-               $ & 0.179 & 0.563 & 0.601 \\
              &        &        &        & $\cp_2    e^-               $ & 0.426 & 0.043 &       \\
   \hline
   $\smur   $ & 1451.7 & 1455.1 & 1455.3 & $\nt_1    \mu^-             $ & 0.308 & 0.973 & 0.995 \\
              &        &        &        & $\nt_3    \mu^-             $ & 0.667 &       &       \\
   \hline
   $\smul   $ & 1456.3 & 1464.9 & 1466.8 & $\nt_1    \mu^-             $ &       & 0.074 &       \\
              &        &        &        & $\nt_2    \mu^-             $ &       & 0.300 & 0.307 \\
              &        &        &        & $\nt_3    \mu^-             $ & 0.190 &       &       \\
              &        &        &        & $\nt_4    \mu^-             $ & 0.207 &       &       \\
              &        &        &        & $\cm_1    \nu_\mu           $ & 0.051 & 0.498 & 0.582 \\
              &        &        &        & $\cm_2    \nu_\mu           $ & 0.551 & 0.104 &       \\
   \hline
   $\snm    $ & 1454.2 & 1462.7 & 1464.6 & $\nt_1    \nu_\mu           $ & 0.103 & 0.112 & 0.101 \\
              &        &        &        & $\nt_2    \nu_\mu           $ &       & 0.231 & 0.285 \\
              &        &        &        & $\nt_4    \nu_\mu           $ & 0.274 & 0.049 &       \\
              &        &        &        & $\cp_1    \mu^-             $ & 0.179 & 0.563 & 0.601 \\
              &        &        &        & $\cp_2    \mu^-             $ & 0.426 & 0.043 &       \\
   \hline
   $\stau_1 $ & 1439.5 & 1441.3 & 1450.0 & $\nt_1    \tau^-            $ & 0.306 & 0.869 & 0.819 \\
              &        &        &        & $\nt_2    \tau^-            $ &       & 0.039 & 0.054 \\
              &        &        &        & $\nt_3    \tau^-            $ & 0.613 &       &       \\
              &        &        &        & $\cm_1    \nu_\tau          $ & 0.045 & 0.055 & 0.102 \\
   \hline
   $\stau_2 $ & 1450.4 & 1458.5 & 1466.1 & $\nt_1    \tau^-            $ &       & 0.079 & 0.108 \\
              &        &        &        & $\nt_2    \tau^-            $ &       & 0.291 & 0.293 \\
              &        &        &        & $\nt_3    \tau^-            $ & 0.190 &       &       \\
              &        &        &        & $\nt_4    \tau^-            $ & 0.206 & 0.033 &       \\
              &        &        &        & $\cm_1    \nu_\tau          $ & 0.046 & 0.478 & 0.554 \\
              &        &        &        & $\cm_2    \nu_\tau          $ & 0.545 & 0.111 &       \\
   \hline
   $\snt    $ & 1448.2 & 1456.0 & 1464.4 & $\nt_1    \nu_\tau          $ & 0.101 & 0.112 & 0.100 \\
              &        &        &        & $\nt_2    \nu_\tau          $ &       & 0.231 & 0.282 \\
              &        &        &        & $\nt_4    \nu_\tau          $ & 0.270 & 0.048 &       \\
              &        &        &        & $\cp_1    \tau^-            $ & 0.190 & 0.563 & 0.595 \\
              &        &        &        & $\cp_2    \tau^-            $ & 0.421 & 0.042 &       \\
   \hline
   \end{tabular}
   \caption{Slepton masses (GeV) and significant branching ratios ($>3\%$) 
            from \isajet~(I), \susygen~(S) and \pythia~(P)}
 \end{table}
  
 \begin{table}[h] \centering
   \begin{tabular}{|l|c c c|c|c c c|}
   \hdick & & & & & & & \\[-2.ex]
   particle & $m_I$ & $m_S$ & $m_P$ & \ decay \ & ${\cal B}_I$ & ${\cal B}_S$ & ${\cal B}_P$ 
   \\[.5ex] \hdick
   $\nt_1   $ &   79.5 & 123.1  & 122.0  & $                           $ & 1.000 & 1.000 & 1.000 \\
   \hline
   $\nt_2   $ &  135.3 & 214.6  & 238.1  & $\cm_1    u        \bar d   $ & 0.033 &       &       \\
              &        &        &        & $\cp_1    d        \bar u   $ & 0.033 &       &       \\
              &        &        &        & $\cm_1    c        \bar s   $ & 0.033 &       &       \\
              &        &        &        & $\cp_1    s        \bar c   $ & 0.033 &       &       \\
              &        &        &        & $\nt_1    u        \bar u   $ & 0.096 &       &       \\
              &        &        &        & $\nt_1    d        \bar d   $ & 0.124 &       &       \\
              &        &        &        & $\nt_1    s        \bar s   $ & 0.124 &       &       \\
              &        &        &        & $\nt_1    c        \bar c   $ & 0.096 &       &       \\
              &        &        &        & $\nt_1    b        \bar b   $ & 0.114 &       &       \\
              &        &        &        & $\nt_1    \nu_e    \anue    $ & 0.056 &       &       \\
              &        &        &        & $\nt_1    \nu_\mu  \anum    $ & 0.056 &       &       \\
              &        &        &        & $\nt_1    \nu_\tau \anut    $ & 0.056 &       &       \\
              &        &        &        & $\nt_1    Z^0               $ &       & 1.000 & 0.223 \\
              &        &        &        & $\nt_1    h^0               $ &       &       & 0.777 \\
   \hline
   $\nt_3   $ &  140.8 & 377.6  & 632.3  & $\cm_1    u        \bar d   $ & 0.155 &       &       \\
              &        &        &        & $\cm_1    \nu_e    e^+      $ & 0.052 &       &       \\
              &        &        &        & $\cm_1    \nu_\mu  \mu^+    $ & 0.052 &       &       \\
              &        &        &        & $\cp_1    d        \bar u   $ & 0.155 &       &       \\
              &        &        &        & $\cp_1    e^-      \anue    $ & 0.052 &       &       \\
              &        &        &        & $\cp_1    \mu^-    \anum    $ & 0.052 &       &       \\
              &        &        &        & $\cm_1    c        \bar s   $ & 0.155 &       &       \\
              &        &        &        & $\cm_1    \nu_\tau \tau^+   $ & 0.052 &       &       \\
              &        &        &        & $\cp_1    s        \bar c   $ & 0.155 &       &       \\
              &        &        &        & $\cp_1    \tau^-   \anut    $ & 0.052 &       &       \\
              &        &        &        & $\nt_1    Z^0               $ &       & 0.094 & 0.075 \\
              &        &        &        & $\nt_2    Z^0               $ &       & 0.151 & 0.253 \\
              &        &        &        & $\nt_2    h^0               $ &       &       & 0.033 \\
              &        &        &        & $\cp_1    W^-               $ &       & 0.369 & 0.309 \\
              &        &        &        & $\cm_1    W^+               $ &       & 0.369 & 0.309 \\
   \hline
   $\nt_4   $ &  269.4 & 399.9  & 640.3  & $\cp_1    W^-               $ & 0.371 & 0.391 & 0.314 \\
              &        &        &        & $\cm_1    W^+               $ & 0.371 & 0.391 & 0.314 \\
              &        &        &        & $\nt_2    Z^0               $ & 0.152 &       & 0.038 \\
              &        &        &        & $\nt_1    h^0               $ & 0.033 & 0.069 & 0.068 \\
              &        &        &        & $\nt_2    h^0               $ &       & 0.119 & 0.242 \\
              &        &        &        & $\nt_3    h^0               $ & 0.050 &       &       \\
   \hline
   \end{tabular}
   \caption{Neutralino masses (GeV) and significant branching ratios ($>3\%$) 
            from \isajet~(I), \susygen~(S) and \pythia~(P)}
 \end{table}
  
 \begin{table}[h] \centering
   \begin{tabular}{|l|c c c|c|c c c|}
   \hdick & & & & & & & \\[-2.ex]
   particle & $m_I$ & $m_S$ & $m_P$ & \ decay \ & ${\cal B}_I$ & ${\cal B}_S$ & ${\cal B}_P$ 
   \\[.5ex] \hdick
   $\cp_1   $ &  104.0 & 213.9  & 238.0  & $\nt_1    u        \bar d   $ & 0.333 &       &       \\
              &        &        &        & $\nt_1    c        \bar s   $ & 0.333 &       &       \\
              &        &        &        & $\nt_1    e^+      \nu_e    $ & 0.111 &       &       \\
              &        &        &        & $\nt_1    \mu^+    \nu_\mu  $ & 0.111 &       &       \\
              &        &        &        & $\nt_1    \tau^+   \nu_\tau $ & 0.111 &       &       \\
              &        &        &        & $\nt_1    W^+               $ &       & 1.000 & 1.000 \\
   \hline
   $\cp_2   $ &  269.0 & 399.8  & 641.2  & $\nt_1    W^+               $ & 0.126 & 0.084 & 0.085 \\
              &        &        &        & $\nt_2    W^+               $ & 0.250 & 0.378 & 0.328 \\
              &        &        &        & $\nt_3    W^+               $ & 0.187 &       &       \\
              &        &        &        & $\cp_1    Z^0               $ & 0.297 & 0.537 & 0.307 \\
              &        &        &        & $\cp_1    h^0               $ & 0.141 &       & 0.280 \\
   \hline
   \end{tabular}
   \caption{Chargino masses (GeV) and significant branching ratios ($>3\%$) 
            from \isajet~(I), \susygen~(S) and \pythia~(P)}
 \end{table}
  
 \begin{table}[h] \centering
   \begin{tabular}{|l|c c c|c|c c c|}
   \hdick & & & & & & & \\[-2.ex]
   particle & $m_I$ & $m_S$ & $m_P$ & \ decay \ & ${\cal B}_I$ & ${\cal B}_S$ & ${\cal B}_P$ 
   \\[.5ex] \hdick
   $h^0     $ &  115.7 & 114.9  &  114.1 & $\tau^-   \tau^+            $ & 0.049 & 0.076 & 0.065 \\
              &        &        &        & $b        \bar b            $ & 0.819 & 0.747 & 0.774 \\
              &        &        &        & $c        \bar c            $ & 0.040 &       & 0.048 \\
              &        &        &        & $g        g                 $ & 0.031 & 0.071 & 0.040 \\
              &        &        &        & $W^+      W^-               $ &       & 0.070 & 0.061 \\
   \hline
   $H^0     $ & 1444.1 & 1490.2 & 1572.3 & $b        \bar b            $ & 0.149 & 0.094 & 0.724 \\
              &        &        &        & $t        \bar t            $ & 0.032 & 0.033 & 0.184 \\
              &        &        &        & $\nt_1    \nt_1             $ & 0.060 &       &       \\
              &        &        &        & $\nt_1    \nt_4             $ & 0.055 &       &       \\
              &        &        &        & $\nt_2    \nt_2             $ &       & 0.033 &       \\
              &        &        &        & $\nt_2    \nt_3             $ &       & 0.114 &       \\
              &        &        &        & $\nt_2    \nt_4             $ & 0.133 &       &       \\
              &        &        &        & $\cp_1    \cm_1             $ & 0.118 & 0.082 &       \\
              &        &        &        & $\cp_1    \cm_2             $ & 0.177 & 0.211 &       \\
              &        &        &        & $\cm_1    \cp_2             $ & 0.177 & 0.211 &       \\
   \hline
   $A^0     $ & 1443.0 & 1489.2 & 1571.6 & $b        \bar b            $ & 0.151 & 0.094 & 0.720 \\
              &        &        &        & $t        \bar t            $ & 0.033 & 0.034 & 0.190 \\
              &        &        &        & $\nt_1    \nt_1             $ & 0.080 &       &       \\
              &        &        &        & $\nt_1    \nt_4             $ & 0.069 &       &       \\
              &        &        &        & $\nt_2    \nt_2             $ &       & 0.046 &       \\
              &        &        &        & $\nt_2    \nt_4             $ & 0.086 & 0.064 &       \\
              &        &        &        & $\nt_3    \nt_2             $ &       & 0.062 &       \\
              &        &        &        & $\nt_3    \nt_4             $ & 0.033 &       &       \\
              &        &        &        & $\nt_4    \nt_4             $ &       & 0.044 &       \\
              &        &        &        & $\cp_1    \cm_1             $ & 0.143 & 0.114 &       \\
              &        &        &        & $\cp_2    \cm_2             $ & 0.035 & 0.044 &       \\
              &        &        &        & $\cp_1    \cm_2             $ & 0.152 & 0.181 &       \\
              &        &        &        & $\cm_1    \cp_2             $ & 0.152 & 0.181 &       \\
   \hline
   $H^+     $ & 1446.2 & 1491.2 & 1574.6 & $t        \bar b            $ & 0.167 & 0.125 & 0.912 \\
              &        &        &        & $\cp_1    \nt_1             $ &       & 0.045 &       \\
              &        &        &        & $\cp_1    \nt_2             $ & 0.030 &       &       \\
              &        &        &        & $\cp_1    \nt_3             $ & 0.098 & 0.219 &       \\
              &        &        &        & $\cp_1    \nt_4             $ & 0.191 & 0.229 &       \\
              &        &        &        & $\cp_2    \nt_1             $ & 0.186 & 0.034 &       \\
              &        &        &        & $\cp_2    \nt_2             $ & 0.186 & 0.295 &       \\
              &        &        &        & $\cp_2    \nt_3             $ & 0.104 & 0.034 &       \\
   \hline
   \end{tabular}
   \caption{Higgs masses (GeV) and significant branching ratios ($>3\%$) 
            from \isajet~(I), \susygen~(S) and \pythia~(P)}
 \end{table}
  
 \begin{table}[h] \centering
   \begin{tabular}{|l|c c c|c|c c c|}
   \hdick & & & & & & & \\[-2.ex]
   particle & $m_I$ & $m_S$ & $m_P$ & \ decay \ & ${\cal B}_I$ & ${\cal B}_S$ & ${\cal B}_P$ 
   \\[.5ex] \hdick
   $\st_1   $ & 1003.9 & 1022.4 & 882.7  & $\sg      t                 $ & 0.119 & 0.233 &       \\
              &        &        &        & $\nt_1    t                 $ & 0.112 & 0.053 & 0.214 \\
              &        &        &        & $\nt_2    t                 $ & 0.225 &       &       \\
              &        &        &        & $\nt_3    t                 $ & 0.089 & 0.172 & 0.178 \\
              &        &        &        & $\nt_4    t                 $ & 0.036 & 0.142 & 0.113 \\
              &        &        &        & $\cp_1    b                 $ & 0.342 & 0.055 & 0.059 \\
              &        &        &        & $\cp_2    b                 $ & 0.077 & 0.321 & 0.409 \\
   \hline
   $\st_2   $ & 1307.4 & 1340.1 & 1274.0 & $\sg      t                 $ & 0.456 & 0.376 & 0.420 \\
              &        &        &        & $\cp_1    b                 $ & 0.034 & 0.140 & 0.183 \\
              &        &        &        & $\cp_2    b                 $ & 0.129 & 0.202 &       \\
              &        &        &        & $\nt_1    t                 $ & 0.051 &       &       \\
              &        &        &        & $\nt_2    t                 $ & 0.145 & 0.051 & 0.088 \\
              &        &        &        & $\nt_3    t                 $ & 0.052 & 0.113 & 0.112 \\
              &        &        &        & $\nt_4    t                 $ & 0.109 & 0.117 & 0.128 \\
   \hline
   $\sb_1   $ & 1296.6 & 1334.9 & 1259.8 & $\nt_2    b                 $ &       & 0.080 & 0.087 \\
              &        &        &        & $\nt_4    b                 $ & 0.063 &       &       \\
              &        &        &        & $\sg      b                 $ & 0.468 & 0.670 & 0.464 \\
              &        &        &        & $\cm_1    t                 $ & 0.207 & 0.151 & 0.167 \\
              &        &        &        & $\cm_2    t                 $ & 0.216 & 0.057 & 0.224 \\
              &        &        &        & $W^-      \st_1             $ &       &       & 0.051 \\
   \hline
   $\sb_2   $ & 1520.1 & 1570.6 & 1530.2 & $\sg      b                 $ & 0.948 & 0.921 & 0.980 \\
   \hline
   \end{tabular}
   \caption{Light squark masses (GeV) and significant branching ratios ($>3\%$) 
            from \isajet~(I), \susygen~(S) and \pythia~(P)}
 \end{table}

\clearpage

\section{SPS 3 -- mSUGRA scenario}
\setcounter{figure}{0}
\setcounter{table}{0}

\large\boldmath
\hspace{20mm}
\begin{tabular}{|l c|}
  \hline
  $m_0$       & $\  90~\GeV$ \\ 
  $m_{1/2}$   & $ 400~\GeV$ \\
  $A_0$       & $ \ \ \ 0~\GeV$ \\
  $\tan\beta$ & $10$       \\
  ${\rm sign}~\mu$ & $+$ \\
  \hline
\end{tabular} \hspace{10mm}
\begin{tabular}{l}
   {\bf model line into} \\
   {\bf 'coannihilation region'} \\
    $m_0 = 0.25\,m_{1/2} - 10~\GeV$ \\
\end{tabular}
\unboldmath\normalsize
\bigskip

\subsection{Spectrum \& parameters of ISAJET 7.58}

\begin{figure}[h] \centering
  \epsfig{file=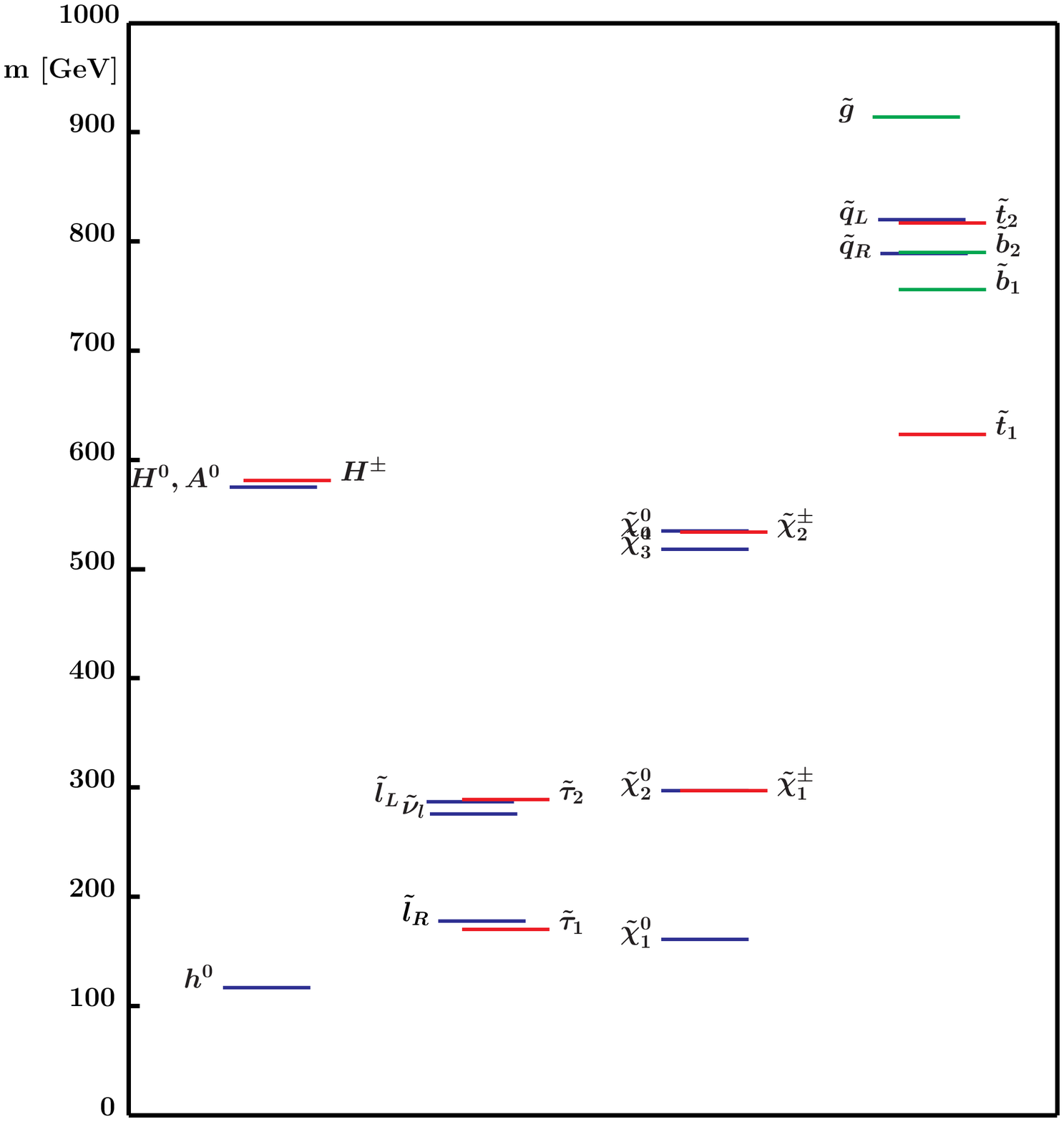,width=.9\textwidth}
  \caption{SPS 3 mass spectrum of \isajet}
\end{figure}

\clearpage
\normalsize
\noindent
{\bf \isajet\ parameters} 
\begin{small}
\begin{verbatim}
 Minimal supergravity (mSUGRA) model:

 M_0,  M_(1/2),  A_0,  tan(beta),  sgn(mu),  M_t =
    90.000   400.000     0.000    10.000     1.0   175.000

 ISASUGRA unification:
 M_GUT      = 0.166E+17   g_GUT          =0.711      alpha_GUT =0.040
 FT_GUT     = 0.473       FB_GUT         = 0.047     FL_GUT = 0.070

 1/alpha_em =  127.72     sin**2(thetaw) =0.2304     alpha_s   =0.121
 M_1        =  162.83     M_2            =  311.38   M_3       =  894.68
 mu(Q)      =  508.59     B(Q)           =   64.52   Q         =  703.81
 M_H1^2     = 0.700E+05   M_H2^2         =-0.248E+06

 ISAJET masses (with signs):
 M(GL)  =   914.26
 M(UL)  =   816.57   M(UR)  =   791.78   M(DL)  =   820.39   M(DR) =   789.34
 M(B1)  =   757.50   M(B2)  =   791.35   M(T1)  =   623.83   M(T2) =   819.54
 M(SN)  =   275.99   M(EL)  =   287.11   M(ER)  =   178.33
 M(NTAU)=   275.11   M(TAU1)=   170.59   M(TAU2)=   289.22
 M(Z1)  =  -160.55   M(Z2)  =  -296.95   M(Z3)  =   512.87   M(Z4) =  -529.57
 M(W1)  =  -296.85   M(W2)  =  -529.51
 M(HL)  =   116.95   M(HH)  =   573.03   M(HA)  =   572.42   M(H+) =   578.30

 theta_t=   1.0446   theta_b=   0.4105   theta_l=   1.3960   alpha_h=   0.1052

 NEUTRALINO MASSES (SIGNED) =  -160.555  -296.949   512.872  -529.573
 EIGENVECTOR 1       =   0.04140  -0.10275  -0.02726   0.99347
 EIGENVECTOR 2       =   0.15060  -0.23308  -0.95905  -0.05670
 EIGENVECTOR 3       =   0.70876   0.70153  -0.06164   0.04133
 EIGENVECTOR 4       =   0.68794  -0.66556   0.27510  -0.08996

 CHARGINO MASSES (SIGNED)  =  -296.846  -529.512
 GAMMAL, GAMMAR             =   1.91245   1.78958


 ISAJET equivalent input:
 MSSMA:   914.26  508.59  572.42   10.00
 MSSMB:   818.29  788.94  792.57  283.27  172.98
 MSSMC:   760.72  785.64  661.24  282.42  170.03 -733.51-1042.16 -246.11
 MSSMD: SAME AS MSSMB (DEFAULT)
 MSSME:   162.83  311.38

\end{verbatim}
\end{small}

\clearpage
\subsection{Spectrum \& parameters of SUSYGEN 3.00/27}

\begin{figure}[h] \centering
  \epsfig{file=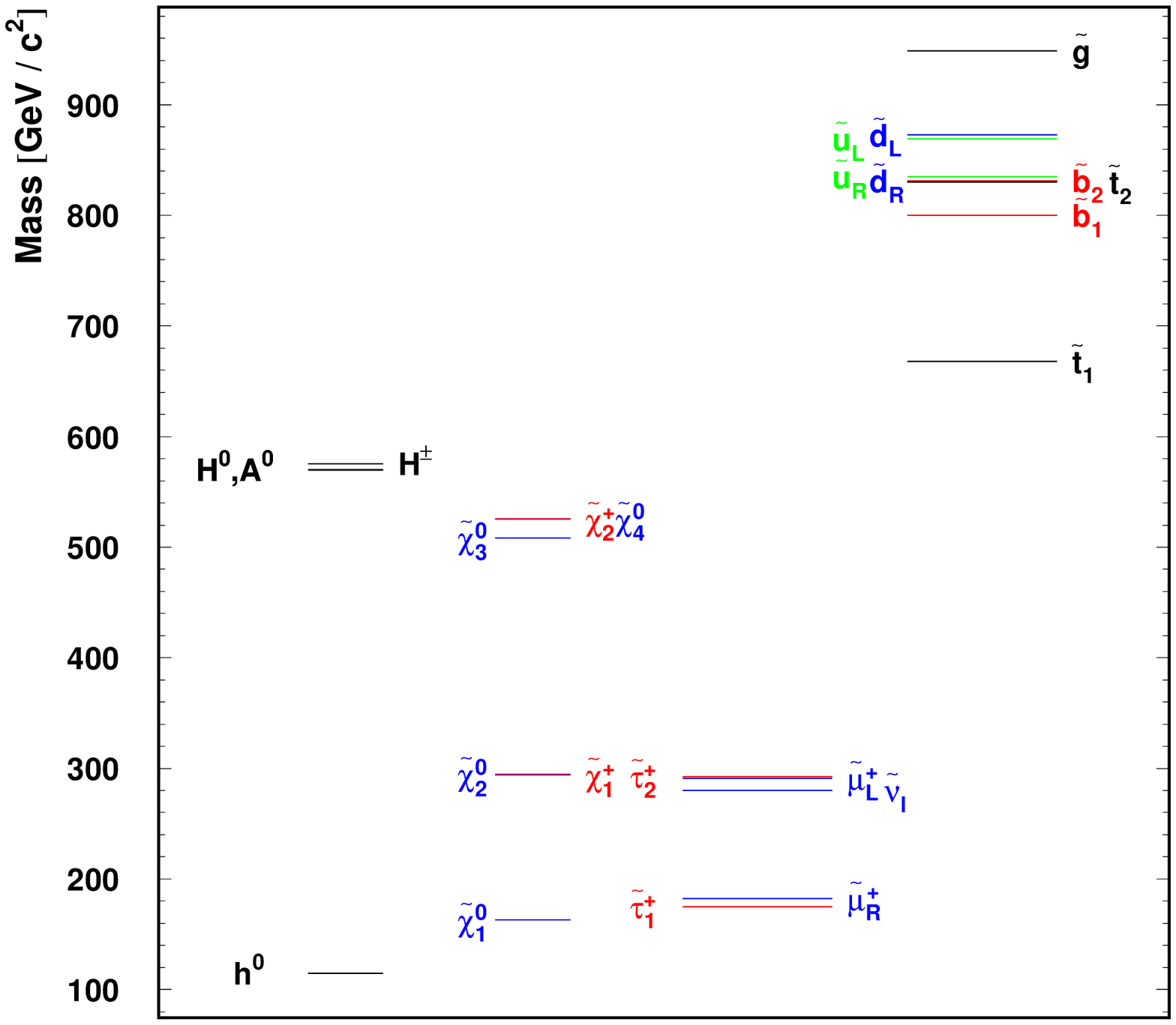,width=.9\textwidth}
  \caption{SPS 3 mass spectrum of \susygen}
\end{figure}
\newpage
\normalsize
\noindent
{\bf \susygen\  parameters} 
\begin{small}
\begin{verbatim}
 Susygen inputs:
 --------------

 m0      =    90.000   TANB     =    10.000
 m1/2    =   400.000    mu/|mu| =     1
 A0      =     0.000

 Sparticle masses: 
 ----------------

 SUPR      835.  SUPL       869.
 SDNR      831.  SDNL       873.
 SELR      182.  SELL       291.
 SNU       280.
 STP1      668.  STP2       830. cosmix =  0.427
 SBT1      800.  SBT2       831. cosmix =  0.929
 STA1      175.  STA2       293. cosmix=   0.170
 SGLU      949.

 Gaugino masses:
 --------------

 M1 =    165.439 M2 =    309.159 M3 =    910.421

 NEUTRALINO m, CP, ph/zi/ha/hb 1 =  163.1  1.  0.857 -0.503  0.053  0.100
 NEUTRALINO m, CP, ph/zi/ha/hb 2 =  294.5  1. -0.513 -0.811  0.175  0.219
 NEUTRALINO m, CP, ph/zi/ha/hb 3 =  508.5 -1. -0.006  0.075 -0.635  0.768
 NEUTRALINO m, CP, ph/zi/ha/hb 4 =  525.5  1. -0.053 -0.288 -0.750 -0.593

 CHARGINO MASSES    =   294.202   525.434
 CHARGINO ETA      =    -1.000     1.000

 U matrix      WINO      HIGGSINO      V matrix      WINO     HIGGSINO 
 W1SS+        -0.941     0.339         W1SS-         0.976    -0.220
 W2SS+         0.339     0.941         W2SS-         0.220     0.976


 Higgses masses: 
 --------------

 Light CP-even Higgs =   114.496
 Heavy CP-even Higgs =   569.934
       CP-odd  Higgs =   570.032
       Charged Higgs =   575.354
       sin(a-b)      =    -0.105
       cos(a-b)      =     0.994

\end{verbatim}
\end{small}
\newpage
\subsection{Spectrum \& parameters of  PYTHIA 6.2/00}
\begin{figure}[h] \centering
  \epsfig{file=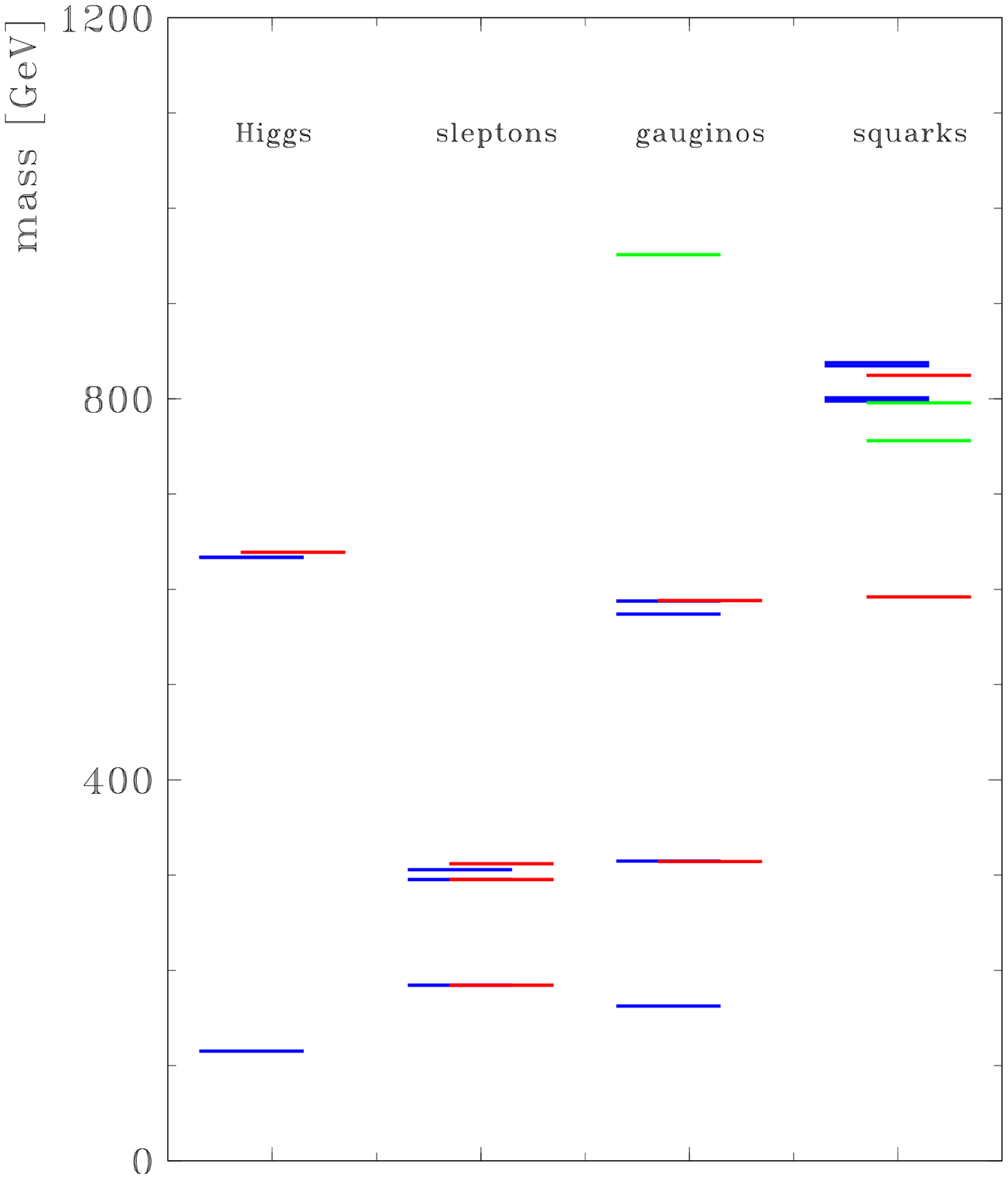,width=.9\textwidth}
  \caption{SPS 3 mass spectrum of \pythia}
\end{figure}
\newpage
\normalsize
\noindent
{\bf \pythia\  parameters} 
\begin{small}
\begin{verbatim}
  SUGRA input parameters
  ----------------------
   m_0         RMSS( 8) =   90.00    
   m_1/2       RMSS( 1) =   400.0    
   A_0         RMSS(16) =   0.000    
   tan_beta    RMSS( 5) =   10.00    
   sign mu     RMSS( 4) =   1.000    

   sparticle masses & widths
   -------------------------
   M_se_R   184.4 ( 0.046)   M_se_L   305.8 ( 0.185)   M_sne_L  295.2 ( 0.196)
   M_sm_R   184.4 ( 0.046)   M_sm_L   305.8 ( 0.185)   M_snm_L  295.2 ( 0.196)
   M_st_1   184.2 ( 0.045)   M_st_2   311.9 ( 0.246)   M_snt_L  295.0 ( 0.227)

   M_ch0_1  162.5 ( 0.000)   M_ch0_2  314.7 ( 0.103)   M_ch0_3  573.9 ( 3.857)
   M_ch0_4  587.5 ( 4.406)
   M_ch+_1  314.4 ( 0.100)   M_ch+_2  587.9 ( 4.398)

   M_h0     114.9 ( 0.004)   M_H0     633.8 ( 1.470)   M_A0     633.7 ( 1.531)
   M_H+     638.9 ( 1.511)

   M_g~     951.2 (21.123)
   M_uL     834.6 ( 7.816)   M_uR     800.8 ( 1.684)   M_dL     837.9 ( 7.677)
   M_dR     797.5 ( 0.419)
   M_b1     756.0 ( 7.559)   M_b2     795.7 ( 1.538)   M_t1     591.7 ( 2.509)
   M_t2     824.7 (12.076)

   parameter settings IMSS, RMSS
   -----------------------------
   IMSS( 1) = 2   IMSS( 4) = 1   IMSS( 7) = 0   IMSS(10) = 0
   IMSS( 2) = 0   IMSS( 5) = 0   IMSS( 8) = 0   IMSS(11) = 0
   IMSS( 3) = 0   IMSS( 6) = 0   IMSS( 9) = 0   IMSS(12) = 0

   RMSS( 1) =   400.0       RMSS( 9) =   700.0       RMSS(17) =   0.000    
   RMSS( 2) =   326.5       RMSS(10) =   757.1       RMSS(18) = -0.1040    
   RMSS( 3) =   914.1       RMSS(11) =   792.1       RMSS(19) =   634.1    
   RMSS( 4) =   570.0       RMSS(12) =   638.0       RMSS(20) =  0.4100E-01
   RMSS( 5) =   10.00       RMSS(13) =   305.6       RMSS(21) =   1.000    
   RMSS( 6) =   305.8       RMSS(14) =   183.8       RMSS(22) =   800.0    
   RMSS( 7) =   184.4       RMSS(15) =   0.000       RMSS(23) =  0.1000E+05
   RMSS( 8) =   90.00       RMSS(16) =  -813.6       RMSS(24) =  0.1000E+05

\end{verbatim}
\end{small}
\newpage

\subsection{Decay modes}
\normalsize
 \begin{table}[h] \centering
   \begin{tabular}{|l|c c c|c|c c c|}
   \hdick & & & \\[-2.ex]
   particle & $m_I$ & $m_S$ & $m_P$ & \ decay \ & ${\cal B}_I$ & ${\cal B}_S$ & ${\cal B}_P$ 
   \\[.5ex] \hdick
   $\ser    $ &  178.3 & 182.1  & 184.4  & $\nt_1    e^-               $ & 1.000 & 1.000 & 1.000 \\
   \hline                                                                                
   $\sel    $ &  287.1 & 290.8  & 305.8  & $\nt_1    e^-               $ & 1.000 & 1.000 & 1.000 \\
   \hline                                                                                
   $\sne    $ &  276.0 & 279.8  & 295.2  & $\nt_1    \nu_e             $ & 1.000 & 1.000 & 1.000 \\
   \hline                                                                                
   $\smur   $ &  178.3 & 182.1  & 184.4  & $\nt_1    \mu^-             $ & 1.000 & 1.000 & 1.000 \\
   \hline                                                                                
   $\smul   $ &  287.1 & 290.8  & 305.8  & $\nt_1    \mu^-             $ & 1.000 & 1.000 & 1.000 \\
   \hline                                                                                
   $\snm    $ &  276.0 & 279.8  & 295.2  & $\nt_1    \nu_\mu           $ & 1.000 & 1.000 & 1.000 \\
   \hline                                                                                
   $\stau_1 $ &  170.6 & 175.0  & 184.2  & $\nt_1    \tau^-            $ & 1.000 & 1.000 & 1.000 \\
   \hline                                                                                
   $\stau_2 $ &  289.2 & 292.7  & 311.9  & $\nt_1    \tau^-            $ & 0.873 & 1.000 & 0.847 \\
              &        &        &        & $Z^0      \stau_1           $ & 0.082 &       & 0.094 \\
              &        &        &        & $h^0      \stau_1           $ & 0.045 &       & 0.059 \\
   \hline
   $\snt    $ &  275.1 &  278.9 & 295.2  & $\nt_1    \nu_\tau          $ & 0.872 & 1.000 & 0.862 \\
              &        &        &        & $W^+      \stau_1           $ & 0.128 &       & 0.138 \\
   \hline
   \end{tabular}
   \caption{Slepton masses (GeV) and significant branching ratios ($>3\%$) 
            from \isajet~(I), \susygen~(S) and \pythia~(P)}
 \end{table}
  
 \begin{table}[h] \centering
   \begin{tabular}{|l|c c c|c|c c c|}
   \hdick & & & \\[-2.ex]
   particle & $m_I$ & $m_S$ & $m_P$ & \ decay \ & ${\cal B}_I$ & ${\cal B}_S$ & ${\cal B}_P$ 
   \\[.5ex] \hdick
   $\nt_1   $ &  160.6 & 163.1  & 162.5  & $                           $ & 1.000 & 1.000 & 1.000 \\
   \hline
   $\nt_2   $ &  296.9 & 294.5  & 314.7  & $\nt_1    h^0               $ & 0.089 & 0.140 & 0.098 \\
              &        &        &        & $\stau_1  \tau^+            $ & 0.121 & 0.177 & 0.126 \\
              &        &        &        & $\staup_1 \tau^-            $ & 0.121 & 0.177 & 0.126 \\
              &        &        &        & $\sne     \anue             $ & 0.083 & 0.067 & 0.088 \\
              &        &        &        & $\bar\sne \nu_e             $ & 0.083 & 0.067 & 0.088 \\
              &        &        &        & $\snm     \anum             $ & 0.083 & 0.067 & 0.088 \\
              &        &        &        & $\bar\snm \nu_\mu           $ & 0.083 & 0.067 & 0.088 \\
              &        &        &        & $\snt     \anut             $ & 0.090 & 0.074 & 0.089 \\
              &        &        &        & $\bar\snt \nu_\tau          $ & 0.090 & 0.074 & 0.089 \\
   \hline
   $\nt_3   $ &  512.9 & 508.5  & 573.9  & $\cp_1    W^-               $ & 0.295 & 0.334 & 0.298 \\
              &        &        &        & $\cm_1    W^+               $ & 0.295 & 0.334 & 0.298 \\
              &        &        &        & $\nt_1    Z^0               $ & 0.105 & 0.086 & 0.092 \\
              &        &        &        & $\nt_2    Z^0               $ & 0.245 & 0.197 & 0.253 \\
   \hline
   $\nt_4   $ &  529.6 & 525.5  & 587.5  & $\cp_1    W^-               $ & 0.272 & 0.306 & 0.278 \\
              &        &        &        & $\cm_1    W^+               $ & 0.272 & 0.306 & 0.278 \\
              &        &        &        & $\nt_1    h^0               $ & 0.074 & 0.063 & 0.069 \\
              &        &        &        & $\nt_2    h^0               $ & 0.176 & 0.150 & 0.206 \\
   \hline
   \end{tabular}
   \caption{Neutralino masses (GeV) and significant branching ratios ($>3\%$) 
            from \isajet~(I), \susygen~(S) and \pythia~(P)}
 \end{table}
  
 \begin{table}[h] \centering
   \begin{tabular}{|l|c c c|c|c c c|}
   \hdick & & & \\[-2.ex]
   particle & $m_I$ & $m_S$ & $m_P$ & \ decay \ & ${\cal B}_I$ & ${\cal B}_S$ & ${\cal B}_P$ 
   \\[.5ex] \hdick
   $\cp_1   $ &  296.8 & 294.2  & 314.4  & $\nt_1    W^+               $ & 0.108 & 0.450 & 0.111 \\
              &        &        &        & $\sne     e^+               $ & 0.185 & 0.169 & 0.189 \\
              &        &        &        & $\snm     \mu^+             $ & 0.185 & 0.169 & 0.189 \\
              &        &        &        & $\snt     \tau^+            $ & 0.202 & 0.189 & 0.194 \\
              &        &        &        & $\selp    \nu_e             $ & 0.039 &       & 0.037 \\
              &        &        &        & $\smulp   \nu_\mu           $ & 0.039 &       & 0.037 \\
              &        &        &        & $\staup_1 \nu_\tau          $ & 0.219 &       & 0.240 \\
   \hline
   $\cp_2   $ &  529.5 & 525.4  & 587.9  & $\nt_1    W^+               $ & 0.080 & 0.087 & 0.081 \\
              &        &        &        & $\nt_2    W^+               $ & 0.291 & 0.322 & 0.296 \\
              &        &        &        & $\selp    \nu_e             $ & 0.035 & 0.038 &       \\
              &        &        &        & $\smulp   \nu_\mu           $ & 0.035 & 0.038 &       \\
              &        &        &        & $\staup_2 \nu_\tau          $ & 0.039 & 0.039 &       \\
              &        &        &        & $\cp_1    Z^0               $ & 0.260 & 0.415 & 0.270 \\
              &        &        &        & $\cp_1    h^0               $ & 0.201 &       & 0.226 \\
   \hline
   \end{tabular}
   \caption{Chargino masses (GeV) and significant branching ratios ($>3\%$) 
            from \isajet~(I), \susygen~(S) and \pythia~(P)}
 \end{table}
  
 \begin{table}[h] \centering
   \begin{tabular}{|l|c c c|c|c c c|}
   \hdick & & & \\[-2.ex]
   particle & $m_I$ & $m_S$ & $m_P$ & \ decay \ & ${\cal B}_I$ & ${\cal B}_S$ & ${\cal B}_P$ 
   \\[.5ex] \hdick
   $h^0     $ &  117.0 & 114.5  & 114.9  & $\tau^-   \tau^+            $ & 0.049 & 0.077 & 0.065 \\
              &        &        &        & $b        \bar b            $ & 0.820 & 0.761 & 0.777 \\
              &        &        &        & $c        \bar c            $ & 0.037 &       & 0.045 \\
              &        &        &        & $g        g                 $ &       & 0.066 & 0.038 \\
              &        &        &        & $W^+      W^-               $ &       & 0.062 & 0.063 \\
   \hline
   $H^0     $ &  573.0 & 569.9  & 633.8  & $\tau^-   \tau^+            $ & 0.061 & 0.096 & 0.090 \\
              &        &        &        & $b        \bar b            $ & 0.799 & 0.693 & 0.764 \\
              &        &        &        & $t        \bar t            $ & 0.113 & 0.164 & 0.139 \\
   \hline
   $A^0     $ &  572.4 & 570.0  & 633.7  & $\tau^-   \tau^+            $ & 0.057 & 0.088 & 0.086 \\
              &        &        &        & $b        \bar b            $ & 0.748 & 0.638 & 0.735 \\
              &        &        &        & $t        \bar t            $ & 0.144 & 0.202 & 0.176 \\
              &        &        &        & $\nt_1    \nt_2             $ & 0.037 &       &       \\
   \hline
   $H^+     $ &  578.3 & 575.4  & 638.9  & $\nu_\tau \tau^+            $ & 0.072 & 0.095 & 0.088 \\
              &        &        &        & $t        \bar b            $ & 0.861 & 0.820 & 0.910 \\
              &        &        &        & $\cp_1    \nt_1             $ & 0.057 & 0.078 &       \\
   \hline
   \end{tabular}
   \caption{Higgs masses (GeV) and significant branching ratios ($>3\%$) 
            from \isajet~(I), \susygen~(S) and \pythia~(P)}
 \end{table}
  
 \begin{table}[h] \centering
   \begin{tabular}{|l|c c c|c|c c c|}
   \hdick & & & \\[-2.ex]
   particle & $m_I$ & $m_S$ & $m_P$ & \ decay \ & ${\cal B}_I$ & ${\cal B}_S$ & ${\cal B}_P$ 
   \\[.5ex] \hdick
   $\st_1   $ &  623.8 & 667.7  & 591.7  & $\nt_1    t                 $ & 0.231 & 0.189 & 0.311 \\
              &        &        &        & $\nt_2    t                 $ & 0.159 & 0.128 & 0.155 \\
              &        &        &        & $\cp_1    b                 $ & 0.483 & 0.390 & 0.534 \\
              &        &        &        & $\cp_2    b                 $ & 0.128 & 0.294 &       \\
   \hline
   $\st_2   $ &  819.5 & 830.4  & 824.8  & $\cp_1    b                 $ & 0.206 & 0.220 & 0.223 \\
              &        &        &        & $\cp_2    b                 $ & 0.157 & 0.330 & 0.107 \\
              &        &        &        & $Z^0      \st_1             $ & 0.143 &       & 0.259 \\
              &        &        &        & $h^0      \st_1             $ & 0.053 &       & 0.069 \\
              &        &        &        & $\nt_2    t                 $ & 0.090 & 0.071 & 0.096 \\
              &        &        &        & $\nt_3    t                 $ & 0.088 & 0.172 & 0.051 \\
              &        &        &        & $\nt_4    t                 $ & 0.238 & 0.195 & 0.166 \\
   \hline
   $\sb_1   $ &  757.5 & 800.2  &  756.0 & $\nt_1    b                 $ &       & 0.037 &       \\
              &        &        &        & $\nt_2    b                 $ & 0.233 & 0.311 & 0.249 \\
              &        &        &        & $\cm_1    t                 $ & 0.373 & 0.555 & 0.395 \\
              &        &        &        & $\cm_2    t                 $ & 0.224 & 0.051 &       \\
              &        &        &        & $W^-      \st_1             $ & 0.127 &       & 0.333 \\
   \hline
   $\sb_2   $ &  791.3 & 831.3  &  795.7 & $\nt_1    b                 $ & 0.125 & 0.168 & 0.255 \\
              &        &        &        & $\nt_2    b                 $ & 0.087 & 0.078 & 0.123 \\
              &        &        &        & $\nt_3    b                 $ & 0.041 & 0.137 &       \\
              &        &        &        & $\nt_4    b                 $ & 0.058 & 0.165 &       \\
              &        &        &        & $\cm_1    t                 $ & 0.143 & 0.130 & 0.203 \\
              &        &        &        & $\cm_2    t                 $ & 0.356 & 0.322 & 0.101 \\
              &        &        &        & $W^-      \st_1             $ & 0.189 &       & 0.315 \\
   \hline
   \end{tabular}
   \caption{Light squark masses (GeV) and significant branching ratios ($>3\%$) 
            from \isajet~(I), \susygen~(S) and \pythia~(P)}
 \end{table}

\clearpage

\section{SPS 4 -- mSUGRA scenario}
\setcounter{figure}{0}
\setcounter{table}{0}

\large\boldmath
\hspace{20mm}
\begin{tabular}{|l c|}
  \hline
  $m_0$       & $  400~\GeV$ \\ 
  $m_{1/2}$   & $  300~\GeV$ \\
  $A_0$       & $\ \ \ \ 0~\GeV$ \\
  $\tan\beta$ & $ 50$       \\
  ${\rm sign}~\mu$ & $+$ \\
  \hline
\end{tabular} \hspace{10mm}
\begin{tabular}{l}
   {\bf `large $\tan\beta$' scenario} \\
\end{tabular}
\unboldmath\normalsize
\bigskip

\subsection{Spectrum \& parameters of ISAJET 7.58}

\begin{figure}[h] \centering
  \epsfig{file=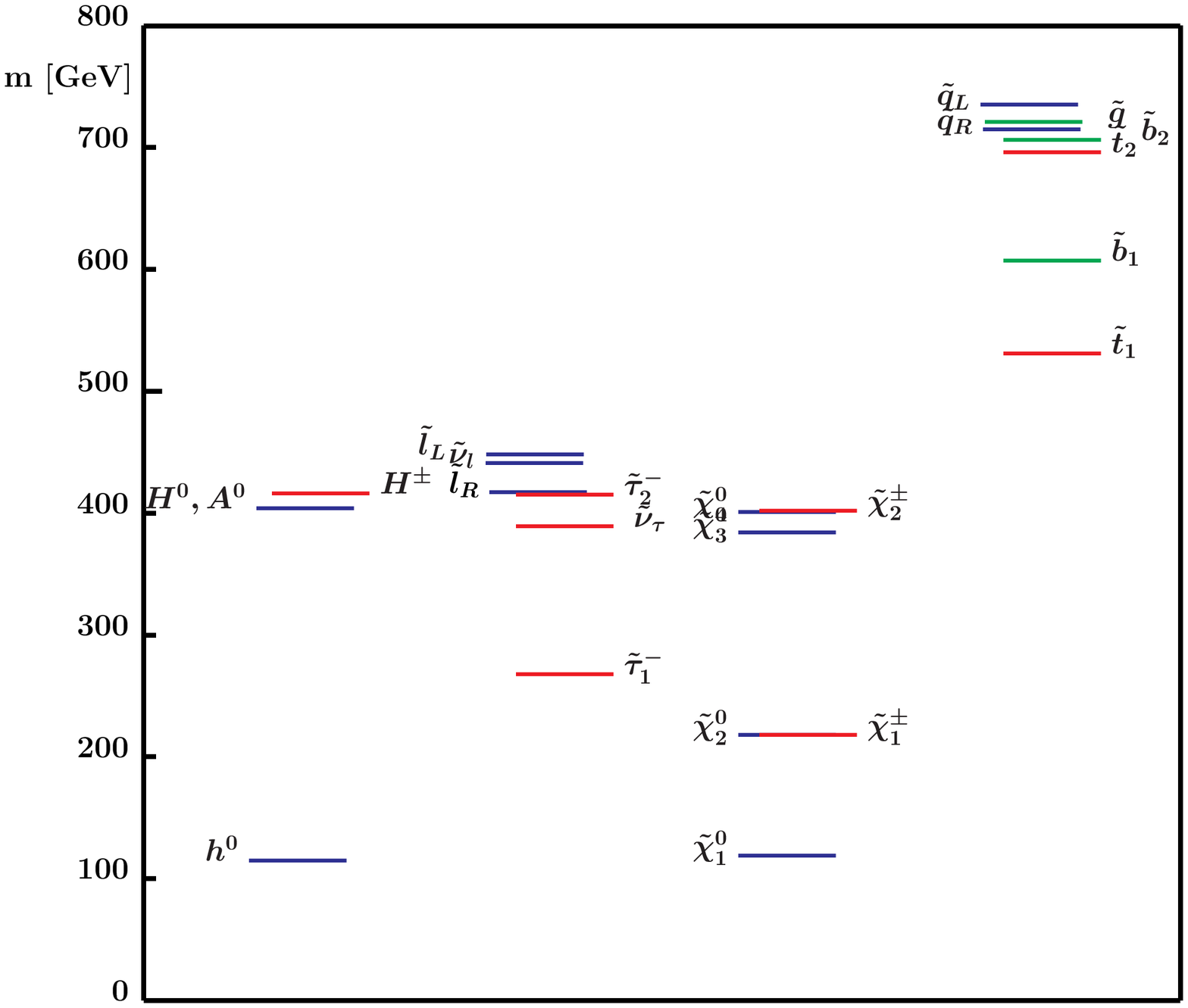,width=.9\textwidth}
  \caption{SPS 4 mass spectrum of \isajet}
\end{figure}

\clearpage
\normalsize
\noindent
{\bf \isajet\ parameters} 
\begin{small}
\begin{verbatim}
 Minimal supergravity (mSUGRA) model:

 M_0,  M_(1/2),  A_0,  tan(beta),  sgn(mu),  M_t =
   400.000   300.000     0.000    50.000     1.0   175.000

 ISASUGRA unification:
 M_GUT      = 0.211E+17   g_GUT          =0.710      alpha_GUT =0.040
 FT_GUT     = 0.486       FB_GUT         = 0.186     FL_GUT = 0.463

 1/alpha_em =  127.73     sin**2(thetaw) =0.2310     alpha_s   =0.119
 M_1        =  120.80     M_2            =  233.17   M_3       =  689.41
 mu(Q)      =  377.03     B(Q)           =   13.05   Q         =  571.25
 M_H1^2     = 0.602E+05   M_H2^2         =-0.140E+06

 ISAJET masses (with signs):
 M(GL)  =   721.03
 M(UL)  =   730.24   M(UR)  =   715.10   M(DL)  =   734.59   M(DR) =   714.32
 M(B1)  =   606.86   M(B2)  =   706.45   M(T1)  =   530.58   M(T2) =   695.88
 M(SN)  =   441.22   M(EL)  =   448.40   M(ER)  =   416.54
 M(NTAU)=   389.43   M(TAU1)=   267.61   M(TAU2)=   414.91
 M(Z1)  =  -118.66   M(Z2)  =  -218.14   M(Z3)  =   383.91   M(Z4) =  -401.08
 M(W1)  =  -218.06   M(W2)  =  -402.28
 M(HL)  =   115.39   M(HH)  =   404.63   M(HA)  =   404.43   M(H+) =   416.28

 theta_t=   1.0387   theta_b=   0.6261   theta_l=   1.1998   alpha_h=   0.0204

 NEUTRALINO MASSES (SIGNED) =  -118.657  -218.137   383.909  -401.079
 EIGENVECTOR 1       =   0.04547  -0.13669  -0.03365   0.98900
 EIGENVECTOR 2       =   0.17072  -0.28852  -0.93876  -0.07966
 EIGENVECTOR 3       =  -0.70951  -0.69629   0.09011  -0.06055
 EIGENVECTOR 4       =  -0.68219   0.64285  -0.33089   0.10895

 CHARGINO MASSES (SIGNED)  =  -218.062  -402.275
 GAMMAL, GAMMAR             =   1.99843   1.81882


 ISAJET equivalent input:
 MSSMA:   721.03  377.03  404.43   50.00
 MSSMB:   732.20  713.87  716.00  445.90  414.23
 MSSMC:   640.09  673.40  556.76  394.72  289.48 -552.20 -729.52 -102.27
 MSSMD: SAME AS MSSMB (DEFAULT)
 MSSME:   120.80  233.17

\end{verbatim}
\end{small}
\clearpage
\subsection{Spectrum \& parameters of SUSYGEN 3.00/27}
\begin{figure}[h] \centering
  \epsfig{file=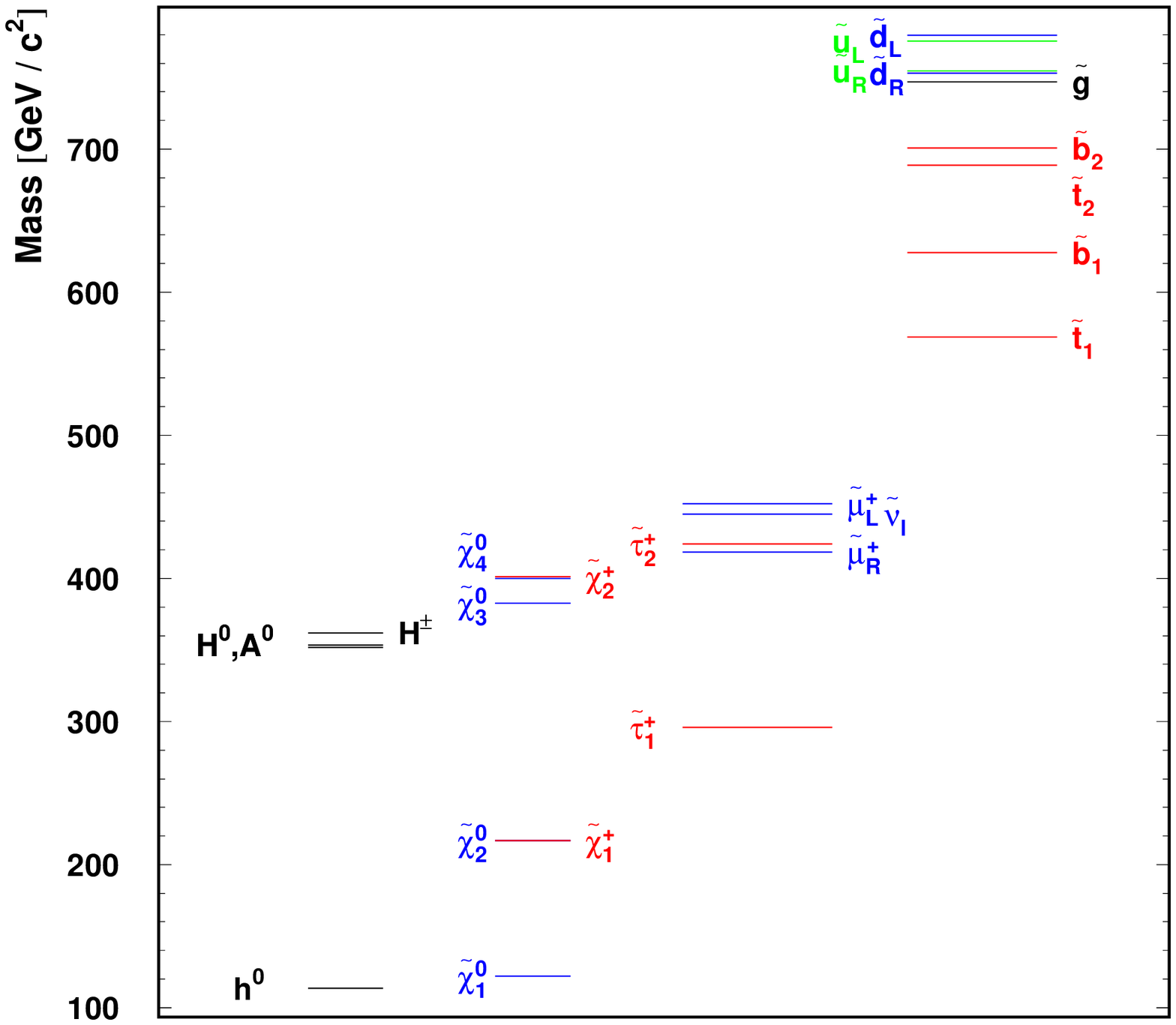,width=.9\textwidth}
  \caption{SPS 4 mass spectrum of \susygen}
\end{figure}
\newpage
\normalsize
\noindent
{\bf \susygen\  parameters} 
\begin{small}
\begin{verbatim}
 Susygen inputs:
 --------------

 m0      =   400.000   TANB     =    50.000
 m1/2    =   300.000    mu/|mu| =     1
 A0      =     0.000

 Sparticle masses: 
 ----------------

 SUPR      755.  SUPL       775.
 SDNR      753.  SDNL       779.
 SELR      418.  SELL       452.
 SNU       445.
 STP1      569.  STP2       689. cosmix =  0.476
 SBT1      628.  SBT2       701. cosmix =  0.762
 STA1      296.  STA2       424. cosmix=   0.372
 SGLU      747.

 Gaugino masses:
 --------------

 M1 =    124.255 M2 =    232.053 M3 =    694.630


 NEUTRALINO m, CP, ph/zi/ha/hb 1 =  122.0  1.  0.849 -0.508  0.050  0.138
 NEUTRALINO m, CP, ph/zi/ha/hb 2 =  216.9  1. -0.525 -0.782  0.177  0.286
 NEUTRALINO m, CP, ph/zi/ha/hb 3 =  382.6 -1. -0.009  0.109 -0.695  0.710
 NEUTRALINO m, CP, ph/zi/ha/hb 4 =  400.0  1. -0.062 -0.345 -0.695 -0.628

 CHARGINO MASSES    =   216.667   401.170
 CHARGINO ETA       =    -1.000     1.000

 U matrix      WINO      HIGGSINO   V matrix      WINO      HIGGSINO 
 W1SS+        -0.909     0.417      W1SS-         0.969    -0.246
 W2SS+         0.417     0.909      W2SS-         0.246     0.969

 Higgses masses: 
 --------------

 Light CP-even Higgs =   113.437
 Heavy CP-even Higgs =   351.833
       CP-odd  Higgs =   353.570
       Charged Higgs =   361.880
       sin(a-b)      =    -0.021
       cos(a-b)      =     1.000

\end{verbatim}
\end{small}

\subsection{Spectrum \& parameters of  PYTHIA 6.2/00}

\begin{figure}[h] \centering
  \epsfig{file=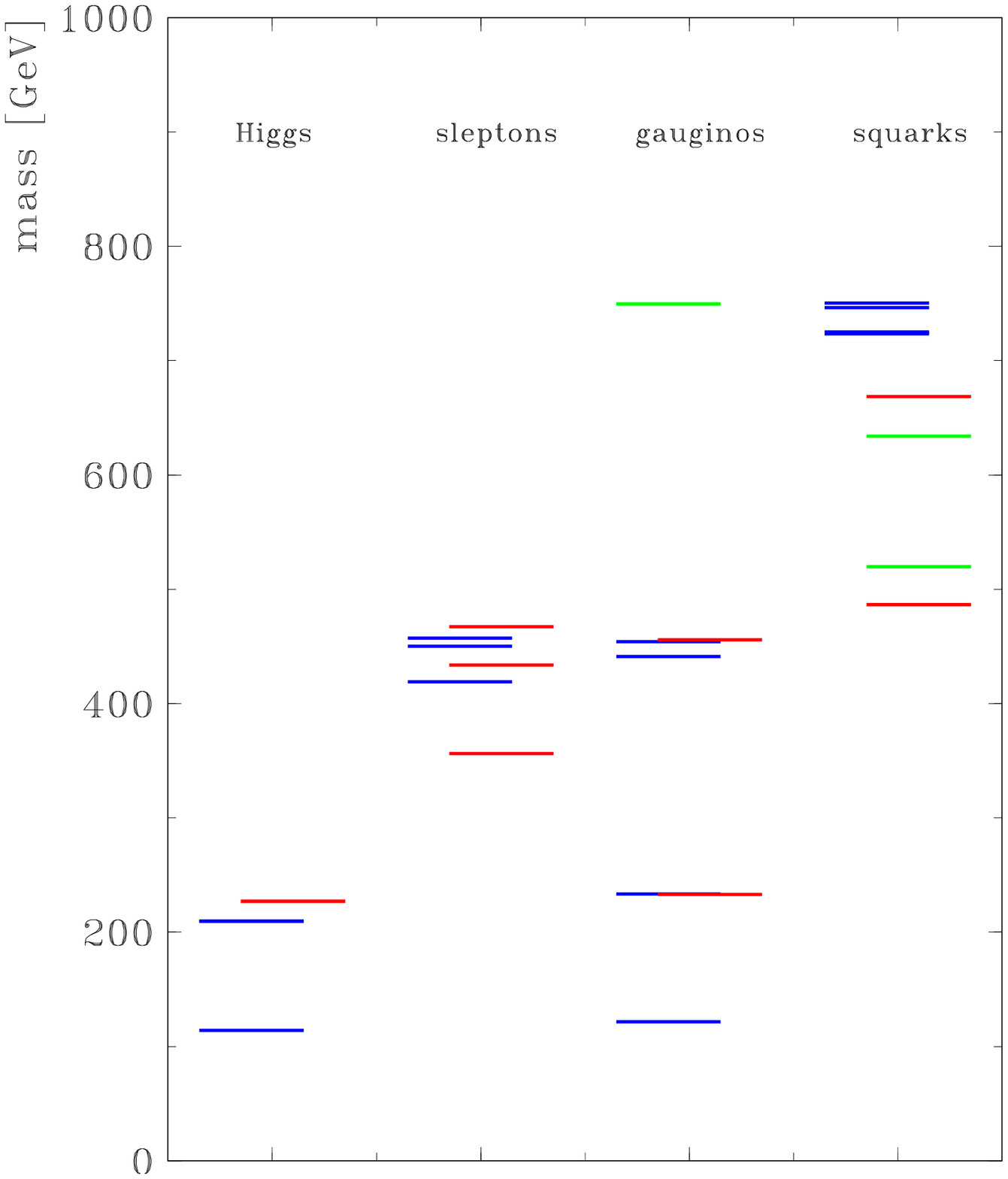,width=.9\textwidth}
  \caption{SPS 4 mass spectrum of \pythia}
\end{figure}
\newpage
\normalsize
\noindent
{\bf \pythia\  parameters} 
\begin{small}
\begin{verbatim}
   SUGRA input parameters
   ----------------------
   m_0         RMSS( 8) =   400.0    
   m_1/2       RMSS( 1) =   300.0    
   A_0         RMSS(16) =   0.000    
   tan_beta    RMSS( 5) =   50.00    
   sign mu     RMSS( 4) =   1.000    

   sparticle masses & widths
   -------------------------
   M_se_R   418.8 ( 1.782)   M_se_L   457.2 ( 3.377)   M_sne_L  450.1 ( 3.396)
   M_sm_R   418.8 ( 1.783)   M_sm_L   457.2 ( 3.377)   M_snm_L  450.1 ( 3.396)
   M_st_1   356.4 ( 1.972)   M_st_2   467.2 ( 2.618)   M_snt_L  433.7 ( 3.243)

   M_ch0_1  121.7 ( 0.000)   M_ch0_2  233.3 ( 0.001)   M_ch0_3  440.9 ( 2.634)
   M_ch0_4  454.0 ( 2.855)
   M_ch+_1  233.1 ( 0.005)   M_ch+_2  455.6 ( 3.020)

   M_h0     114.1 ( 0.004)   M_H0     209.6 ( 7.886)   M_A0     209.8 ( 7.909)
   M_H+     226.9 ( 3.453)

   M_g~     749.4 ( 7.937)
   M_uL     746.3 ( 7.718)   M_uR     725.0 ( 1.563)   M_dL     750.4 ( 7.534)
   M_dR     723.4 ( 0.390)
   M_b1     519.8 ( 1.335)   M_b2     634.0 ( 4.829)   M_t1     486.7 ( 2.672)
   M_t2     668.7 (10.050)

   parameter settings IMSS, RMSS
   -----------------------------
   IMSS( 1) = 2   IMSS( 4) = 1   IMSS( 7) = 0   IMSS(10) = 0
   IMSS( 2) = 0   IMSS( 5) = 0   IMSS( 8) = 0   IMSS(11) = 0
   IMSS( 3) = 0   IMSS( 6) = 0   IMSS( 9) = 0   IMSS(12) = 0

   RMSS( 1) =   300.0       RMSS( 9) =   700.0       RMSS(17) =   0.000    
   RMSS( 2) =   244.9       RMSS(10) =   584.6       RMSS(18) = -0.2098E-01
   RMSS( 3) =   706.3       RMSS(11) =   571.2       RMSS(19) =   210.0    
   RMSS( 4) =   434.8       RMSS(12) =   540.7       RMSS(20) =  0.4100E-01
   RMSS( 5) =   50.00       RMSS(13) =   441.1       RMSS(21) =   1.000    
   RMSS( 6) =   457.2       RMSS(14) =   382.8       RMSS(22) =   800.0    
   RMSS( 7) =   418.8       RMSS(15) =   0.000       RMSS(23) =  0.1000E+05
   RMSS( 8) =   400.0       RMSS(16) =  -614.3       RMSS(24) =  0.1000E+05

\end{verbatim}
\end{small}

\clearpage

\subsection{Decay modes}
\normalsize
\begin{table}[h!] \centering
   \begin{tabular}{|l|c c c|c|c c c|}
   \hdick & & & \\[-2.ex]
   particle & $m_I$ & $m_S$ & $m_P$ & \ decay \ & ${\cal B}_I$ & ${\cal B}_S$ & ${\cal B}_P$ 
   \\[.5ex] \hdick
   $\ser    $ &  416.5 &  418.3 & 418.8  & $\nt_1    e^-               $ & 0.996 & 0.995 & 0.996 \\
   \hline
   $\sel    $ &  448.4 &  452.0 & 457.2  & $\nt_1    e^-               $ & 0.127 & 0.122 & 0.136 \\
              &        &        &        & $\nt_2    e^-               $ & 0.318 & 0.320 & 0.309 \\
              &        &        &        & $\cm_1    \nu_e             $ & 0.546 & 0.547 & 0.555 \\
   \hline
   $\sne    $ &  441.2 &  444.9 & 450.1  & $\nt_1    \nu_e             $ & 0.157 & 0.155 & 0.157 \\
              &        &        &        & $\nt_2    \nu_e             $ & 0.250 & 0.250 & 0.261 \\
              &        &        &        & $\cp_1    e^-               $ & 0.588 & 0.590 & 0.582 \\
   \hline
   $\smur   $ &  416.5 &  418.3 & 418.8  & $\nt_1    \mu^-             $ & 0.996 & 0.995 & 0.996 \\
   \hline
   $\smul   $ &  448.4 &  452.0 & 457.2  & $\nt_1    \mu^-             $ & 0.127 & 0.122 & 0.136 \\
              &        &        &        & $\nt_2    \mu^-             $ & 0.318 & 0.320 & 0.309 \\
              &        &        &        & $\cm_1    \nu_\mu           $ & 0.546 & 0.547 & 0.555 \\
   \hline
   $\snm    $ &  441.2 &  444.9 & 450.1  & $\nt_1    \nu_\mu           $ & 0.157 & 0.155 & 0.157 \\
              &        &        &        & $\nt_2    \nu_\mu           $ & 0.250 & 0.250 & 0.261 \\
              &        &        &        & $\cp_1    \mu^-             $ & 0.588 & 0.590 & 0.582 \\
   \hline
   $\stau_1 $ &  267.6 &  296.0 & 356.4  & $\nt_1    \tau^-            $ & 0.836 & 0.742 & 0.631 \\
              &        &        &        & $\nt_2    \tau^-            $ & 0.057 & 0.090 & 0.129 \\
              &        &        &        & $\cm_1    \nu_\tau          $ & 0.107 & 0.168 & 0.240 \\
   \hline
   $\stau_2 $ &  414.9 &  424.1 & 467.2  & $\nt_1    \tau^-            $ & 0.196 & 0.222 & 0.292 \\
              &        &        &        & $\nt_2    \tau^-            $ & 0.263 & 0.308 & 0.261 \\
              &        &        &        & $\cm_1    \nu_\tau          $ & 0.378 & 0.437 & 0.407 \\
              &        &        &        & $Z^0      \stau_1           $ & 0.088 &       & 0.031 \\
              &        &        &        & $h^0      \stau_1           $ & 0.063 &       &       \\
   \hline
   $\snt    $ &  389.4 &  400.8 & 433.7  & $\nt_1    \nu_\tau          $ & 0.152 & 0.171 & 0.157 \\
              &        &        &        & $\nt_2    \nu_\tau          $ & 0.209 & 0.246 & 0.248 \\
              &        &        &        & $\cp_1    \tau^-            $ & 0.542 & 0.583 & 0.595 \\
              &        &        &        & $W^+      \stau_1           $ & 0.097 &       &       \\
   \hline
   \end{tabular}
   \caption{Slepton masses (GeV) and significant branching ratios ($>3\%$) 
            from \isajet~(I), \susygen~(S) and \pythia~(P)}
 \end{table}
  
 \begin{table}[h] \centering
   \begin{tabular}{|l|c c c|c|c c c|}
   \hdick & & & \\[-2.ex]
   particle & $m_I$ & $m_S$ & $m_P$ & \ decay \ & ${\cal B}_I$ & ${\cal B}_S$ & ${\cal B}_P$ 
   \\[.5ex] \hdick
   $\nt_1   $ &  118.7 & 122.0  & 121.7  & $                           $ & 1.000 & 1.000 & 1.000 \\
   \hline
   $\nt_2   $ &  218.1 & 216.9  & 233.3  & $\nt_1    Z^0               $ & 0.989 & 0.995 & 1.000 \\
   \hline
   $\nt_3   $ &  383.9 & 382.6  & 440.9  & $\cp_1    W^-               $ & 0.273 & 0.345 & 0.279 \\
              &        &        &        & $\cm_1    W^+               $ & 0.273 & 0.345 & 0.279 \\
              &        &        &        & $\nt_1    Z^0               $ & 0.107 & 0.081 & 0.087 \\
              &        &        &        & $\nt_2    Z^0               $ & 0.199 & 0.148 & 0.217 \\
              &        &        &        & $\stau_1  \tau^+            $ & 0.058 &       &       \\
              &        &        &        & $\staup_1 \tau^-            $ & 0.058 &       &       \\
   \hline
   $\nt_4   $ &  401.1 & 400.0  & 454.0  & $\cp_1    W^-               $ & 0.312 & 0.364 & 0.285 \\
              &        &        &        & $\cm_1    W^+               $ & 0.312 & 0.364 & 0.285 \\
              &        &        &        & $\nt_1    h^0               $ & 0.080 & 0.063 & 0.067 \\
              &        &        &        & $\nt_2    h^0               $ & 0.160 & 0.125 & 0.184 \\
              &        &        &        & $\stau_1  \tau^+            $ & 0.042 &       &       \\
              &        &        &        & $\staup_1 \tau^-            $ & 0.042 &       &       \\
   \hline
   \end{tabular}
   \caption{Neutralino masses (GeV) and significant branching ratios ($>3\%$) 
            from \isajet~(I), \susygen~(S) and \pythia~(P)}
 \end{table}
  
 \begin{table}[h] \centering
   \begin{tabular}{|l|c c c|c|c c c|}
   \hdick & & & \\[-2.ex]
   particle & $m_I$ & $m_S$ & $m_P$ & \ decay \ & ${\cal B}_I$ & ${\cal B}_S$ & ${\cal B}_P$ 
   \\[.5ex] \hdick
   $\cp_1   $ &  218.1 & 216.7  & 233.1  & $\nt_1    W^+               $ & 0.998 & 1.000 & 1.000 \\
   \hline
   $\cp_2   $ &  402.3 & 401.2  & 455.6  & $\nt_1    W^+               $ & 0.100 & 0.103 & 0.082 \\
              &        &        &        & $\nt_2    W^+               $ & 0.346 & 0.367 & 0.284 \\
              &        &        &        & $\staup_1 \nu_\tau          $ & 0.071 &       &       \\
              &        &        &        & $\cp_1    Z^0               $ & 0.287 & 0.525 & 0.248 \\
              &        &        &        & $\cp_1    h^0               $ & 0.194 &       & 0.194 \\
   \hline
   \end{tabular}
   \caption{Chargino masses (GeV) and significant branching ratios ($>3\%$) 
            from \isajet~(I), \susygen~(S) and \pythia~(P)}
 \end{table}
  
 \begin{table}[h] \centering
   \begin{tabular}{|l|c c c|c|c c c|}
   \hdick & & & \\[-2.ex]
   particle & $m_I$ & $m_S$ & $m_P$ & \ decay \ & ${\cal B}_I$ & ${\cal B}_S$ & ${\cal B}_P$ 
   \\[.5ex] \hdick
   $h^0     $ &  115.4 & 113.4  & 114.1  & $\tau^-   \tau^+            $ & 0.050 & 0.078 & 0.066 \\
              &        &        &        & $b        \bar b            $ & 0.822 & 0.766 & 0.780 \\
              &        &        &        & $c        \bar c            $ & 0.040 &       & 0.046 \\
              &        &        &        & $g        g                 $ & 0.031 & 0.068 & 0.038 \\
              &        &        &        & $W^+      W^-               $ &       & 0.055 & 0.058 \\
   \hline
   $H^0     $ &  404.6 & 351.8  & 209.6  & $\tau^-   \tau^+            $ & 0.067 & 0.112 & 0.137 \\
              &        &        &        & $b        \bar b            $ & 0.928 & 0.885 & 0.861 \\
   \hline
   $A^0     $ &  404.4 & 353.6  & 209.8  & $\tau^-   \tau^+            $ & 0.067 & 0.112 & 0.137 \\
              &        &        &        & $b        \bar b            $ & 0.926 & 0.883 & 0.861 \\
   \hline
   $H^+     $ &  416.3 & 361.9  & 226.9  & $\nu_\tau \tau^+            $ & 0.105 & 0.149 & 0.339 \\
              &        &        &        & $t        \bar b            $ & 0.886 & 0.844 & 0.658 \\
   \hline
   \end{tabular}
   \caption{Higgs masses (GeV) and significant branching ratios ($>3\%$) 
            from \isajet~(I), \susygen~(S) and \pythia~(P)}
 \end{table}
  
 \begin{table}[h] \centering
   \begin{tabular}{|l|c c c|c|c c c|}
   \hdick & & & \\[-2.ex]
   particle & $m_I$ & $m_S$ & $m_P$ & \ decay \ & ${\cal B}_I$ & ${\cal B}_S$ & ${\cal B}_P$ 
   \\[.5ex] \hdick
   $\st_1   $ &  530.6 & 568.5  & 486.7  & $\nt_1    t                 $ & 0.150 & 0.127 & 0.192 \\
              &        &        &        & $\nt_2    t                 $ & 0.127 & 0.116 & 0.154 \\
              &        &        &        & $\cp_1    b                 $ & 0.468 & 0.396 & 0.631 \\
              &        &        &        & $\cp_2    b                 $ & 0.256 & 0.343 &       \\
   \hline
   $\st_2   $ &  695.9 & 688.8  & 668.7  & $\cp_1    b                 $ & 0.209 & 0.207 & 0.156 \\
              &        &        &        & $\cp_2    b                 $ & 0.343 & 0.344 & 0.174 \\
              &        &        &        & $W^+      \sb_1             $ &       &       & 0.190 \\
              &        &        &        & $Z^0      \st_1             $ & 0.057 &       & 0.158 \\
              &        &        &        & $\nt_1    t                 $ &       &       & 0.040 \\
              &        &        &        & $\nt_2    t                 $ & 0.058 & 0.054 & 0.063 \\
              &        &        &        & $\nt_3    t                 $ & 0.079 & 0.175 & 0.036 \\
              &        &        &        & $\nt_4    t                 $ & 0.210 & 0.209 & 0.175 \\
   \hline
   $\sb_1   $ &  606.9 & 627.5  & 519.8  & $\nt_1    b                 $ & 0.072 & 0.060 & 0.139 \\
              &        &        &        & $\nt_2    b                 $ & 0.273 & 0.206 & 0.411 \\
              &        &        &        & $\nt_3    b                 $ & 0.166 & 0.204 &       \\
              &        &        &        & $\nt_4    b                 $ & 0.099 & 0.133 &       \\
              &        &        &        & $\cm_1    t                 $ & 0.343 & 0.333 & 0.447 \\
              &        &        &        & $\cm_2    t                 $ & 0.047 & 0.064 &       \\
   \hline
   $\sb_2   $ &  706.5 & 700.7  & 634.0  & $\nt_1    b                 $ &       &       & 0.042 \\
              &        &        &        & $\nt_2    b                 $ & 0.213 & 0.243 & 0.215 \\
              &        &        &        & $\nt_4    b                 $ & 0.208 & 0.220 &       \\
              &        &        &        & $\cm_1    t                 $ &       &       & 0.319 \\
              &        &        &        & $\cm_2    t                 $ & 0.450 & 0.488 & 0.073 \\
              &        &        &        & $W^-      \st_1             $ & 0.086 &       & 0.310 \\
              &        &        &        & $Z^0      \sb_1             $ &       &       & 0.033 \\
   \hline
   \end{tabular}
   \caption{Light squark masses (GeV) and significant branching ratios ($>3\%$) 
            from \isajet~(I), \susygen~(S) and \pythia~(P)}
 \end{table}

\clearpage

\section{SPS 5 -- mSUGRA scenario}
\setcounter{figure}{0}
\setcounter{table}{0}

\large\boldmath
\hspace{20mm}
\begin{tabular}{|l c|}
  \hline
  $m_0$       & $\ \ \ 150~\GeV$ \\ 
  $m_{1/2}$   & $\ \ \ 300~\GeV$ \\
  $A_0$       & $-1000~\GeV$ \\
  $\tan\beta$ & $5$       \\
  ${\rm sign}~\mu$ & $+$ \\
  \hline
\end{tabular} \hspace{10mm}
\begin{tabular}{l}
   {\bf `light stop' scenario} \\
\end{tabular}
\unboldmath\normalsize
\bigskip

\subsection{Spectrum \& parameters of ISAJET 7.58}

\begin{figure}[h] \centering
  \epsfig{file=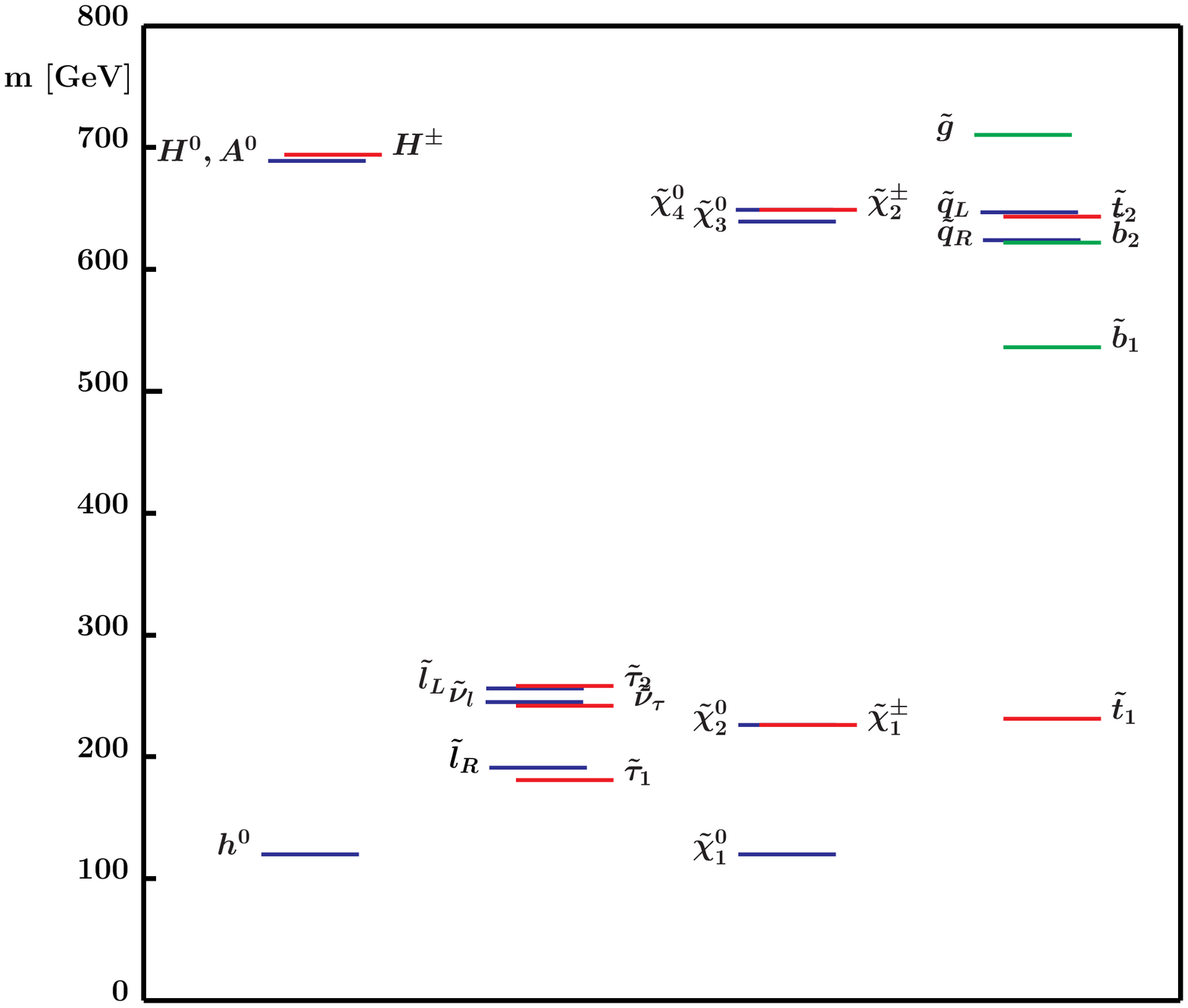,width=.9\textwidth}
  \caption{SPS 5 mass spectrum of \isajet}
\end{figure}

\clearpage
\normalsize
\noindent
{\bf \isajet\ parameters} 
\begin{small}
\begin{verbatim}
 Minimal supergravity (mSUGRA) model:

 M_0,  M_(1/2),  A_0,  tan(beta),  sgn(mu),  M_t =
   150.000   300.000 -1000.000     5.000     1.0   175.000

 ISASUGRA unification:
 M_GUT      = 0.190E+17   g_GUT          =0.713      alpha_GUT =0.040
 FT_GUT     = 0.487       FB_GUT         = 0.023     FL_GUT = 0.035

 1/alpha_em =  127.66     sin**2(thetaw) =0.2309     alpha_s   =0.119
 M_1        =  121.39     M_2            =  234.56   M_3       =  694.68
 mu(Q)      =  639.80     B(Q)           =  144.90   Q         =  449.79
 M_H1^2     = 0.572E+05   M_H2^2         =-0.393E+06

 ISAJET masses (with signs):
 M(GL)  =   710.31
 M(UL)  =   641.82   M(UR)  =   624.49   M(DL)  =   646.40   M(DR) =   623.38
 M(B1)  =   535.86   M(B2)  =   622.99   M(T1)  =   220.74   M(T2) =   644.65
 M(SN)  =   244.52   M(EL)  =   256.30   M(ER)  =   191.45
 M(NTAU)=   242.34   M(TAU1)=   180.67   M(TAU2)=   257.86
 M(Z1)  =  -119.51   M(Z2)  =  -226.33   M(Z3)  =   642.83   M(Z4) =  -652.95
 M(W1)  =  -226.33   M(W2)  =  -652.68
 M(HL)  =   119.79   M(HH)  =   694.03   M(HA)  =   693.86   M(H+) =   698.49

 theta_t=   1.0031   theta_b=   0.1580   theta_l=   1.3356   alpha_h=   0.2069

 NEUTRALINO MASSES (SIGNED) =  -119.510  -226.329   642.833  -652.949
 EIGENVECTOR 1       =   0.02835  -0.07603  -0.02968   0.99626
 EIGENVECTOR 2       =   0.07465  -0.14434  -0.98579  -0.04251
 EIGENVECTOR 3       =  -0.70833  -0.70332   0.05073  -0.03201
 EIGENVECTOR 4       =   0.70135  -0.69190   0.15735  -0.06807

 CHARGINO MASSES (SIGNED)  =  -226.327  -652.683
 GAMMAL, GAMMAR             =   1.77997   1.67887


 ISAJET equivalent input:
 MSSMA:   710.31  639.80  693.86    5.00
 MSSMB:   643.88  622.91  625.44  252.24  186.76
 MSSMC:   535.16  620.50  360.54  250.13  180.89 -905.63-1671.36-1179.34
 MSSMD: SAME AS MSSMB (DEFAULT)
 MSSME:   121.39  234.56

\end{verbatim}
\end{small}

\clearpage
\subsection{Spectrum \& parameters of SUSYGEN 3.00/27}
\begin{figure}[h] \centering
  \epsfig{file=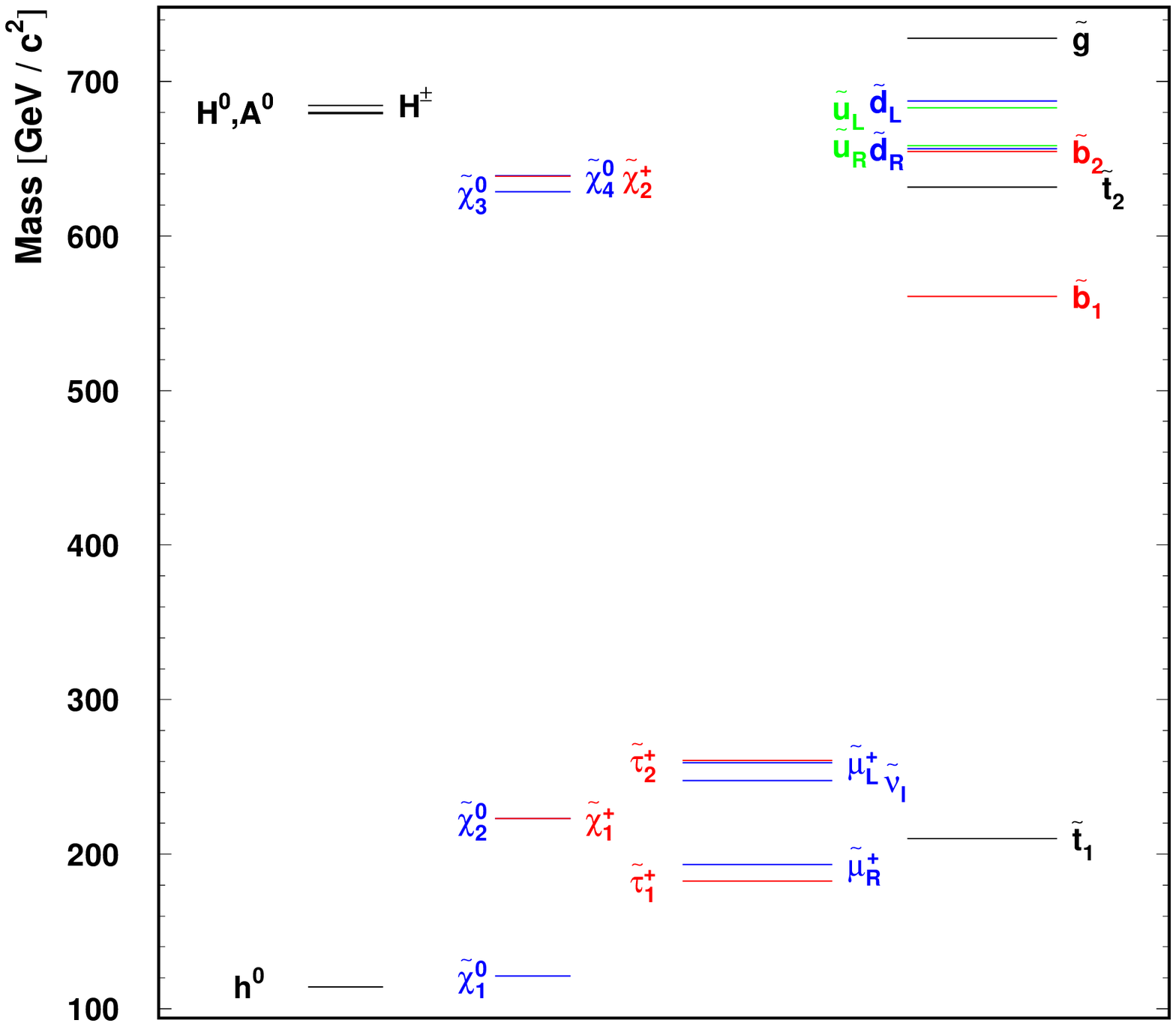,width=.9\textwidth}
  \caption{SPS 5 mass spectrum of \susygen}
\end{figure}
\newpage
\normalsize
\noindent
{\bf \susygen\  parameters} 
\begin{small}
\begin{verbatim}
 Susygen inputs:
 --------------

 m0      =   150.000   TANB     =     5.000
 m1/2    =   300.000    mu/|mu| =     1
 A0      = -1000.000

 Sparticle masses: 
 ----------------

 SUPR      658.  SUPL       683.
 SDNR      656.  SDNL       687.
 SELR      193.  SELL       259.
 SNU       248.
 STP1      210.  STP2       632. cosmix =  0.513
 SBT1      561.  SBT2       654. cosmix =  0.992
 STA1      183.  STA2       261. cosmix=   0.226
 SGLU      728.


 Gaugino masses:
 --------------

 M1 =    123.173 M2 =    231.684 M3 =    712.143

 NEUTRALINO m, CP, ph/zi/ha/hb 1 =  121.2  1.  0.858 -0.507  0.045  0.071
 NEUTRALINO m, CP, ph/zi/ha/hb 2 =  223.1  1. -0.514 -0.841  0.105  0.130
 NEUTRALINO m, CP, ph/zi/ha/hb 3 =  628.5 -1. -0.004  0.061 -0.557  0.828
 NEUTRALINO m, CP, ph/zi/ha/hb 4 =  639.0  1. -0.016 -0.176 -0.823 -0.540

 CHARGINO MASSES    =   222.964   638.753
 CHARGINO ETA      =     -1.000     1.000

 U matrix      WINO      HIGGSINO      V matrix      WINO     HIGGSINO 
 W1SS+        -0.977     0.214         W1SS-         0.994    -0.112
 W2SS+         0.214     0.977         W2SS-         0.112     0.994

 Higgses masses: 
 --------------

 Light CP-even Higgs =   113.925
 Heavy CP-even Higgs =   679.220
       CP-odd  Higgs =   679.964
       Charged Higgs =   684.484
       sin(a-b)      =    -0.205
       cos(a-b)      =     0.979
\end{verbatim}
\end{small}
\newpage

\subsection{Spectrum \& parameters of  PYTHIA 6.2/00}

\begin{figure}[h] \centering
  \epsfig{file=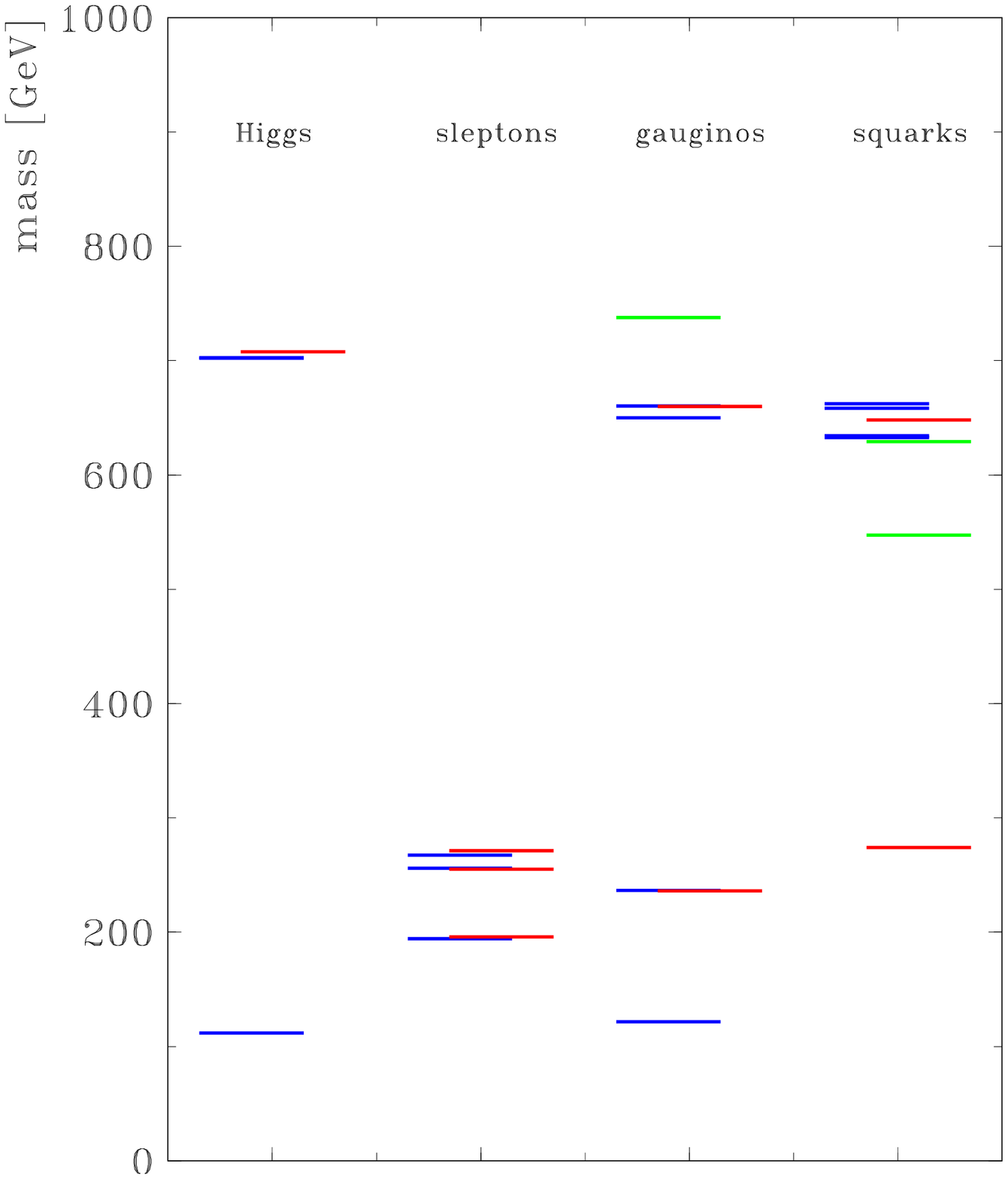,width=.9\textwidth}
  \caption{SPS 5 mass spectrum of \pythia}
\end{figure}
\newpage
\normalsize
\noindent
{\bf \pythia\  parameters} 
\begin{small}
\begin{verbatim}
   SUGRA input parameters
   ----------------------
   m_0         RMSS( 8) =   150.0    
   m_1/2       RMSS( 1) =   300.0    
   A_0         RMSS(16) =  -1000.    
   tan_beta    RMSS( 5) =   5.000    
   sign mu     RMSS( 4) =   1.000    

   sparticle masses & widths
   -------------------------
   M_se_R   194.4 ( 0.366)   M_se_L   267.1 ( 0.349)   M_sne_L  255.7 ( 0.281)
   M_sm_R   194.4 ( 0.366)   M_sm_L   267.1 ( 0.349)   M_snm_L  255.7 ( 0.281)
   M_st_1   195.9 ( 0.373)   M_st_2   271.3 ( 0.399)   M_snt_L  255.1 ( 0.276)

   M_ch0_1  121.4 ( 0.000)   M_ch0_2  236.5 ( 0.004)   M_ch0_3  649.9 (12.037)
   M_ch0_4  660.1 (24.943)
   M_ch+_1  236.2 ( 0.005)   M_ch+_2  660.0 (15.581)

   M_h0     111.9 ( 0.004)   M_H0     702.4 ( 1.405)   M_A0     702.0 ( 1.610)
   M_H+     707.6 ( 1.624)

   M_g~     737.6 (19.386)
   M_uL     658.1 ( 6.389)   M_uR     634.3 ( 1.344)   M_dL     662.1 ( 6.306)
   M_dR     632.5 ( 0.335)
   M_b1     547.2 (14.603)   M_b2     629.3 ( 0.466)   M_t1     274.0 ( 0.068)
   M_t2     647.9 (15.171)

   parameter settings IMSS, RMSS
   -----------------------------
   IMSS( 1) = 2   IMSS( 4) = 1   IMSS( 7) = 0   IMSS(10) = 0
   IMSS( 2) = 0   IMSS( 5) = 0   IMSS( 8) = 0   IMSS(11) = 0
   IMSS( 3) = 0   IMSS( 6) = 0   IMSS( 9) = 0   IMSS(12) = 0

   RMSS( 1) =   300.0       RMSS( 9) =   700.0       RMSS(17) =   1000.    
   RMSS( 2) =   244.9       RMSS(10) =   544.6       RMSS(18) = -0.2049    
   RMSS( 3) =   706.3       RMSS(11) =   628.4       RMSS(19) =   703.5    
   RMSS( 4) =   646.9       RMSS(12) =   384.8       RMSS(20) =  0.4100E-01
   RMSS( 5) =   5.000       RMSS(13) =   266.5       RMSS(21) =   1.000    
   RMSS( 6) =   267.1       RMSS(14) =   192.7       RMSS(22) =   800.0    
   RMSS( 7) =   194.4       RMSS(15) =   1000.       RMSS(23) =  0.1000E+05
   RMSS( 8) =   150.0       RMSS(16) =  -819.2       RMSS(24) =  0.1000E+05

\end{verbatim}
\end{small}
\clearpage

\subsection{Decay modes}
\normalsize
 \begin{table}[h] \centering
   \begin{tabular}{|l|c c c|c|c c c|}
   \hdick & & & \\[-2.ex]
   particle & $m_I$ & $m_S$ & $m_P$ & \ decay \ & ${\cal B}_I$ & ${\cal B}_S$ & ${\cal B}_P$ 
   \\[.5ex] \hdick
   $\ser    $ &  191.4 & 193.5  & 194.4  & $\nt_1    e^-               $ & 1.000 & 1.000 & 1.000 \\
   \hline
   $\sel    $ &  256.3 & 259.3  & 267.1  & $\nt_1    e^-               $ & 0.536 & 0.448 & 0.552 \\
              &        &        &        & $\nt_2    e^-               $ & 0.161 & 0.191 & 0.154 \\
              &        &        &        & $\cm_1    \nu_e             $ & 0.303 & 0.361 & 0.294 \\
   \hline
   $\sne    $ &  244.5 & 247.7  & 255.7  & $\nt_1    \nu_e             $ & 0.764 & 0.651 & 0.762 \\
              &        &        &        & $\nt_2    \nu_e             $ & 0.075 & 0.110 & 0.075 \\
              &        &        &        & $\cp_1    e^-               $ & 0.161 & 0.239 & 0.163 \\
   \hline
   $\smur   $ &  191.4 & 193.5  & 194.4  & $\nt_1    \mu^-             $ & 1.000 & 1.000 & 1.000 \\
   \hline
   $\smul   $ &  256.3 & 259.3  & 267.1  & $\nt_1    \mu^-             $ & 0.536 & 0.448 & 0.552 \\
              &        &        &        & $\nt_2    \mu^-             $ & 0.161 & 0.191 & 0.154 \\
              &        &        &        & $\cm_1    \nu_\mu           $ & 0.303 & 0.361 & 0.294 \\
   \hline
   $\snm    $ &  244.5 & 247.7  & 255.7  & $\nt_1    \nu_\mu           $ & 0.764 & 0.651 & 0.762 \\
              &        &        &        & $\nt_2    \nu_\mu           $ & 0.075 & 0.110 & 0.075 \\
              &        &        &        & $\cp_1    \mu^-             $ & 0.161 & 0.239 & 0.163 \\
   \hline
   $\stau_1 $ &  180.7 & 182.7  & 195.9  & $\nt_1    \tau^-            $ & 1.000 & 1.000 & 1.000 \\
   \hline
   $\stau_2 $ &  257.9 & 260.5  & 271.3  & $\nt_1    \tau^-            $ & 0.573 & 0.490 & 0.517 \\
              &        &        &        & $\nt_2    \tau^-            $ & 0.148 & 0.176 & 0.166 \\
              &        &        &        & $\cm_1    \nu_\tau          $ & 0.279 & 0.334 & 0.317 \\
   \hline
   $\snt    $ &  242.3 & 245.4  & 255.1  & $\nt_1    \nu_\tau          $ & 0.801 & 0.686 & 0.773 \\
              &        &        &        & $\nt_2    \nu_\tau          $ & 0.064 & 0.099 & 0.072 \\
              &        &        &        & $\cp_1    \tau^-            $ & 0.134 & 0.215 & 0.155 \\
   \hline
   \end{tabular}
   \caption{Slepton masses (GeV) and significant branching ratios ($>3\%$) 
            from \isajet~(I), \susygen~(S) and \pythia~(P)}
 \end{table}
  
 \begin{table}[h] \centering
   \begin{tabular}{|l|c c c|c|c c c|}
   \hdick & & & \\[-2.ex]
   particle & $m_I$ & $m_S$ & $m_P$ & \ decay \ & ${\cal B}_I$ & ${\cal B}_S$ & ${\cal B}_P$ 
   \\[.5ex] \hdick
   $\nt_1   $ &  119.5 & 121.2  & 121.4  & $                           $ & 1.000 & 1.000 & 1.000 \\
   \hline
   $\nt_2   $ &  226.3 & 223.1  & 236.5  & $\stau_1  \tau^+            $ & 0.466 & 0.465 & 0.224 \\
              &        &        &        & $\staup_1 \tau^-            $ & 0.466 & 0.465 & 0.224 \\
              &        &        &        & $\nt_1    h^0               $ &       &       & 0.416 \\
   \hline
   $\nt_3   $ &  642.8 & 628.5  & 649.9  & $\cp_1    W^-               $ & 0.125 & 0.079 & 0.139 \\
              &        &        &        & $\cm_1    W^+               $ & 0.125 & 0.079 & 0.139 \\
              &        &        &        & $\nt_1    Z^0               $ & 0.033 &       & 0.036 \\
              &        &        &        & $\nt_2    Z^0               $ & 0.106 & 0.062 & 0.118 \\
              &        &        &        & $\st_1    \bar t            $ & 0.294 & 0.373 & 0.270 \\
              &        &        &        & $\bar\st_1    t             $ & 0.294 & 0.373 & 0.270 \\
   \hline
   $\nt_4   $ &  652.9 & 639.0  & 660.1  & $\cp_1    W^-               $ & 0.064 & 0.089 & 0.068 \\
              &        &        &        & $\cm_1    W^+               $ & 0.064 & 0.089 & 0.068 \\
              &        &        &        & $\nt_2    h^0               $ & 0.049 & 0.064 & 0.054 \\
              &        &        &        & $\st_1    \bar t            $ & 0.391 & 0.352 & 0.383 \\
              &        &        &        & $\bar\st_1    t             $ & 0.391 & 0.352 & 0.383 \\
   \hline
   \end{tabular}
   \caption{Neutralino masses (GeV) and significant branching ratios ($>3\%$) 
            from \isajet~(I), \susygen~(S) and \pythia~(P)}
 \end{table}
  
 \begin{table}[h] \centering
   \begin{tabular}{|l|c c c|c|c c c|}
   \hdick & & & \\[-2.ex]
   particle & $m_I$ & $m_S$ & $m_P$ & \ decay \ & ${\cal B}_I$ & ${\cal B}_S$ & ${\cal B}_P$ 
   \\[.5ex] \hdick
   $\cp_1   $ &  226.3 & 223.0  & 236.2  & $\nt_1    W^+               $ & 0.205 & 0.369 & 0.667 \\
              &        &        &        & $\st_1    \bar b            $ & 0.163 & 0.548 &       \\
              &        &        &        & $\staup_1 \nu_\tau          $ & 0.622 & 0.080 & 0.333 \\
   \hline
   $\cp_2   $ &  652.7 & 638.8  & 660.0  & $\nt_2    W^+               $ & 0.109 & 0.112 & 0.117 \\
              &        &        &        & $\st_1    \bar b            $ & 0.647 & 0.723 & 0.623 \\
              &        &        &        & $\cp_1    Z^0               $ & 0.103 & 0.115 & 0.109 \\
              &        &        &        & $\cp_1    h^0               $ & 0.093 &       & 0.099 \\
   \hline
   \end{tabular}
   \caption{Chargino masses (GeV) and significant branching ratios ($>3\%$) 
            from \isajet~(I), \susygen~(S) and \pythia~(P)}
 \end{table}
  
 \begin{table}[h] \centering
   \begin{tabular}{|l|c c c|c|c c c|}
   \hdick & & & \\[-2.ex]
   particle & $m_I$ & $m_S$ & $m_P$ & \ decay \ & ${\cal B}_I$ & ${\cal B}_S$ & ${\cal B}_P$ 
   \\[.5ex] \hdick
   $h^0     $ &  119.8 & 113.9  & 111.9  & $\tau^-   \tau^+            $ & 0.048 & 0.080 & 0.066 \\
              &        &        &        & $b        \bar b            $ & 0.798 & 0.790 & 0.796 \\
              &        &        &        & $c        \bar c            $ & 0.036 &       & 0.047 \\
              &        &        &        & $g         g                $ &       & 0.034 & 0.037 \\
              &        &        &        & $W^+      W^-               $ &       & 0.060 & 0.045 \\
   \hline
   $H^0     $ &  694.0 & 679.2  & 702.4  & $b        \bar b            $ & 0.058 & 0.030 & 0.237 \\
              &        &        &        & $t        \bar t            $ & 0.162 & 0.139 & 0.708 \\
              &        &        &        & $\st_1    \bar\st_1         $ & 0.735 & 0.790 &       \\
   \hline
   $A^0     $ &  693.9 & 680.0  & 702.0  & $b        \bar b            $ & 0.065 & 0.121 & 0.208 \\
              &        &        &        & $t        \bar t            $ & 0.214 & 0.641 & 0.764 \\
              &        &        &        & $\st_1    \bar\st_1         $ & 0.661 &       &       \\
              &        &        &        & $\cp_1        \cm_1         $ &       & 0.111 &       \\
   \hline
   $H^+     $ &  698.5 & 684.5  & 707.6  & $t        \bar b            $ & 0.885 & 0.891 & 0.973 \\
              &        &        &        & $\cp_1    \nt_1             $ & 0.051 & 0.058 &       \\
              &        &        &        & $\staup_1 \snt              $ & 0.036 &       &       \\
   \hline
   \end{tabular}
   \caption{Higgs masses (GeV) and significant branching ratios ($>3\%$) 
            from \isajet~(I), \susygen~(S) and \pythia~(P)}
 \end{table}
  
 \begin{table}[h] \centering
   \begin{tabular}{|l|c c c|c|c c c|}
   \hdick & & & \\[-2.ex]
   particle & $m_I$ & $m_S$ & $m_P$ & \ decay \ & ${\cal B}_I$ & ${\cal B}_S$ & ${\cal B}_P$ 
   \\[.5ex] \hdick
   $\st_1   $ &  220.7 & 210.0  & 274.0  & $\nt_1    c                 $ & 1.000 & 1.000 &       \\
              &        &        &        & $\cp_1    b                 $ &       &       & 1.000 \\
   \hline
   $\st_2   $ &  644.6 & 631.6  & 647.9  & $\cp_1    b                 $ & 0.125 & 0.704 & 0.155 \\
              &        &        &        & $Z^0      \st_1             $ & 0.612 &       & 0.611 \\
              &        &        &        & $h^0      \st_1             $ & 0.139 &       & 0.112 \\
              &        &        &        & $W^+      \sb_1             $ & 0.060 &       & 0.043 \\
              &        &        &        & $\nt_1    t                 $ &       & 0.061 &       \\
              &        &        &        & $\nt_2    t                 $ & 0.048 & 0.235 & 0.060 \\
   \hline
   $\sb_1   $ &  535.9 & 560.8  & 547.2  & $\nt_2    b                 $ & 0.080 & 0.389 & 0.101 \\
              &        &        &        & $\cm_1    t                 $ & 0.101 & 0.584 & 0.129 \\
              &        &        &        & $W^-      \st_1             $ & 0.813 &       & 0.764 \\
   \hline
   $\sb_2   $ &  623.0 & 654.5  & 629.3  & $\nt_1    b                 $ & 0.285 & 0.894 & 0.711 \\
              &        &        &        & $\nt_2    b                 $ &       & 0.041 &       \\
              &        &        &        & $\cm_1    t                 $ & 0.042 & 0.062 &       \\
              &        &        &        & $W^-      \st_1             $ & 0.644 &       & 0.238 \\
   \hline
   \end{tabular}
   \caption{Light squark masses (GeV) and significant branching ratios ($>3\%$) 
            from \isajet~(I), \susygen~(S) and \pythia~(P)}
 \end{table}

\clearpage

\section{SPS 6 -- MSSM scenario}
\setcounter{figure}{0}
\setcounter{table}{0}

\large\boldmath
\hspace{20mm}
\begin{tabular}{|l c|}
  \hline
  $m_0$       & $ 150~\GeV$ \\ 
  $m_{1/2}$   & $ 300~\GeV$ \\
  $A_0$       & $ \ \ \ 0~\GeV$ \\
  $\tan\beta$ & $10$       \\
  ${\rm sign}~\mu$ & $+$ \\
  \hline
\end{tabular} \hspace{10mm}
\begin{tabular}{l}
   {\bf `non-unified gaugino masses'} \\
    $M_1 = 480~\GeV, \ M_2 = M_3 = 300~\GeV$ \\
    $m_0 = 0.5\,M_2$ \\ 
\end{tabular}
\unboldmath\normalsize
\bigskip

\subsection{Spectrum \& parameters of ISAJET 7.58}

\begin{figure}[h] \centering
  \epsfig{file=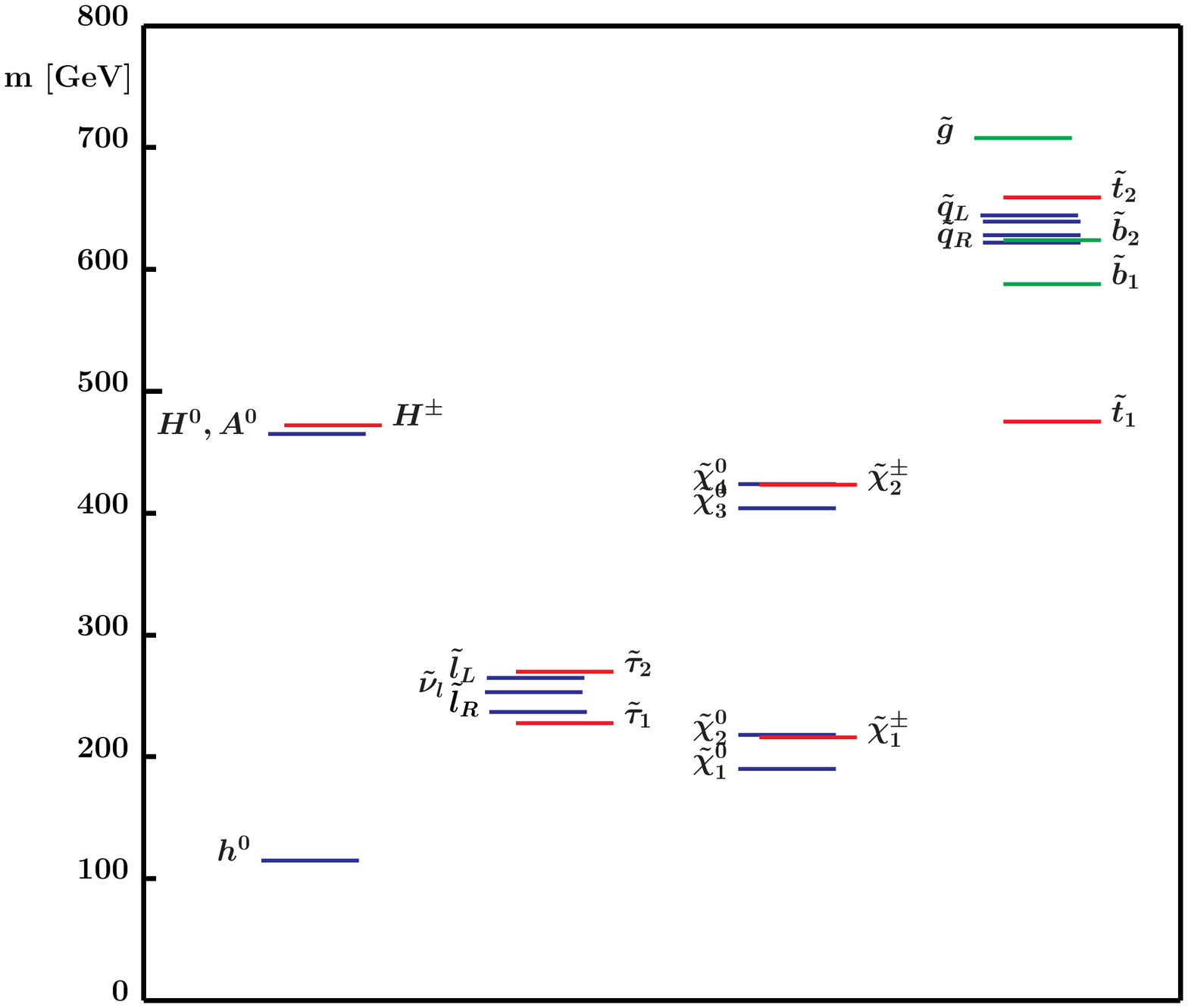,width=.9\textwidth}
  \caption{SPS 6 mass spectrum of \isajet}
\end{figure}

\clearpage
\normalsize
\noindent
{\bf \isajet\ parameters} 
\begin{small}
\begin{verbatim}
 Minimal supergravity (mSUGRA) model:

 M_0,  M_(1/2),  A_0,  tan(beta),  sgn(mu),  M_t =
   150.000   300.000     0.000    10.000     1.0   175.000

 M_1(GUT)=   480.00    M_2(GUT)=   300.00    M_3(GUT)=   300.00

 ISASUGRA unification:
 M_GUT      = 0.190E+17   g_GUT          =0.713      alpha_GUT =0.040
 FT_GUT     = 0.478       FB_GUT         = 0.047     FL_GUT = 0.069

 1/alpha_em =  127.65     sin**2(thetaw) =0.2308     alpha_s   =0.119
 M_1        =  195.89     M_2            =  232.06   M_3       =  691.24
 mu(Q)      =  393.89     B(Q)           =   54.73   Q         =  548.25
 M_H1^2     = 0.612E+05   M_H2^2         =-0.151E+06

 ISAJET masses (with signs):
 M(GL)  =   708.45
 M(UL)  =   639.13   M(UR)  =   628.29   M(DL)  =   644.01   M(DR) =   622.27
 M(B1)  =   588.93   M(B2)  =   624.45   M(T1)  =   476.16   M(T2) =   660.67
 M(SN)  =   252.77   M(EL)  =   264.88   M(ER)  =   236.76
 M(NTAU)=   251.75   M(TAU1)=   227.92   M(TAU2)=   269.64
 M(Z1)  =  -189.37   M(Z2)  =  -217.91   M(Z3)  =   399.31   M(Z4) =  -419.98
 M(W1)  =  -215.34   M(W2)  =  -418.91
 M(HL)  =   114.83   M(HH)  =   463.62   M(HA)  =   463.04   M(H+) =   470.11

 theta_t=   0.9831   theta_b=   0.3882   theta_l=   1.1799   alpha_h=   0.1079

 NEUTRALINO MASSES (SIGNED) =  -189.375  -217.905   399.307  -419.978
 EIGENVECTOR 1       =  -0.11796   0.21288   0.25953  -0.93456
 EIGENVECTOR 2       =   0.13928  -0.22503  -0.90923  -0.32133
 EIGENVECTOR 3       =  -0.70987  -0.69812   0.08065  -0.04702
 EIGENVECTOR 4       =  -0.68027   0.64550  -0.31533   0.14534

 CHARGINO MASSES (SIGNED)  =  -215.342  -418.910
 GAMMAL, GAMMAR             =   1.97743   1.81705


 ISAJET equivalent input:
 MSSMA:   708.45  393.89  463.04   10.00
 MSSMB:   641.33  621.76  629.29  260.71  232.75
 MSSMC:   591.24  618.96  516.96  259.71  230.48 -569.95 -811.30 -213.39
 MSSMD: SAME AS MSSMB (DEFAULT)
 MSSME:   195.89  232.06

\end{verbatim}
\end{small}
\clearpage

\subsection{Spectrum \& parameters of SUSYGEN 3.00/25}

\begin{figure}[h] \centering
  \epsfig{file=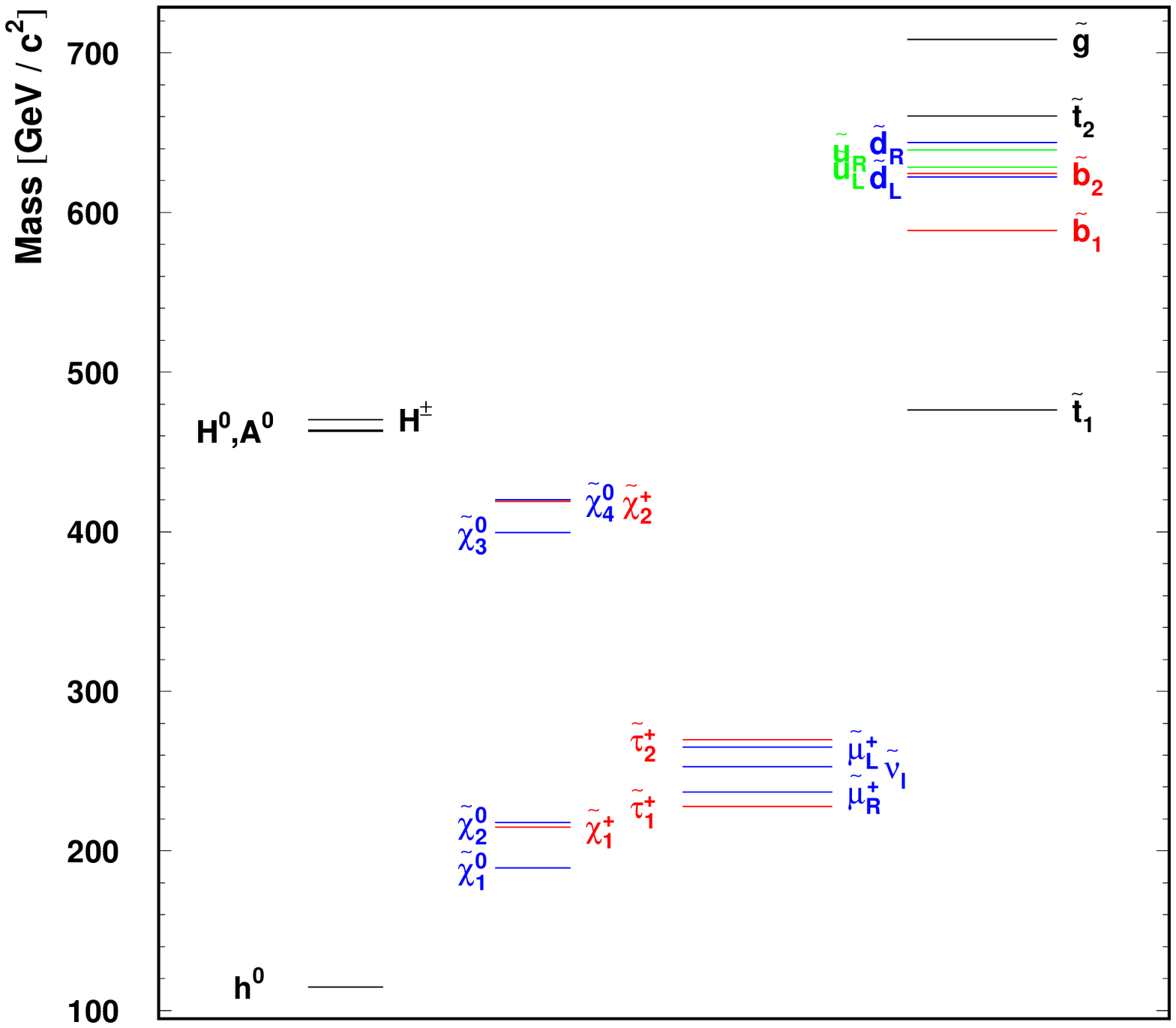,width=.9\textwidth}
  \caption{SPS 6 mass spectrum of \susygen}
\end{figure}
\newpage
\normalsize
\noindent
{\bf \susygen\  parameters} 
\begin{small}
\begin{verbatim}
 Susygen inputs:
 --------------

 m0      =   150.000   TANB     =    10.000
 m1/2    =   300.000    mu/|mu| =     1
 A0      =     0.000

 Sparticle masses: 
 ----------------

 SUPR      639.  SUPL       628.
 SDNR      644.  SDNL       622.
 SELR      237.  SELL       265.
 SNU       253.
 STP1      476.  STP2       661. cosmix =  0.554
 SBT1      589.  SBT2       624. cosmix =  0.926
 STA1      228.  STA2       270. cosmix=   0.381
 SGLU      708.


 Gaugino masses:
 --------------

 M1 =    195.890 M2 =    232.060 M3 =    691.240

 NEUTRALINO m, CP, ph/zi/ha/hb 1 =  189.3  1. -0.693  0.678 -0.139 -0.201
 NEUTRALINO m, CP, ph/zi/ha/hb 2 =  217.8  1.  0.720  0.641 -0.161 -0.210
 NEUTRALINO m, CP, ph/zi/ha/hb 3 =  399.3 -1. -0.002  0.094 -0.637  0.765
 NEUTRALINO m, CP, ph/zi/ha/hb 4 =  420.1  1. -0.025 -0.347 -0.741 -0.574

 CHARGINO MASSES    =   215.088   419.050
 CHARGINO ETA       =    -1.000     1.000

 U matrix      WINO      HIGGSINO   V matrix      WINO      HIGGSINO 
 W1SS+        -0.918     0.396      W1SS-         0.970    -0.244
 W2SS+         0.396     0.918      W2SS-         0.244     0.970

 Higgses masses: 
 --------------

 Light CP-even Higgs =   114.830
 Heavy CP-even Higgs =   463.620
       CP-odd  Higgs =   463.040
       Charged Higgs =   470.110
       sin(a-b)      =     0.108
       cos(a-b)      =     0.994

\end{verbatim}
\end{small}
\clearpage

\subsection{Spectrum \& parameters of  PYTHIA 6.2/00}

\begin{figure}[h] \centering
  \epsfig{file=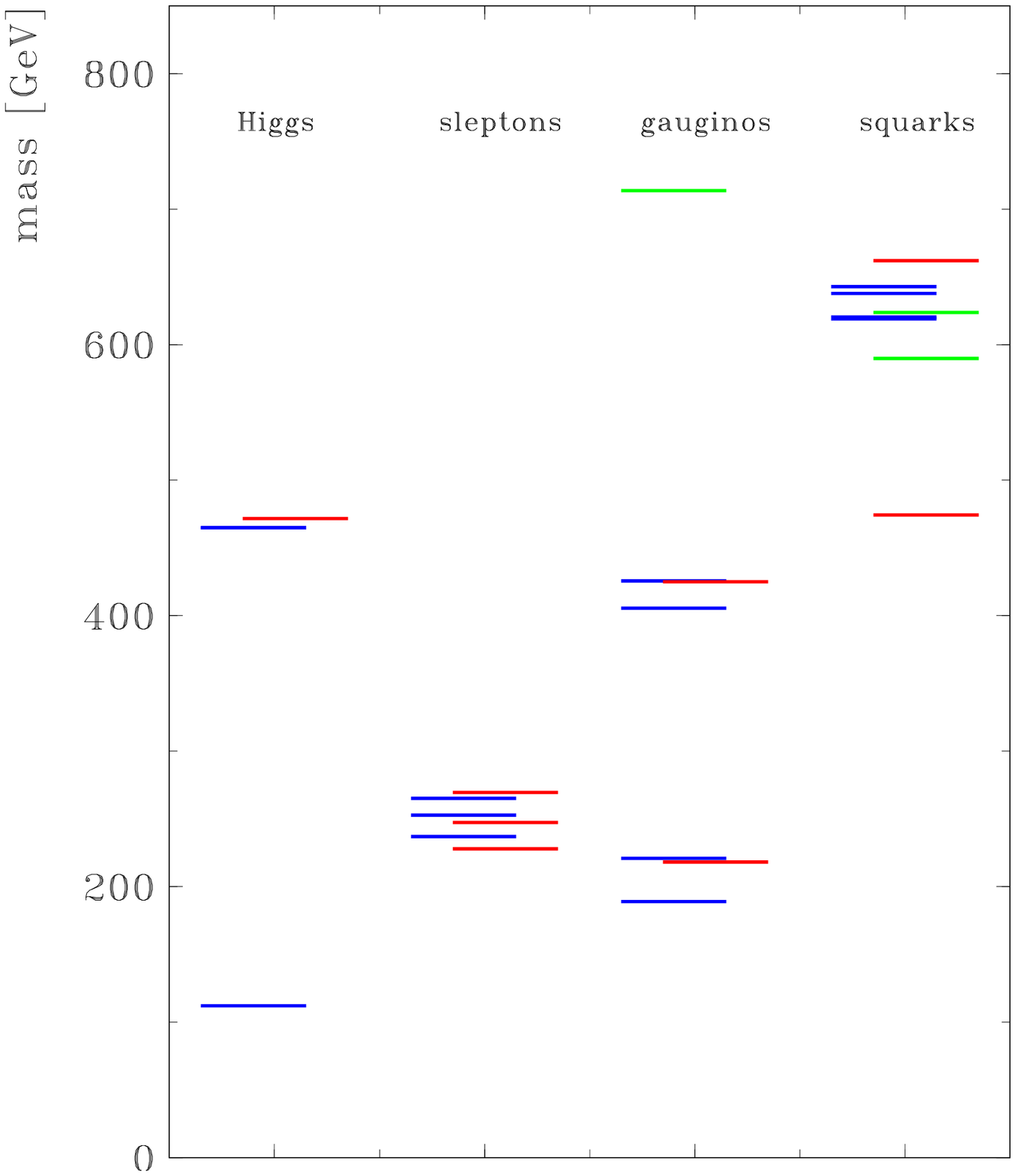,width=.9\textwidth}
  \caption{SPS 6 mass spectrum of \pythia}
\end{figure}
\newpage
\normalsize
\noindent
{\bf \pythia\  parameters} 
\begin{small}
\begin{verbatim}
 MSSM input parameters
 ---------------------
   M_1         RMSS( 1) =   195.0    
   M_2         RMSS( 2) =   235.0    
   M_3         RMSS( 3) =   680.0    
   tan_beta    RMSS( 5) =   10.00    
   mu          RMSS( 4) =   400.0    

   sparticle masses & widths
   -------------------------
   M_se_R   237.0 ( 0.146)   M_se_L   265.0 ( 0.343)   M_sne_L  252.7 ( 0.281)
   M_sm_R   237.0 ( 0.146)   M_sm_L   265.0 ( 0.343)   M_snm_L  252.7 ( 0.281)
   M_st_1   227.8 ( 0.095)   M_st_2   269.6 ( 0.369)   M_snt_L  247.5 ( 0.224)

   M_ch0_1  189.1 ( 0.000)   M_ch0_2  220.7 ( 0.000)   M_ch0_3  405.3 ( 2.256)
   M_ch0_4  425.6 ( 2.918)
   M_ch+_1  218.2 ( 0.000)   M_ch+_2  424.9 ( 2.828)

   M_h0     112.2 ( 0.004)   M_H0     465.0 ( 1.051)   M_A0     464.8 ( 1.124)
   M_H+     471.5 ( 1.022)

   M_g~     713.5 (10.510)
   M_uL     637.8 ( 6.177)   M_uR     619.0 ( 1.145)   M_dL     642.7 ( 5.991)
   M_dR     620.5 ( 0.287)
   M_b1     589.7 ( 4.359)   M_b2     623.6 ( 1.080)   M_t1     474.3 ( 2.734)
   M_t2     662.0 ( 9.357)

   parameter settings IMSS, RMSS
   -----------------------------
   IMSS( 1) = 1   IMSS( 4) = 1   IMSS( 7) = 0   IMSS(10) = 0
   IMSS( 2) = 0   IMSS( 5) = 0   IMSS( 8) = 0   IMSS(11) = 0
   IMSS( 3) = 0   IMSS( 6) = 0   IMSS( 9) = 0   IMSS(12) = 0

   RMSS( 1) =   195.0       RMSS( 9) =   620.0       RMSS(17) =   0.000    
   RMSS( 2) =   235.0       RMSS(10) =   590.0       RMSS(18) = -0.1078    
   RMSS( 3) =   680.0       RMSS(11) =   620.0       RMSS(19) =   465.0    
   RMSS( 4) =   400.0       RMSS(12) =   515.0       RMSS(20) =  0.4100E-01
   RMSS( 5) =   10.00       RMSS(13) =   260.0       RMSS(21) =   1.000    
   RMSS( 6) =   265.0       RMSS(14) =   230.0       RMSS(22) =   620.0    
   RMSS( 7) =   237.0       RMSS(15) =   0.000       RMSS(23) =  0.1000E+05
   RMSS( 8) =   640.0       RMSS(16) =  -560.0       RMSS(24) =  0.1000E+05
\end{verbatim}
\end{small}
\clearpage

\subsection{Branching ratios}
\normalsize
 \begin{table}[h] \centering
   \begin{tabular}{|l|c c c|c|c c c|}
   \hdick & & & \\[-2.ex]
   particle & $m_I$ & $m_S$ & $m_P$ & \ decay \ & ${\cal B}_I$ & ${\cal B}_S$ & ${\cal B}_P$ 
   \\[.5ex] \hdick
   $\ser    $ &  236.8 & 236.8  & 237.0  & $\nt_1    e^-               $ & 0.979 & 0.979  & 0.988\\
   \hline
   $\sel    $ &  264.9 & 264.9  & 265.0  & $\nt_1    e^-               $ & 0.046 & 0.045  & 0.070\\
              &        &        &        & $\nt_2    e^-               $ & 0.370 & 0.370  & 0.355\\
              &        &        &        & $\cm_1    \nu_e             $ & 0.583 & 0.585  & 0.575\\
   \hline
   $\sne    $ &  252.8 & 252.8  & 252.7  & $\nt_1    \nu_e             $ & 0.394 & 0.392 & 0.409 \\
              &        &        &        & $\nt_2    \nu_e             $ & 0.121 & 0.121 & 0.126 \\
              &        &        &        & $\cp_1    e^-               $ & 0.485 & 0.487 & 0.465 \\
   \hline
   $\smur   $ &  236.8 & 236.8  & 237.0  & $\nt_1    \mu^-             $ & 0.979 & 0.979 & 0.988 \\
   \hline
   $\smul   $ &  264.9 & 264.9  & 265.0  & $\nt_1    \mu^-             $ & 0.046 & 0.045 & 0.070 \\
              &        &        &        & $\nt_2    \mu^-             $ & 0.370 & 0.370 & 0.355 \\
              &        &        &        & $\cm_1    \nu_\mu           $ & 0.583 & 0.585 & 0.575 \\
   \hline
   $\snm    $ &  252.8 & 252.8  & 252.7  & $\nt_1    \nu_\mu           $ & 0.394 & 0.392 & 0.409 \\
              &        &        &        & $\nt_2    \nu_\mu           $ & 0.121 & 0.121 & 0.126 \\
              &        &        &        & $\cp_1    \mu^-             $ & 0.485 & 0.487 & 0.465 \\
   \hline
   $\stau_1 $ &  227.9 & 227.9  & 227.8  & $\nt_1    \tau^-            $ & 0.933 & 0.932 & 0.966 \\
              &        &        &        & $\cm_1    \nu_\tau          $ & 0.040 & 0.041 &       \\
   \hline
   $\stau_2 $ &  269.6 & 269.6  & 269.6  & $\nt_1    \tau^-            $ & 0.135 & 0.133 & 0.159 \\
              &        &        &        & $\nt_2    \tau^-            $ & 0.346 & 0.346 & 0.329 \\
              &        &        &        & $\cm_1    \nu_\tau          $ & 0.519 & 0.521 & 0.511 \\
   \hline
   $\snt    $ &  251.7 & 251.8  & 247.5  & $\nt_1    \nu_\tau          $ & 0.400 & 0.398 & 0.450 \\
              &        &        &        & $\nt_2    \nu_\tau          $ & 0.120 & 0.119 & 0.115 \\
              &        &        &        & $\cp_1    \tau^-            $ & 0.480 & 0.483 & 0.435 \\
   \hline
   \end{tabular}
   \caption{Slepton masses (GeV) and significant branching ratios ($>3\%$) 
            from \isajet~(I), \susygen~(S) and \pythia~(P)}
 \end{table}
  
 \begin{table}[h] \centering
   \begin{tabular}{|l|c c c|c|c c c|}
   \hdick & & & \\[-2.ex]
   particle & $m_I$ & $m_S$ & $m_P$ & \ decay \ & ${\cal B}_I$ & ${\cal B}_S$ & ${\cal B}_P$ 
   \\[.5ex] \hdick
   $\nt_1   $ &  189.4 & 189.3  & 189.1  & $                           $ & 1.000 & 1.000 & 1.000 \\
   \hline
   $\nt_2   $ &  217.9 & 217.8  & 220.7  & $\nt_1    e^-      e^+      $ & 0.131 & 0.169 & 0.123 \\
              &        &        &        & $\nt_1    \mu^-    \mu^+    $ & 0.131 & 0.169 & 0.123 \\
              &        &        &        & $\nt_1    \tau^-   \tau^+   $ & 0.375 & 0.195 & 0.272 \\
              &        &        &        & $\nt_1    \nu_e    \anue    $ & 0.093 & 0.119 & 0.116 \\
              &        &        &        & $\nt_1    \nu_\mu  \anum    $ & 0.093 & 0.119 & 0.116 \\
              &        &        &        & $\nt_1    \nu_\tau \anut    $ & 0.099 & 0.127 & 0.160 \\
   \hline
   $\nt_3   $ &  399.3 & 399.3  & 405.3  & $\cp_1    W^-               $ & 0.305 & 0.350 & 0.305 \\
              &        &        &        & $\cm_1    W^+               $ & 0.305 & 0.350 & 0.305 \\
              &        &        &        & $\nt_1    Z^0               $ & 0.192 & 0.144 & 0.176 \\
              &        &        &        & $\nt_2    Z^0               $ & 0.152 & 0.114 & 0.165 \\
   \hline
   $\nt_4   $ &  420.0 & 420.1  & 425.6  & $\cp_1    W^-               $ & 0.279 & 0.337 & 0.276 \\
              &        &        &        & $\cm_1    W^+               $ & 0.279 & 0.337 & 0.276 \\
              &        &        &        & $\nt_1    h^0               $ & 0.121 & 0.072 & 0.111 \\
              &        &        &        & $\nt_2    h^0               $ & 0.099 & 0.059 & 0.114 \\
   \hline
   \end{tabular}
   \caption{Neutralino masses (GeV) and significant branching ratios ($>3\%$) 
            from \isajet~(I), \susygen~(S) and \pythia~(P)}
 \end{table}
  
 \begin{table}[h] \centering
   \begin{tabular}{|l|c c c|c|c c c|}
   \hdick & & & \\[-2.ex]
   particle & $m_I$ & $m_S$ & $m_P$ & \ decay \ & ${\cal B}_I$ & ${\cal B}_S$ & ${\cal B}_P$ 
   \\[.5ex] \hdick
   $\cp_1   $ &  215.3 & 215.1  & 218.2  & $\nt_1    u        \bar d   $ & 0.295 &  0.303& 0.292 \\
              &        &        &        & $\nt_1    c        \bar s   $ & 0.295 &  0.296& 0.292 \\
              &        &        &        & $\nt_1    e^+      \nu_e    $ & 0.133 &  0.137& 0.136 \\
              &        &        &        & $\nt_1    \mu^+    \nu_\mu  $ & 0.133 &  0.137& 0.136 \\
              &        &        &        & $\nt_1    \tau^+   \nu_\tau $ & 0.144 &  0.128& 0.146 \\
   \hline
   $\cp_2   $ &  418.9 & 419.0  & 424.9  & $\nt_2    W^+               $ & 0.363 & 0.376 & 0.356 \\
              &        &        &        & $\selp    \nu_e             $ & 0.038 & 0.040 & 0.038 \\
              &        &        &        & $\smulp   \nu_\mu           $ & 0.038 & 0.040 & 0.038 \\
              &        &        &        & $\staup_2 \nu_\tau          $ & 0.041 & 0.037 & 0.042 \\
              &        &        &        & $\cp_1    Z^0               $ & 0.265 & 0.433 & 0.263 \\
              &        &        &        & $\cp_1    h^0               $ & 0.185 &       & 0.188 \\
   \hline
   \end{tabular}
   \caption{Chargino masses (GeV) and significant branching ratios ($>3\%$) 
            from \isajet~(I), \susygen~(S) and \pythia~(P)}
 \end{table}
  
 \begin{table}[h] \centering
   \begin{tabular}{|l|c c c|c|c c c|}
   \hdick & & & \\[-2.ex]
   particle & $m_I$ & $m_S$ & $m_P$ & \ decay \ & ${\cal B}_I$ & ${\cal B}_S$ & ${\cal B}_P$ 
   \\[.5ex] \hdick
   $h^0     $ &  114.8 & 114.8  & 112.2  & $\tau^-   \tau^+            $ & 0.050 & 0.069 & 0.067 \\
              &        &        &        & $b        \bar b            $ & 0.837 & 0.799 & 0.802 \\
              &        &        &        & $c        \bar c            $ & 0.036 &       & 0.044 \\
              &        &        &        & $g         g                $ &       & 0.060 & 0.034 \\
              &        &        &        & $W^+      W^-               $ &       & 0.047 & 0.043 \\
   \hline
   $H^0     $ &  463.6 & 463.6  & 464.9  & $\tau^-   \tau^+            $ & 0.060 & 0.066 & 0.092 \\
              &        &        &        & $b        \bar b            $ & 0.817 & 0.560 & 0.818 \\
              &        &        &        & $t        \bar t            $ & 0.074 & 0.111 & 0.078 \\
              &        &        &        & $\cp_1        \cm_1         $ &       & 0.090 &       \\
              &        &        &        & $\nt_1        \nt_3         $ &       & 0.064 &       \\
              &        &        &        & $\nt_2        \nt_2         $ &       & 0.036 &       \\
   \hline
   $A^0     $ &  463.0 & 463.0  &  464.8 & $\tau^-   \tau^+            $ & 0.051 & 0.062 & 0.086 \\
              &        &        &        & $b        \bar b            $ & 0.693 & 0.529 & 0.767 \\
              &        &        &        & $t        \bar t            $ & 0.118 & 0.141 & 0.144 \\
              &        &        &        & $\nt_1    \nt_1             $ & 0.036 &       &       \\
              &        &        &        & $\nt_1    \nt_2             $ & 0.060 &       &       \\
              &        &        &        & $\cp_1        \cm_1         $ &       & 0.263 &       \\
    \hline
   $H^+     $ &  470.1 & 470.1  & 471.5  & $\nu_\tau \tau^+            $ & 0.076 & 0.077 & 0.096 \\
              &        &        &        & $t        \bar b            $ & 0.839 & 0.746 & 0.902 \\
              &        &        &        & $\cp_1    \nt_1             $ & 0.074 & 0.138 &       \\
              &        &        &        & $\cp_2    \nt_1             $ &       & 0.032 &       \\
   \hline
   \end{tabular}
   \caption{Higgs masses (GeV) and significant branching ratios ($>3\%$) 
            from \isajet~(I), \susygen~(S) and \pythia~(P)}
 \end{table}
  
 \begin{table}[h] \centering
   \begin{tabular}{|l|c c c|c|c c c|}
   \hdick & & & \\[-2.ex]
   particle & $m_I$ & $m_S$ & $m_P$ & \ decay \ & ${\cal B}_I$ & ${\cal B}_S$ & ${\cal B}_P$ 
   \\[.5ex] \hdick
   $\st_1   $ &  476.2 & 476.2  & 474.3  & $\nt_1    t                 $ & 0.134 & 0.115 & 0.140 \\
              &        &        &        & $\nt_2    t                 $ & 0.166 & 0.143 & 0.163 \\
              &        &        &        & $\cp_1    b                 $ & 0.624 & 0.639 & 0.635 \\
              &        &        &        & $\cp_2    b                 $ & 0.075 & 0.103 & 0.062 \\
   \hline
   $\st_2   $ &  660.7 & 660.7  & 662.0  & $\cp_1    b                 $ & 0.182 & 0.200 & 0.180 \\
              &        &        &        & $\cp_2    b                 $ & 0.210 & 0.359 & 0.203 \\
              &        &        &        & $Z^0      \st_1             $ & 0.155 &       & 0.169 \\
              &        &        &        & $h^0      \st_1             $ & 0.036 &       &       \\
              &        &        &        & $\nt_2    t                 $ & 0.088 & 0.060 & 0.089 \\
              &        &        &        & $\nt_3    t                 $ & 0.072 & 0.181 & 0.076 \\
              &        &        &        & $\nt_4    t                 $ & 0.249 & 0.196 & 0.247 \\
   \hline
   $\sb_1   $ &  588.9 & 588.9  & 589.7  & $\nt_1    b                 $ & 0.102 & 0.113 & 0.072 \\
              &        &        &        & $\nt_2    b                 $ & 0.281 & 0.282 & 0.293 \\
              &        &        &        & $\cm_1    t                 $ & 0.479 & 0.559 & 0.481 \\
              &        &        &        & $W^-      \st_1             $ & 0.115 &       & 0.140 \\
   \hline
   $\sb_2   $ &  624.5 & 624.5  & 623.6  & $\nt_1    b                 $ & 0.121 & 0.133 & 0.253 \\
              &        &        &        & $\nt_2    b                 $ & 0.099 & 0.108 & 0.152 \\
              &        &        &        & $\nt_3    b                 $ & 0.056 & 0.179 &       \\
              &        &        &        & $\nt_4    b                 $ & 0.078 & 0.201 &       \\
              &        &        &        & $\cm_1    t                 $ & 0.127 & 0.111 & 0.235 \\
              &        &        &        & $\cm_2    t                 $ & 0.325 & 0.268 & 0.159 \\
              &        &        &        & $W^-      \st_1             $ & 0.193 &       & 0.191 \\
   \hline
   \end{tabular}
   \caption{Light squark masses (GeV) and significant branching ratios ($>3\%$) 
            from \isajet~(I), \susygen~(S) and \pythia~(P)}
 \end{table}

\clearpage

\section{SPS 7 -- GMSB scenario}
\setcounter{figure}{0}
\setcounter{table}{0}

\large\boldmath
\hspace{20mm}
\begin{tabular}{|l c|}
  \hline
  $\Lambda$   & $ 40~\TeV$ \\ 
  $M_{mess}$  & $ 80~\TeV$ \\
  $N_{mess}$  & $ 3$ \\
  $\tan\beta$ & $ 15$       \\
  ${\rm sign}~\mu$ & $+$ \\
  \hline
\end{tabular} \hspace{10mm}
\begin{tabular}{l}
   {\bf NLSP = $\tilde\tau$} \\
    $M_{mess} = 2\,\Lambda$  \\
\end{tabular}
\unboldmath\normalsize
\bigskip

\subsection{Spectrum \& parameters of ISAJET 7.58}

\begin{figure}[h] \centering
  \epsfig{file=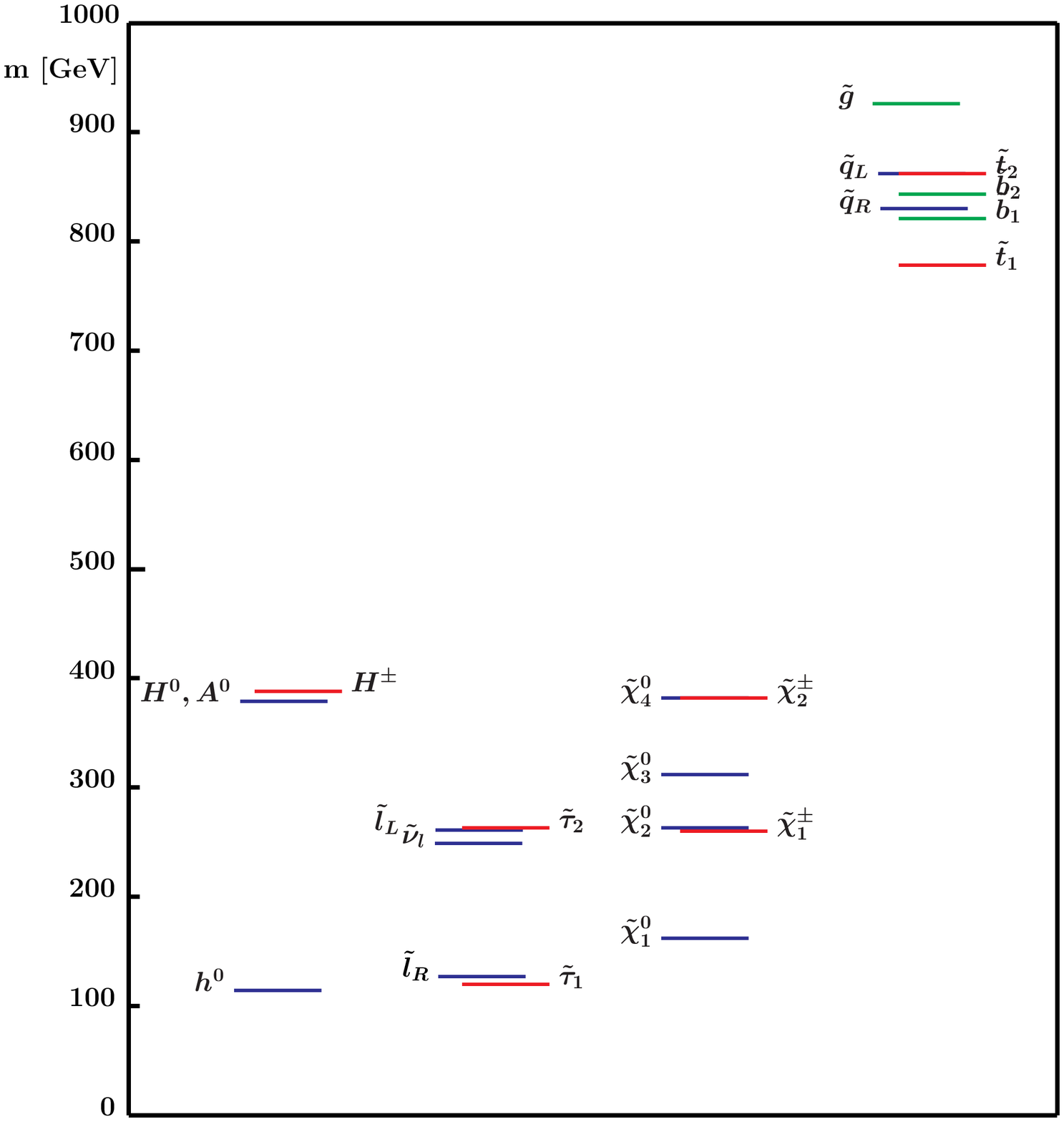,width=.9\textwidth}
  \caption{SPS 7 mass spectrum of \isajet}
\end{figure}

\clearpage
\normalsize
\noindent
{\bf \isajet\ parameters} 
\begin{small}
\begin{verbatim}
 Minimal gauge mediated (GMSB) model:

 Lambda,  M_mes,  N_5,  tan(beta),  sgn(mu),  M_t,  C_grav=
 0.400E+05 0.800E+05     3.000    15.000     1.0   175.000  0.100E+01

 GMSB2 model input:
 Rsl,    dmH_d^2,   dmH_u^2,     d_Y,     N5_1,  N5_2,  N5_3=
  1.000  0.000E+00  0.000E+00  0.000E+00    3.000  3.000  3.000

 M(gravitino)= 0.770E-09

 1/alpha_em =  127.86     sin**2(thetaw) =0.2311     alpha_s   =0.118
 M_1        =  168.59     M_2            =  326.81   M_3       =  895.45
 mu(Q)      =  300.03     B(Q)           =   31.35   Q         =  839.57
 M_H1^2     = 0.574E+05   M_H2^2         =-0.746E+05

 ISAJET masses (with signs):
 M(GL)  =   926.04
 M(UL)  =   859.66   M(UR)  =   830.54   M(DL)  =   863.34   M(DR) =   828.93
 M(B1)  =   822.17   M(B2)  =   843.35   M(T1)  =   779.09   M(T2) =   863.00
 M(SN)  =   249.06   M(EL)  =   261.47   M(ER)  =   127.43
 M(NTAU)=   248.62   M(TAU1)=   120.45   M(TAU2)=   263.40
 M(Z1)  =  -161.65   M(Z2)  =  -260.06   M(Z3)  =   306.26   M(Z4) =  -379.94
 M(W1)  =  -256.33   M(W2)  =  -379.45
 M(HL)  =   113.57   M(HH)  =   378.37   M(HA)  =   377.89   M(H+) =   386.70

 theta_t=   1.1366   theta_b=   1.0603   theta_l=   1.4237   alpha_h=   0.0765

 NEUTRALINO MASSES (SIGNED) =  -161.653  -260.057   306.255  -379.941
 EIGENVECTOR 1       =   0.13633  -0.23326  -0.07337   0.96001
 EIGENVECTOR 2       =   0.48781  -0.55279  -0.62706  -0.25151
 EIGENVECTOR 3       =   0.71148   0.69498  -0.08382   0.06142
 EIGENVECTOR 4       =  -0.48709   0.39625  -0.77096   0.10653

 CHARGINO MASSES (SIGNED)  =  -256.330  -379.452
 GAMMAL, GAMMAR             =   2.54887   2.38244


 ISAJET equivalent input:
 MSSMA:   926.04  300.03  377.89   15.00
 MSSMB:   861.32  828.55  831.31  257.19  119.73
 MSSMC:   836.27  826.88  780.14  256.77  117.61 -319.43 -350.48  -38.97
 MSSMD: SAME AS MSSMB (DEFAULT)
 MSSME:   168.59  326.81

\end{verbatim}
\end{small}

\clearpage
\subsection{Spectrum \& parameters of SUSYGEN 3.00/27}

\begin{figure}[h] \centering
  \epsfig{file=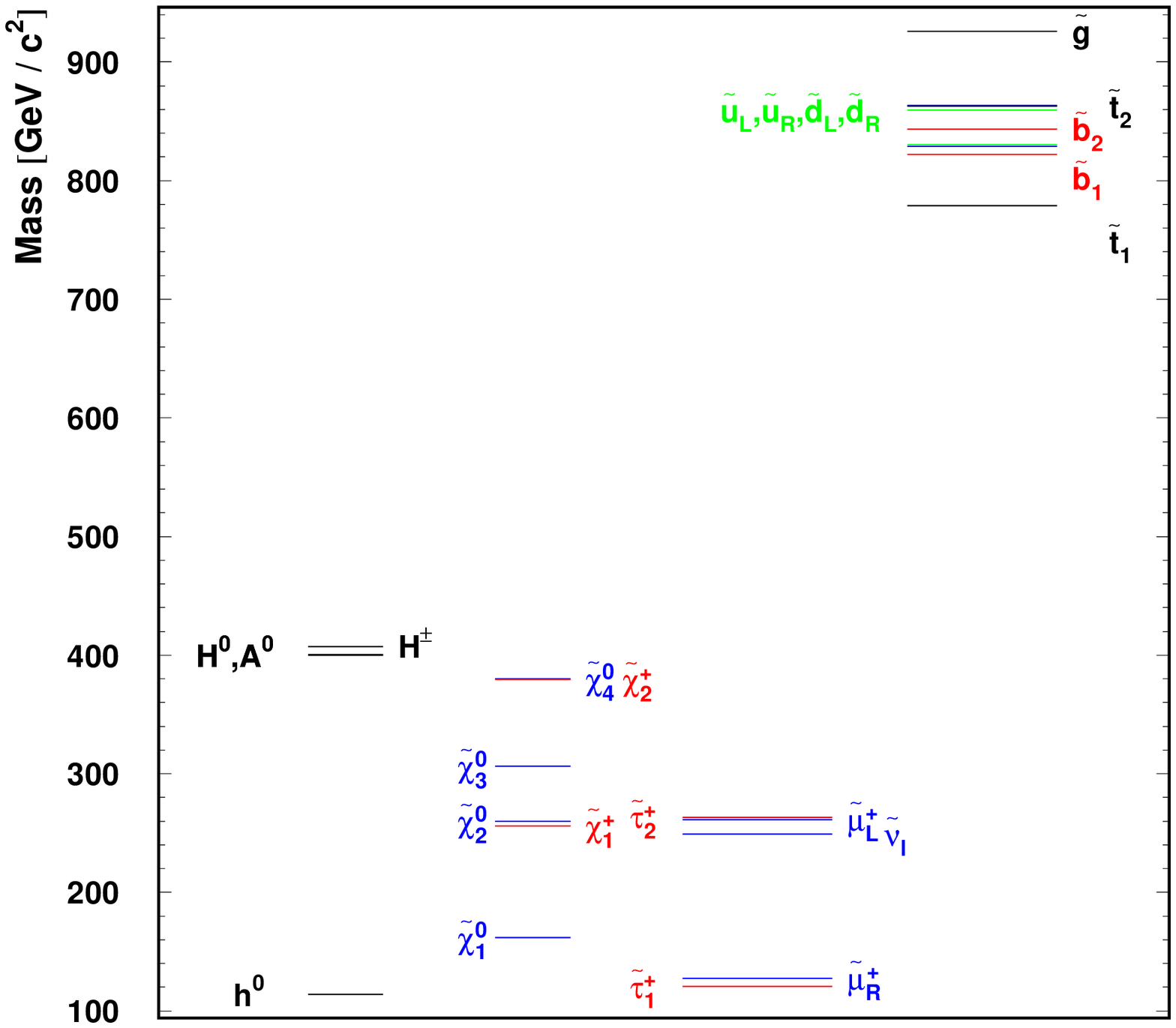,width=.9\textwidth}
  \caption{SPS 7 mass spectrum of \susygen}
\end{figure}
\clearpage
\normalsize
\noindent
{\bf \susygen\  parameters} 
\begin{small}
\begin{verbatim}
 Susygen inputs:
 --------------

 M2      =   326.810   mu       =   300.030
 M1      =   168.590   M3       =   895.450
 XRFSUSY =    56.568   NFAM     =     3
 TANB    =    15.000

 Sparticle masses: (input from Isajet)
 ----------------

 SUPR      831.  SUPL       860.
 SDNR      829.  SDNL       863.
 SELR      127.  SELL       261.
 SNU       249.
 STP1      779.  STP2       863. cosmix =  0.421
 SBT1      822.  SBT2       843. cosmix =  0.489
 STA1      120.  STA2       263. cosmix=   0.147
 SGLU      926.

 Gravitino mass: 0.769799202  eV  
 --------------

 Gaugino masses:
 --------------

 M1 =    168.590 M2 =    326.810 M3 =    895.450

 NEUTRALINO m, CP, ph/zi/ha/hb 1 =  161.6  1.  0.806 -0.526  0.152  0.224
 NEUTRALINO m, CP, ph/zi/ha/hb 2 =  259.9  1. -0.523 -0.429  0.523  0.519
 NEUTRALINO m, CP, ph/zi/ha/hb 3 =  306.3 -1. -0.013  0.103 -0.664  0.741
 NEUTRALINO m, CP, ph/zi/ha/hb 4 =  380.1  1. -0.277 -0.727 -0.513 -0.363

 CHARGINO MASSES    =   256.014   379.662
 CHARGINO ETA       =    -1.000     1.000

 U matrix      WINO      HIGGSINO      V matrix      WINO     HIGGSINO 
 W1SS+        -0.559     0.829         W1SS-         0.689   -0.725
 W2SS+         0.829     0.559         W2SS-         0.725    0.689

 Higgses masses: 
 --------------

 Light CP-even Higgs =   113.895
 Heavy CP-even Higgs =   400.218
       CP-odd  Higgs =   400.000
       Charged Higgs =   407.406
       sin(a-b)      =    -0.076
       cos(a-b)      =     0.997  
\end{verbatim}
\end{small}

\clearpage

\subsection{Decay modes}
\normalsize
 \begin{table}[h] \centering
   \begin{tabular}{|l|c c c|c|c c c|}
   \hdick & & & \\[-2.ex]
   particle & $m_I$ & $m_S$ & $m_P$ & \ decay \ & ${\cal B}_I$ & ${\cal B}_S$ & ${\cal B}_P$ 
   \\[.5ex] \hdick
   $\ser    $ &  127.4 & 127.4  &        & $\stau_1  e^-      \tau^+   $ & 0.342 &       &       \\
              &        &        &        & $\staup_1 e^-      \tau^-   $ & 0.591 &       &       \\
              &        &        &        & $e^-      \tilde{G}         $ & 0.067 & 1.000 &       \\
   \hline
   $\sel    $ &  261.5 & 261.5  &        & $\nt_1    e^-               $ & 0.987 & 0.986 &       \\
   \hline
   $\sne    $ &  249.1 & 249.1  &        & $\nt_1    \nu_e             $ & 1.000 & 1.000 &       \\
   \hline
   $\smur   $ &  127.4 & 127.4  &        & $\stau_1  \mu^-    \tau^+   $ & 0.342 &       &       \\
              &        &        &        & $\staup_1 \mu^-    \tau^-   $ & 0.591 &       &       \\
              &        &        &        & $\mu^-    \tilde{G}         $ & 0.067 & 1.000 &       \\
   \hline
   $\smul   $ &  261.5 & 261.5  &        & $\nt_1    \mu^-             $ & 0.987 & 0.986 &       \\
   \hline
   $\snm    $ &  249.1 & 249.1  &        & $\nt_1    \nu_\mu           $ & 1.000 & 1.000 &       \\
   \hline
   $\stau_1 $ &  120.4 & 120.4  &        & $\tau^-   \tilde{G}         $ & 1.000 & 1.000 &       \\
   \hline
   $\stau_2 $ &  263.4 & 263.4  &        & $\nt_1    \tau^-            $ & 0.604 & 0.978 &       \\
              &        &        &        & $Z^0      \stau_1           $ & 0.168 &       &       \\
              &        &        &        & $h^0      \stau_1           $ & 0.215 &       &       \\
   \hline
   $\snt    $ &  248.6 & 248.6  &        & $\nt_1    \nu_\tau          $ & 0.746 & 1.000 &       \\
              &        &        &        & $W^+      \stau_1           $ & 0.254 &       &       \\
   \hline
   \end{tabular}
   \caption{Slepton masses (GeV) and significant branching ratios ($>3\%$) 
            from \isajet~(I), \susygen~(S) and \pythia~(P)}
 \end{table}
  
 \begin{table}[h] \centering
   \begin{tabular}{|l|c c c|c|c c c|}
   \hdick & & & \\[-2.ex]
   particle & $m_I$ & $m_S$ & $m_P$ & \ decay \ & ${\cal B}_I$ & ${\cal B}_S$ & ${\cal B}_P$ 
   \\[.5ex] \hdick
   $\nt_1   $ &  161.7 & 161.6  &        & $\ser     e^+               $ & 0.148 & 0.148 &       \\
              &        &        &        & $\serp    e^-               $ & 0.148 & 0.148 &       \\
              &        &        &        & $\smur    \mu^+             $ & 0.148 & 0.148 &       \\
              &        &        &        & $\smurp   \mu^-             $ & 0.148 & 0.148 &       \\
              &        &        &        & $\stau_1  \tau^+            $ & 0.203 & 0.204 &       \\
              &        &        &        & $\staup_1 \tau^-            $ & 0.203 & 0.204 &       \\
   \hline
   $\nt_2   $ &  260.1 & 259.9  &        & $\ser     e^+               $ & 0.118 & 0.117 &       \\
              &        &        &        & $\serp    e^-               $ & 0.118 & 0.117 &       \\
              &        &        &        & $\smur    \mu^+             $ & 0.118 & 0.117 &       \\
              &        &        &        & $\smurp   \mu^-             $ & 0.118 & 0.117 &       \\
              &        &        &        & $\stau_1  \tau^+            $ & 0.237 & 0.241 &       \\
              &        &        &        & $\staup_1 \tau^-            $ & 0.237 & 0.241 &       \\
   \hline
   $\nt_3   $ &  306.3 & 306.3  &        & $\nt_1    Z^0               $ & 0.627 & 0.622 &       \\
              &        &        &        & $\stau_1  \tau^+            $ & 0.120 & 0.122 &       \\
              &        &        &        & $\staup_1 \tau^-            $ & 0.120 & 0.122 &       \\
   \hline
   $\nt_4   $ &  379.9 & 380.1  &        & $\cp_1    W^-               $ & 0.131 & 0.249 &       \\
              &        &        &        & $\cm_1    W^+               $ & 0.131 & 0.249 &       \\
              &        &        &        & $\sel     e^+               $ & 0.044 & 0.030 &       \\
              &        &        &        & $\selp    e^-               $ & 0.044 & 0.030 &       \\
              &        &        &        & $\smul    \mu^+             $ & 0.044 & 0.030 &       \\
              &        &        &        & $\smulp   \mu^-             $ & 0.044 & 0.030 &       \\
              &        &        &        & $\stau_2  \tau^+            $ & 0.046 & 0.032 &       \\
              &        &        &        & $\staup_2 \tau^-            $ & 0.046 & 0.032 &       \\
              &        &        &        & $\sne     \anue             $ & 0.071 & 0.048 &       \\
              &        &        &        & $\bar\sne \nu_e             $ & 0.071 & 0.048 &       \\
              &        &        &        & $\snm     \anum             $ & 0.071 & 0.048 &       \\
              &        &        &        & $\bar\snm \nu_\mu           $ & 0.071 & 0.048 &       \\
              &        &        &        & $\snt     \anut             $ & 0.071 & 0.048 &       \\
              &        &        &        & $\bar\snt \nu_\tau          $ & 0.071 & 0.048 &       \\
   \hline
   \end{tabular}
   \caption{Neutralino masses (GeV) and significant branching ratios ($>3\%$) 
            from \isajet~(I), \susygen~(S) and \pythia~(P)}
 \end{table}
  
 \begin{table}[h] \centering
   \begin{tabular}{|l|c c c|c|c c c|}
   \hdick & & & \\[-2.ex]
   particle & $m_I$ & $m_S$ & $m_P$ & \ decay \ & ${\cal B}_I$ & ${\cal B}_S$ & ${\cal B}_P$ 
   \\[.5ex] \hdick
   $\cp_1   $ &  256.3 & 256.0  &        & $\nt_1    W^+               $ & 0.292 & 0.954 &       \\
              &        &        &        & $\staup_1 \nu_\tau          $ & 0.638 &       &       \\
   \hline
   $\cp_2   $ &  379.5 & 379.7  &        & $\nt_2    W^+               $ & 0.143 & 0.109 &       \\
              &        &        &        & $\sne     e^+               $ & 0.118 & 0.090 &       \\
              &        &        &        & $\snm     \mu^+             $ & 0.118 & 0.090 &       \\
              &        &        &        & $\snt     \tau^+            $ & 0.122 & 0.094 &       \\
              &        &        &        & $\selp    \nu_e             $ & 0.131 & 0.100 &       \\
              &        &        &        & $\smulp   \nu_\mu           $ & 0.131 & 0.100 &       \\
              &        &        &        & $\staup_2 \nu_\tau          $ & 0.131 & 0.103 &       \\
              &        &        &        & $\cp_1    Z^0               $ & 0.092 & 0.304 &       \\
   \hline
   \end{tabular}
   \caption{Chargino masses (GeV) and significant branching ratios ($>3\%$) 
            from \isajet~(I), \susygen~(S) and \pythia~(P)}
 \end{table}
  
 \begin{table}[h] \centering
   \begin{tabular}{|l|c c c|c|c c c|}
   \hdick & & & \\[-2.ex]
   particle & $m_I$ & $m_S$ & $m_P$ & \ decay \ & ${\cal B}_I$ & ${\cal B}_S$ & ${\cal B}_P$ 
   \\[.5ex] \hdick
   $h^0     $ &  113.6 & 113.9  &        & $\tau^-   \tau^+            $ & 0.051 & 0.070 &       \\
              &        &        &        & $b        \bar b            $ & 0.854 & 0.810 &       \\
              &        &        &        & $c        \bar c            $ & 0.032 &       &       \\
              &        &        &        & $g         g                $ &       & 0.052 &       \\
              &        &        &        & $W^+      W^-               $ &       & 0.045 &       \\
   \hline
   $H^0     $ &  378.4 & 400.2  &        & $\tau^-   \tau^+            $ & 0.066 & 0.097 &       \\
              &        &        &        & $b        \bar b            $ & 0.921 & 0.877 &       \\
   \hline
   $A^0     $ &  377.9 & 400.0  &        & $\tau^-   \tau^+            $ & 0.065 & 0.096 &       \\
              &        &        &        & $b        \bar b            $ & 0.898 & 0.868 &       \\
              &        &        &        & $t        \bar t            $ &       & 0.034 &       \\
   \hline
   $H^+     $ &  386.7 & 407.4  &        & $\nu_\tau \tau^+            $ & 0.107 & 0.118 &       \\
              &        &        &        & $t        \bar b            $ & 0.887 & 0.878 &       \\
   \hline
   \end{tabular}
   \caption{Higgs masses (GeV) and significant branching ratios ($>3\%$) 
            from \isajet~(I), \susygen~(S) and \pythia~(P)}
 \end{table}
  
 \begin{table}[h] \centering
   \begin{tabular}{|l|c c c|c|c c c|}
   \hdick & & & \\[-2.ex]
   particle & $m_I$ & $m_S$ & $m_P$ & \ decay \ & ${\cal B}_I$ & ${\cal B}_S$ & ${\cal B}_P$ 
   \\[.5ex] \hdick
   $\st_1   $ &  779.1 & 779.1  &        & $\nt_1    t                 $ & 0.083 & 0.071 &       \\
              &        &        &        & $\nt_2    t                 $ & 0.156 & 0.152 &       \\
              &        &        &        & $\nt_3    t                 $ & 0.263 & 0.190 &       \\
              &        &        &        & $\cp_1    b                 $ & 0.409 & 0.463 &       \\
              &        &        &        & $\cp_2    b                 $ & 0.065 & 0.095 &       \\
   \hline
   $\st_2   $ &  863.0 & 863.0  &        & $\cp_1    b                 $ & 0.047 & 0.126 &       \\
              &        &        &        & $\cp_2    b                 $ & 0.310 & 0.253 &       \\
              &        &        &        & $\nt_2    t                 $ & 0.133 & 0.066 &       \\
              &        &        &        & $\nt_3    t                 $ & 0.221 & 0.173 &       \\
              &        &        &        & $\nt_4    t                 $ & 0.282 & 0.144 &       \\
              &        &        &        & $\tilde{g} t                $ &       & 0.214 &       \\
   \hline
   $\sb_1   $ &  822.2 & 822.2  &        & $\nt_1    b                 $ & 0.115 & 0.121 &       \\
              &        &        &        & $\nt_2    b                 $ & 0.167 & 0.226 &       \\
              &        &        &        & $\nt_3    b                 $ & 0.059 & 0.130 &       \\
              &        &        &        & $\cm_1    t                 $ & 0.438 & 0.481 &       \\
              &        &        &        & $\cm_2    t                 $ & 0.205 & 0.034 &       \\
   \hline
   $\sb_2   $ &  843.4 & 843.4  &        & $\nt_2    b                 $ & 0.048 & 0.089 &       \\
              &        &        &        & $\nt_3    b                 $ & 0.041 & 0.132 &       \\
              &        &        &        & $\nt_4    b                 $ & 0.115 & 0.196 &       \\
              &        &        &        & $\cm_1    t                 $ & 0.297 & 0.130 &       \\
              &        &        &        & $\cm_2    t                 $ & 0.492 & 0.448 &       \\
   \hline
   \end{tabular}
   \caption{Light squark masses (GeV) and significant branching ratios ($>3\%$) 
            from \isajet~(I), \susygen~(S) and \pythia~(P)}
 \end{table}

\clearpage

\section{SPS 8 -- GMSB scenario}
\setcounter{figure}{0}
\setcounter{table}{0}

\large\boldmath
\hspace{20mm}
\begin{tabular}{|l c|}
  \hline
  $\Lambda$   & $ 100~\TeV$ \\ 
  $M_{mess}$  & $ 200~\TeV$ \\
  $N_{mess}$  & $ 1$ \\
  $\tan\beta$ & $ 15$       \\
  ${\rm sign}~\mu$ & $+$ \\
  \hline
\end{tabular} \hspace{10mm}
\begin{tabular}{l}
   {\bf NLSP = $\tilde\chi^0_1$} \\
    $M_{mess} = 2\,\Lambda$  \\
\end{tabular}
\unboldmath\normalsize
\bigskip

\subsection{Spectrum \& parameters of ISAJET 7.58}

\begin{figure}[h] \centering
  \epsfig{file=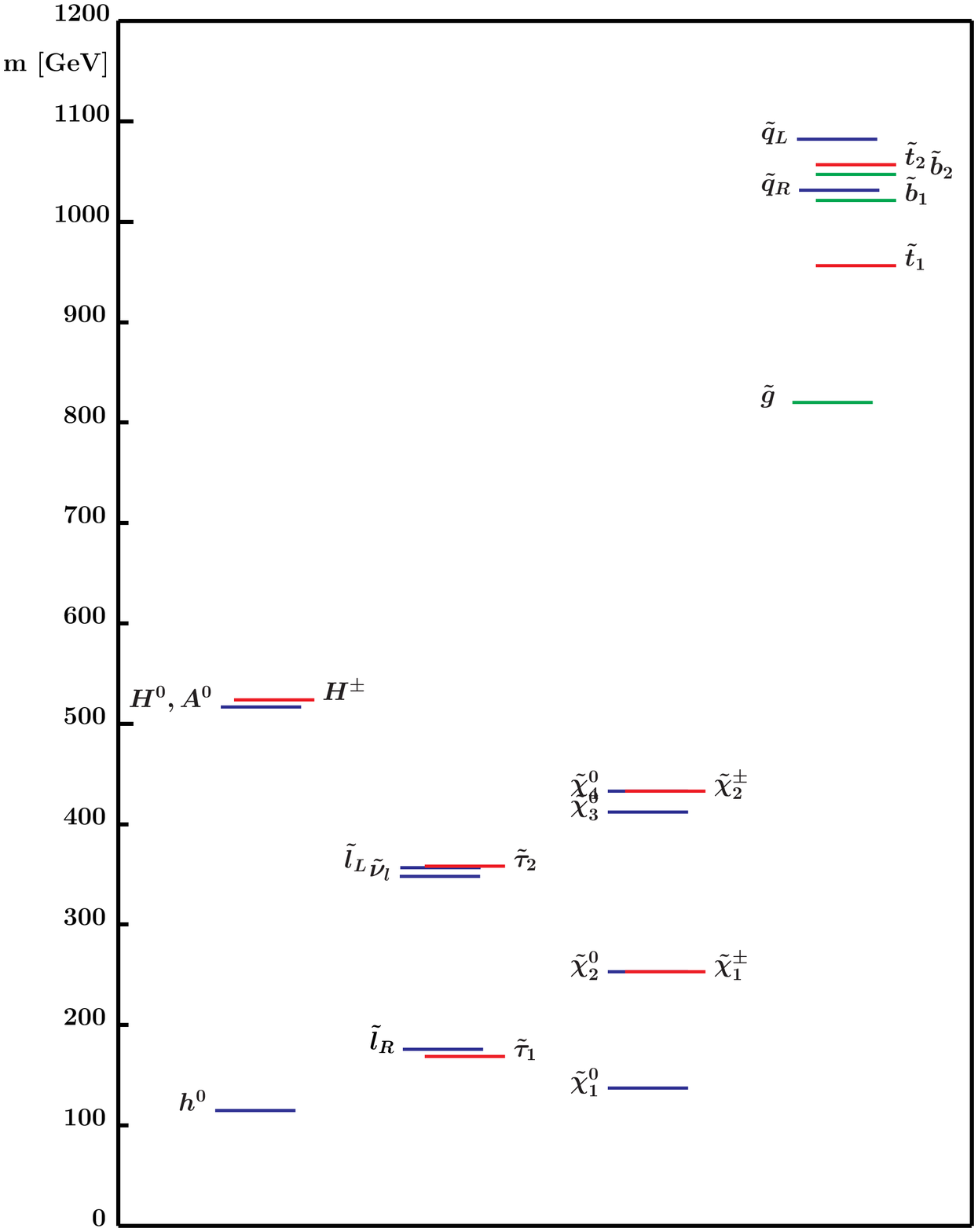,width=.75\textwidth}
  \caption{SPS 8 mass spectrum of \isajet}
\end{figure}

\clearpage
\normalsize
\noindent
{\bf \isajet\ parameters} 
\begin{small}
\begin{verbatim}
 Minimal gauge mediated (GMSB) model:

 Lambda,  M_mes,  N_5,  tan(beta),  sgn(mu),  M_t,  C_grav=
 0.100E+06 0.200E+06     1.000    15.000     1.0   175.000  0.100E+01

 GMSB2 model input:
 Rsl,    dmH_d^2,   dmH_u^2,     d_Y,     N5_1,  N5_2,  N5_3=
  1.000  0.000E+00  0.000E+00  0.000E+00    1.000  1.000  1.000

 M(gravitino)= 0.481E-08

 1/alpha_em =  127.87     sin**2(thetaw) =0.2311     alpha_s   =0.118
 M_1        =  140.00     M_2            =  271.80   M_3       =  755.00
 mu(Q)      =  398.31     B(Q)           =   44.32   Q         =  987.76
 M_H1^2     = 0.111E+06   M_H2^2         =-0.135E+06

 ISAJET masses (with signs):
 M(GL)  =   820.50
 M(UL)  =  1080.25   M(UR)  =  1033.16   M(DL)  =  1083.17   M(DR) =  1029.29
 M(B1)  =  1021.90   M(B2)  =  1048.26   M(T1)  =   957.65   M(T2) =  1058.68
 M(SN)  =   347.61   M(EL)  =   356.61   M(ER)  =   175.87
 M(NTAU)=   346.94   M(TAU1)=   169.42   M(TAU2)=   357.59
 M(Z1)  =  -137.19   M(Z2)  =  -252.33   M(Z3)  =   404.00   M(Z4) =  -426.28
 M(W1)  =  -252.03   M(W2)  =  -426.47
 M(HL)  =   114.83   M(HH)  =   515.01   M(HA)  =   514.49   M(H+) =   521.17

 theta_t=   1.3169   theta_b=   1.1767   theta_l=   1.4639   alpha_h=   0.0719

 NEUTRALINO MASSES (SIGNED) =  -137.192  -252.334   404.003  -426.276
 EIGENVECTOR 1       =   0.05427  -0.13503  -0.03750   0.98864
 EIGENVECTOR 2       =   0.20544  -0.30581  -0.92547  -0.08815
 EIGENVECTOR 3       =   0.70961   0.69819  -0.07827   0.05344
 EIGENVECTOR 4       =   0.67179  -0.63307   0.36873  -0.10936

 CHARGINO MASSES (SIGNED)  =  -252.028  -426.469
 GAMMAL, GAMMAR             =   2.02894   1.87308


 ISAJET equivalent input:
 MSSMA:   820.50  398.31  514.49   15.00
 MSSMB:  1081.56 1028.98 1033.78  353.48  170.37
 MSSMC:  1042.74 1025.51  952.74  352.82  167.23 -296.71 -330.28  -36.69
 MSSMD: SAME AS MSSMB (DEFAULT)
 MSSME:   140.00  271.80
 WARNING IN SSXINT: BAD CONVERGENCE FOR  508 INTERVALS.

\end{verbatim}
\end{small}
\clearpage

\subsection{Spectrum \& parameters of SUSYGEN 3.00/27}

\begin{figure}[h] \centering
  \epsfig{file=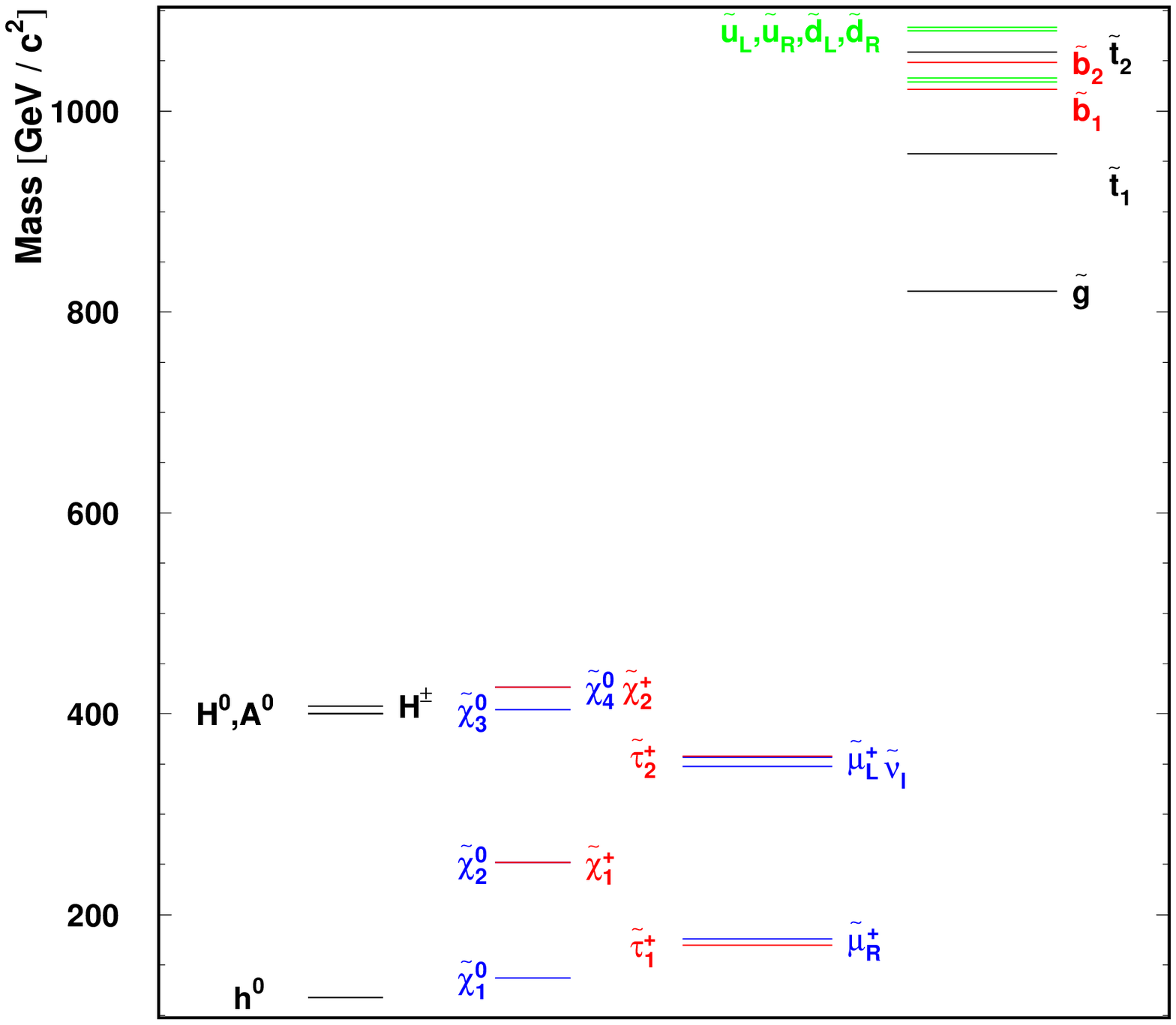,width=.9\textwidth}
  \caption{SPS 8 mass spectrum of \susygen}
\end{figure}
\clearpage
\normalsize
\noindent
{\bf \susygen\  parameters} 
\begin{small}
\begin{verbatim}
 Susygen inputs:
 --------------

 M2      =   271.800   mu       =   398.310
 M1      =   140.000   M3       =   755.000
 XRFSUSY =   141.421   NFAM     =     1
 TANB    =    15.000

 Sparticle masses:  (input from Isajet)
 ----------------

 SUPR     1033.  SUPL      1080.
 SDNR     1029.  SDNL      1083.
 SELR      176.  SELL       357.
 SNU       348.
 STP1      958.  STP2      1059. cosmix =  0.425
 SBT1     1022.  SBT2      1048. cosmix =  0.384
 STA1      169.  STA2       358. cosmix=   0.107
 SGLU      820.

 Gravitino mass: 4.81125246  eV  
 --------------

 Gaugino masses:
 --------------

 M1 =    140.000 M2 =    271.800 M3 =    755.000

 NEUTRALINO m, CP, ph/zi/ha/hb 1 =  137.2  1.  0.849 -0.508  0.063  0.131
 NEUTRALINO m, CP, ph/zi/ha/hb 2 =  252.2  1. -0.522 -0.769  0.226  0.292
 NEUTRALINO m, CP, ph/zi/ha/hb 3 =  404.0 -1. -0.009  0.095 -0.662  0.744
 NEUTRALINO m, CP, ph/zi/ha/hb 4 =  426.4  1. -0.082 -0.377 -0.712 -0.587

 CHARGINO MASSES    =   251.761   426.622
 CHARGINO ETA      =    -1.000     1.000

 U matrix      WINO      HIGGSINO      V matrix      WINO     HIGGSINO 
 W1SS+        -0.896     0.443         W1SS-         0.955   -0.298
 W2SS+         0.443     0.896         W2SS-         0.298    0.955

 Higgses masses: 
 --------------

 Light CP-even Higgs =   116.282
 Heavy CP-even Higgs =   400.221
       CP-odd  Higgs =   400.000
       Charged Higgs =   407.301
       sin(a-b)      =    -0.076
       cos(a-b)      =     0.997
\end{verbatim}
\end{small}
\clearpage

\subsection{Spectrum \& parameters of  PYTHIA 6.2/00}

\begin{figure}[h] \centering
  \epsfig{file=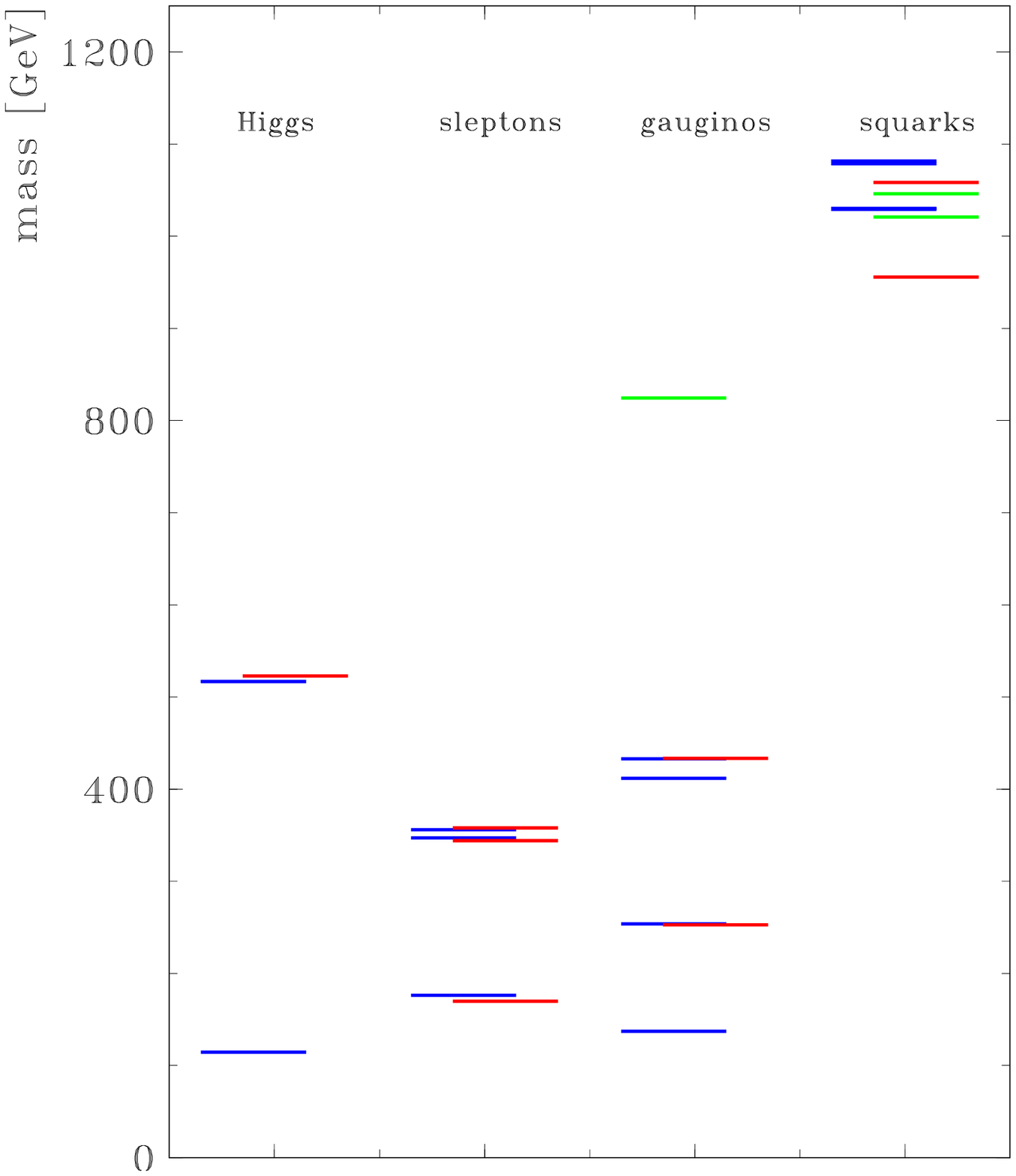,width=.9\textwidth}
  \caption{SPS 8 mass spectrum of \pythia}
\end{figure}
\newpage
\normalsize
\noindent
{\bf \pythia\  parameters} 
\begin{small}
\begin{verbatim}
   MSSM input parameters
   ---------------------
   M_1         RMSS( 1) =   140.0    
   M_2         RMSS( 2) =   272.0    
   M_3         RMSS( 3) =   740.0    
   tan_beta    RMSS( 5) =   15.00    
   mu          RMSS( 4) =   406.0    

   sparticle masses & widths
   -------------------------
   M_se_R   176.0 ( 0.134)   M_se_L   356.0 ( 1.239)   M_sne_L  346.9 ( 1.197)
   M_sm_R   176.0 ( 0.134)   M_sm_L   356.0 ( 1.239)   M_snm_L  346.9 ( 1.197)
   M_st_1   169.8 ( 0.101)   M_st_2   357.9 ( 1.342)   M_snt_L  343.8 ( 1.248)

   M_ch0_1  137.3 ( 0.000)   M_ch0_2  253.5 ( 0.028)   M_ch0_3  411.6 ( 1.832)
   M_ch0_4  432.9 ( 2.306)
   M_ch+_1  252.8 ( 0.026)   M_ch+_2  433.4 ( 2.192)

   M_h0     114.3 ( 0.004)   M_H0     517.0 ( 2.425)   M_A0     516.9 ( 2.456)
   M_H+     522.9 ( 2.256)

   M_g~     824.3 ( 0.021)
   M_uL    1078.7 (23.774)   M_uR    1029.4 (10.449)   M_dL    1081.6 (23.823)
   M_dR    1030.3 ( 8.794)
   M_b1    1021.0 (12.139)   M_b2    1045.9 (29.128)   M_t1     955.9 (23.923)
   M_t2    1058.4 (28.340)

   parameter settings IMSS, RMSS
   -----------------------------
   IMSS( 1) = 1   IMSS( 4) = 1   IMSS( 7) = 0   IMSS(10) = 0
   IMSS( 2) = 0   IMSS( 5) = 0   IMSS( 8) = 0   IMSS(11) = 1
   IMSS( 3) = 0   IMSS( 6) = 0   IMSS( 9) = 0   IMSS(12) = 0

   RMSS( 1) =   140.0       RMSS( 9) =   1030.       RMSS(17) =  -35.00    
   RMSS( 2) =   272.0       RMSS(10) =   1040.       RMSS(18) = -0.7168E-01
   RMSS( 3) =   740.0       RMSS(11) =   1025.       RMSS(19) =   517.0    
   RMSS( 4) =   406.0       RMSS(12) =   950.0       RMSS(20) =  0.4100E-01
   RMSS( 5) =   15.00       RMSS(13) =   353.0       RMSS(21) =   1.000    
   RMSS( 6) =   356.0       RMSS(14) =   168.0       RMSS(22) =   1030.    
   RMSS( 7) =   176.0       RMSS(15) =  -330.0       RMSS(23) =  0.1000E+05
   RMSS( 8) =   1080.       RMSS(16) =  -295.0       RMSS(24) =  0.1000E+05

\end{verbatim}
\end{small}

\clearpage

\subsection{Decay modes}
\normalsize
 \begin{table}[h] \centering
   \begin{tabular}{|l|c c c|c|c c c|}
   \hdick & & & \\[-2.ex]
   particle & $m_I$ & $m_S$ & $m_P$ & \ decay \ & ${\cal B}_I$ & ${\cal B}_S$ & ${\cal B}_P$ 
   \\[.5ex] \hdick
   $\ser    $ &  175.9 & 175.9  & 176.0  & $\nt_1    e^-               $ & 1.000 & 1.000 & 1.000 \\
   \hline
   $\sel    $ &  356.6 & 356.6  & 356.0  & $\nt_1    e^-               $ & 0.225 & 0.224 & 0.228 \\
              &        &        &        & $\nt_2    e^-               $ & 0.287 & 0.287 & 0.283 \\
              &        &        &        & $\cm_1    \nu_e             $ & 0.488 & 0.490 & 0.489 \\
   \hline
   $\sne    $ &  347.6 & 347.6  & 346.9  & $\nt_1    \nu_e             $ & 0.292 & 0.291 & 0.294 \\
              &        &        &        & $\nt_2    \nu_e             $ & 0.209 & 0.209 & 0.209 \\
              &        &        &        & $\cp_1    e^-               $ & 0.499 & 0.500 & 0.497 \\
   \hline
   $\smur   $ &  175.9 & 175.9  & 176.0  & $\nt_1    \mu^-             $ & 1.000 & 1.000 & 1.000 \\
   \hline
   $\smul   $ &  356.6 & 356.6  & 356.0  & $\nt_1    \mu^-             $ & 0.225 & 0.224 & 0.228 \\
              &        &        &        & $\nt_2    \mu^-             $ & 0.287 & 0.287 & 0.283 \\
              &        &        &        & $\cm_1    \nu_\mu           $ & 0.488 & 0.490 & 0.489 \\
   \hline
   $\snm    $ &  347.6 & 347.6  & 346.9  & $\nt_1    \nu_\mu           $ & 0.292 & 0.291 & 0.294 \\
              &        &        &        & $\nt_2    \nu_\mu           $ & 0.209 & 0.209 & 0.209 \\
              &        &        &        & $\cp_1    \mu^-             $ & 0.499 & 0.500 & 0.497 \\
   \hline
   $\stau_1 $ &  169.4 & 169.4  & 169.8  & $\nt_1    \tau^-            $ & 1.000 & 1.000 & 1.000 \\
   \hline
   $\stau_2 $ &  357.6 & 357.6  & 357.9  & $\nt_1    \tau^-            $ & 0.215 & 0.232 & 0.218 \\
              &        &        &        & $\nt_2    \tau^-            $ & 0.262 & 0.286 & 0.262 \\
              &        &        &        & $\cm_1    \nu_\tau          $ & 0.441 & 0.482 & 0.447 \\
              &        &        &        & $Z^0      \stau_1           $ & 0.038 &       & 0.041 \\
              &        &        &        & $h^0      \stau_1           $ & 0.044 &       & 0.032 \\
   \hline
   $\snt    $ &  346.9 & 346.9  & 343.8  & $\nt_1    \nu_\tau          $ & 0.271 & 0.292 & 0.278 \\
              &        &        &        & $\nt_2    \nu_\tau          $ & 0.193 & 0.209 & 0.191 \\
              &        &        &        & $\cp_1    \tau^-            $ & 0.463 & 0.499 & 0.456 \\
              &        &        &        & $W^+      \stau_1           $ & 0.073 &       & 0.075 \\
   \hline
   \end{tabular}
   \caption{Slepton masses (GeV) and significant branching ratios ($>3\%$) 
            from \isajet~(I), \susygen~(S) and \pythia~(P)}
 \end{table}
  
 \begin{table}[h] \centering
   \begin{tabular}{|l|c c c|c|c c c|}
   \hdick & & & \\[-2.ex]
   particle & $m_I$ & $m_S$ & $m_P$ & \ decay \ & ${\cal B}_I$ & ${\cal B}_S$ & ${\cal B}_P$ 
   \\[.5ex] \hdick
   $\nt_1   $ &  137.2 & 137.2  & 137.3  & $\tilde{G}     \gamma       $ & 0.947 & 0.974 & 0.968 \\
              &        &        &        & $\tilde{G}     Z^0          $ & 0.033 &       & 0.032 \\
   \hline
   $\nt_2   $ &  252.3 & 252.2  & 253.5  & $\nt_1    Z^0               $ & 0.086 & 0.091 & 0.075 \\
              &        &        &        & $\nt_1    h^0               $ & 0.078 &       & 0.168 \\
              &        &        &        & $\ser     e^+               $ & 0.053 & 0.057 & 0.043 \\
              &        &        &        & $\serp    e^-               $ & 0.053 & 0.057 & 0.043 \\
              &        &        &        & $\smur    \mu^+             $ & 0.053 & 0.057 & 0.043 \\
              &        &        &        & $\smurp   \mu^-             $ & 0.053 & 0.057 & 0.043 \\
              &        &        &        & $\stau_1  \tau^+            $ & 0.311 & 0.340 & 0.292 \\
              &        &        &        & $\staup_1 \tau^-            $ & 0.311 & 0.340 & 0.292 \\
   \hline
   $\nt_3   $ &  404.0 & 404.0  & 411.7  & $\cp_1    W^-               $ & 0.289 & 0.363 & 0.291 \\
              &        &        &        & $\cm_1    W^+               $ & 0.289 & 0.363 & 0.291 \\
              &        &        &        & $\nt_1    Z^0               $ & 0.142 & 0.092 & 0.133 \\
              &        &        &        & $\nt_2    Z^0               $ & 0.207 & 0.134 & 0.211 \\
   \hline
   $\nt_4   $ &  426.3 & 426.4  & 432.9  & $\cp_1    W^-               $ & 0.307 & 0.361 & 0.303 \\
              &        &        &        & $\cm_1    W^+               $ & 0.307 & 0.361 & 0.303 \\
              &        &        &        & $\nt_1    h^0               $ & 0.087 & 0.063 & 0.082 \\
              &        &        &        & $\nt_2    h^0               $ & 0.142 & 0.102 & 0.153 \\
   \hline
   \end{tabular}
   \caption{Neutralino masses (GeV) and significant branching ratios ($>3\%$) 
            from \isajet~(I), \susygen~(S) and \pythia~(P)}
 \end{table}
  
 \begin{table}[h] \centering
   \begin{tabular}{|l|c c c|c|c c c|}
   \hdick & & & \\[-2.ex]
   particle & $m_I$ & $m_S$ & $m_P$ & \ decay \ & ${\cal B}_I$ & ${\cal B}_S$ & ${\cal B}_P$ 
   \\[.5ex] \hdick
   $\cp_1   $ &  252.0 & 251.8  & 252.8  & $\nt_1    W^+               $ & 0.508 & 0.899 & 0.480 \\
              &        &        &        & $\staup_1 \nu_\tau          $ & 0.491 & 0.099 & 0.520 \\
   \hline
   $\cp_2   $ &  426.5 & 426.6  & 433.4  & $\nt_1    W^+               $ & 0.097 & 0.089 & 0.093 \\
              &        &        &        & $\nt_2    W^+               $ & 0.348 & 0.323 & 0.343 \\
              &        &        &        & $\cp_1    Z^0               $ & 0.283 & 0.509 & 0.281 \\
              &        &        &        & $\cp_1    h^0               $ & 0.173 &       & 0.180 \\
   \hline
   \end{tabular}
   \caption{Chargino masses (GeV) and significant branching ratios ($>3\%$) 
            from \isajet~(I), \susygen~(S) and \pythia~(P)}
 \end{table}
  
 \begin{table}[h] \centering
   \begin{tabular}{|l|c c c|c|c c c|}
   \hdick & & & \\[-2.ex]
   particle & $m_I$ & $m_S$ & $m_P$ & \ decay \ & ${\cal B}_I$ & ${\cal B}_S$ & ${\cal B}_P$ 
   \\[.5ex] \hdick
   $h^0     $ &  114.8 & 117.3  & 114.3  & $\tau^-   \tau^+            $ & 0.050 & 0.077 & 0.066 \\
              &        &        &        & $b        \bar b            $ & 0.837 & 0.758 & 0.789 \\
              &        &        &        & $c        \bar c            $ & 0.036 &       & 0.043 \\
              &        &        &        & $g         g                $ &       & 0.057 & 0.035 \\
              &        &        &        & $W^+      W^-               $ &       & 0.076 & 0.055 \\
   \hline
   $H^0     $ &  515.0 & 400.1  & 517.0  & $\tau^-   \tau^+            $ & 0.067 & 0.112 & 0.100 \\
              &        &        &        & $b        \bar b            $ & 0.888 & 0.860 & 0.874 \\
   \hline
   $A^0     $ &  514.5 & 400.0  & 516.9  & $\tau^-   \tau^+            $ & 0.063 & 0.107 & 0.099 \\
              &        &        &        & $b        \bar b            $ & 0.845 & 0.822 & 0.864 \\
              &        &        &        & $t        \bar t            $ & 0.031 & 0.038 & 0.036 \\
              &        &        &        & $\nt_1    \nt_2             $ & 0.033 &       &       \\
   \hline
   $H^+     $ &  521.2 & 407.3  & 523.0  & $\nu_\tau \tau^+            $ & 0.086 & 0.131 & 0.109 \\
              &        &        &        & $t        \bar b            $ & 0.854 & 0.833 & 0.890 \\
              &        &        &        & $\cp_1    \nt_1             $ & 0.056 & 0.033 &       \\
   \hline
   \end{tabular}
   \caption{Higgs masses (GeV) and significant branching ratios ($>3\%$) 
            from \isajet~(I), \susygen~(S) and \pythia~(P)}
 \end{table}
  
 \begin{table}[h] \centering
   \begin{tabular}{|l|c c c|c|c c c|}
   \hdick & & & \\[-2.ex]
   particle & $m_I$ & $m_S$ & $m_P$ & \ decay \ & ${\cal B}_I$ & ${\cal B}_S$ & ${\cal B}_P$ 
   \\[.5ex] \hdick
   $\st_1   $ &  957.6 & 957.7  & 955.9  & $\nt_1    t                 $ & 0.089 & 0.066 & 0.081 \\
              &        &        &        & $\nt_2    t                 $ & 0.057 & 0.077 & 0.056 \\
              &        &        &        & $\nt_3    t                 $ & 0.245 & 0.195 & 0.248 \\
              &        &        &        & $\nt_4    t                 $ & 0.140 & 0.122 & 0.144 \\
              &        &        &        & $\cp_1    b                 $ & 0.131 & 0.197 & 0.127 \\
              &        &        &        & $\cp_2    b                 $ & 0.339 & 0.343 & 0.345 \\
   \hline
   $\st_2   $ & 1058.7 & 1058.7 & 1058.4 & $\sg      t                 $ & 0.269 & 0.230 & 0.140 \\
              &        &        &        & $\cp_1    b                 $ & 0.174 & 0.125 & 0.189 \\
              &        &        &        & $\cp_2    b                 $ & 0.098 & 0.282 & 0.092 \\
              &        &        &        & $\nt_2    t                 $ & 0.088 & 0.041 & 0.102 \\
              &        &        &        & $\nt_3    t                 $ & 0.158 & 0.195 & 0.204 \\
              &        &        &        & $\nt_4    t                 $ & 0.209 & 0.122 & 0.267 \\
   \hline
   $\sb_1   $ & 1021.9 & 1021.9 & 1021.5 & $\nt_1    b                 $ & 0.049 & 0.038 & 0.041 \\
              &        &        &        & $\nt_2    b                 $ & 0.075 & 0.069 & 0.046 \\
              &        &        &        & $\nt_3    b                 $ & 0.030 & 0.050 &       \\
              &        &        &        & $\sg      b                 $ & 0.594 & 0.651 & 0.639 \\
              &        &        &        & $\cm_1    t                 $ & 0.144 & 0.132 & 0.093 \\
              &        &        &        & $\cm_2    t                 $ & 0.098 & 0.037 & 0.175 \\
   \hline
   $\sb_2   $ & 1048.3 & 1048.3 & 1045.9 & $\nt_2    b                 $ & 0.082 & 0.082 & 0.095 \\
              &        &        &        & $\nt_3    b                 $ &       & 0.046 &       \\
              &        &        &        & $\nt_4    b                 $ & 0.038 & 0.063 &       \\
              &        &        &        & $\sg      b                 $ & 0.322 & 0.509 & 0.312 \\
              &        &        &        & $\cm_1    t                 $ & 0.164 & 0.144 & 0.197 \\
              &        &        &        & $\cm_2    t                 $ & 0.367 & 0.152 & 0.373 \\
   \hline
   \end{tabular}
   \caption{Light squark masses (GeV) and significant branching ratios ($>3\%$) 
            from \isajet~(I), \susygen~(S) and \pythia~(P)}
 \end{table}

\clearpage

\section{SPS 9 -- AMSB scenario}
\setcounter{figure}{0}
\setcounter{table}{0}

\large\boldmath
\hspace{20mm}
\begin{tabular}{|l c|}
  \hline
  $m_0$       & $ 400~\GeV$ \\ 
  $m_{3/2}$   & $\ 60~\TeV$ \\
  $\tan\beta$ & $10$       \\
  ${\rm sign}~\mu$ & $+$ \\
  \hline
\end{tabular} \hspace{10mm}
\begin{tabular}{l}
    $m_0 = 0.0075\,m_{3/2} $ \\
\end{tabular}
\unboldmath\normalsize
\bigskip

\subsection{Spectrum \& parameters of ISAJET 7.58}

\begin{figure}[h] \centering
  \epsfig{file=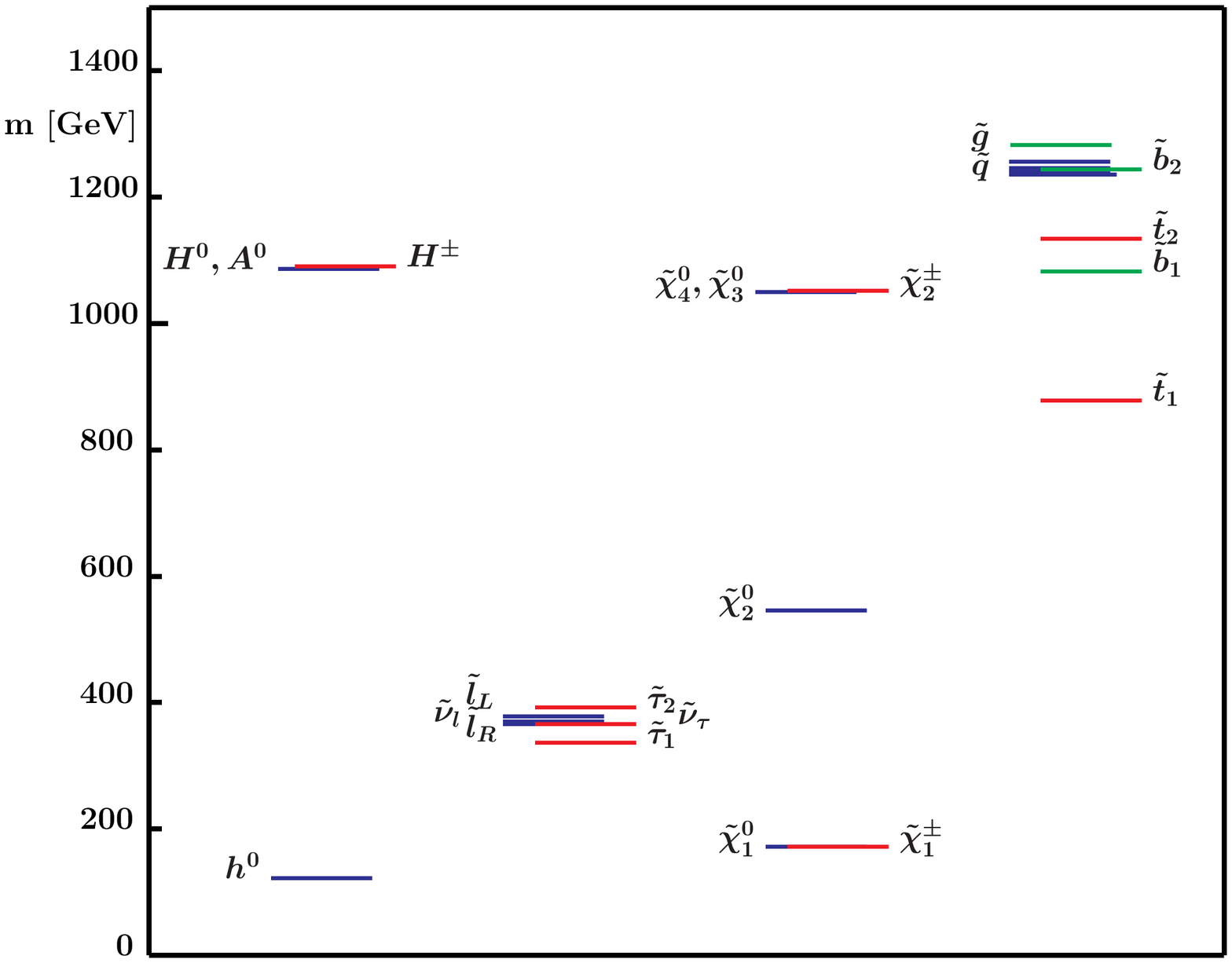,width=.9\textwidth}
  \caption{SPS 9 mass spectrum of \isajet}
\end{figure}

\clearpage
\normalsize
\noindent
{\bf \isajet\ parameters} 
\begin{small}
\begin{verbatim}
 WARNING: TACHYONIC SLEPTONS AT GUT SCALE
          POINT MAY BE INVALID

 Supergravity model with right-handed neutrinos:

 M_0,  M_(3/2),  tan(beta),  sgn(mu),  M_t =
   400.000 60000.000    10.000     1.0   175.000

 ISASUGRA unification:
 M_GUT      = 0.101E+17   g_GUT          =0.711      alpha_GUT =0.040
 FT_GUT     = 0.476       FB_GUT         = 0.046     FL_GUT = 0.069

 1/alpha_em =  127.81     sin**2(thetaw) =0.2306     alpha_s   =0.120
 M_1        = -550.60     M_2            = -175.53   M_3       = 1232.27
 mu(Q)      =  869.90     B(Q)           =   94.70   Q         = 1076.05
 M_H1^2     = 0.714E+05   M_H2^2         =-0.720E+06

 ISAJET masses (with signs):
 M(GL)  =  1275.18
 M(UL)  =  1218.09   M(UR)  =  1227.35   M(DL)  =  1220.65   M(DR) =  1237.81
 M(B1)  =  1112.07   M(B2)  =  1232.88   M(T1)  =  1005.17   M(T2) =  1128.80
 M(SN)  =   309.71   M(EL)  =   319.66   M(ER)  =   303.01
 M(NTAU)=   300.71   M(TAU1)=   271.28   M(TAU2)=   322.54
 M(Z1)  =   175.51   M(Z2)  =   549.03   M(Z3)  =  -874.37   M(Z4) =   875.97
 M(W1)  =   175.67   M(W2)  =   877.22
 M(HL)  =   114.83   M(HH)  =   912.56   M(HA)  =   911.74   M(H+) =   915.83

 theta_t=   1.3055   theta_b=   0.0919   theta_l=   1.0546   alpha_h=   0.1020

 NEUTRALINO MASSES (SIGNED) =   175.505   549.027  -874.365   875.970
 EIGENVECTOR 1       =  -0.00966  -0.09301  -0.99562   0.00004
 EIGENVECTOR 2       =   0.04358   0.07687  -0.00765  -0.99606
 EIGENVECTOR 3       =   0.70715  -0.70420   0.05892  -0.02386
 EIGENVECTOR 4       =   0.70565   0.69968  -0.07220   0.08543

 CHARGINO MASSES (SIGNED)  =   175.672   877.221
 GAMMAL, GAMMAR             =   1.70212   1.55731


 ISAJET equivalent input:
 MSSMA:  1275.18  869.90  911.74   10.00
 MSSMB:  1219.24 1237.55 1227.86  316.22  299.89
 MSSMC:  1111.59 1231.65 1003.22  307.41  281.16 -350.26  216.41 1162.39
 MSSMD: SAME AS MSSMB (DEFAULT)
 MSSME:  -550.60 -175.53
 WARNING IN SSXINT: BAD CONVERGENCE FOR  264 INTERVALS.
\end{verbatim}  
\clearpage

\subsection{Spectrum \& parameters of SUSYGEN 3.00/27}
\begin{figure}[h] \centering
  \epsfig{file=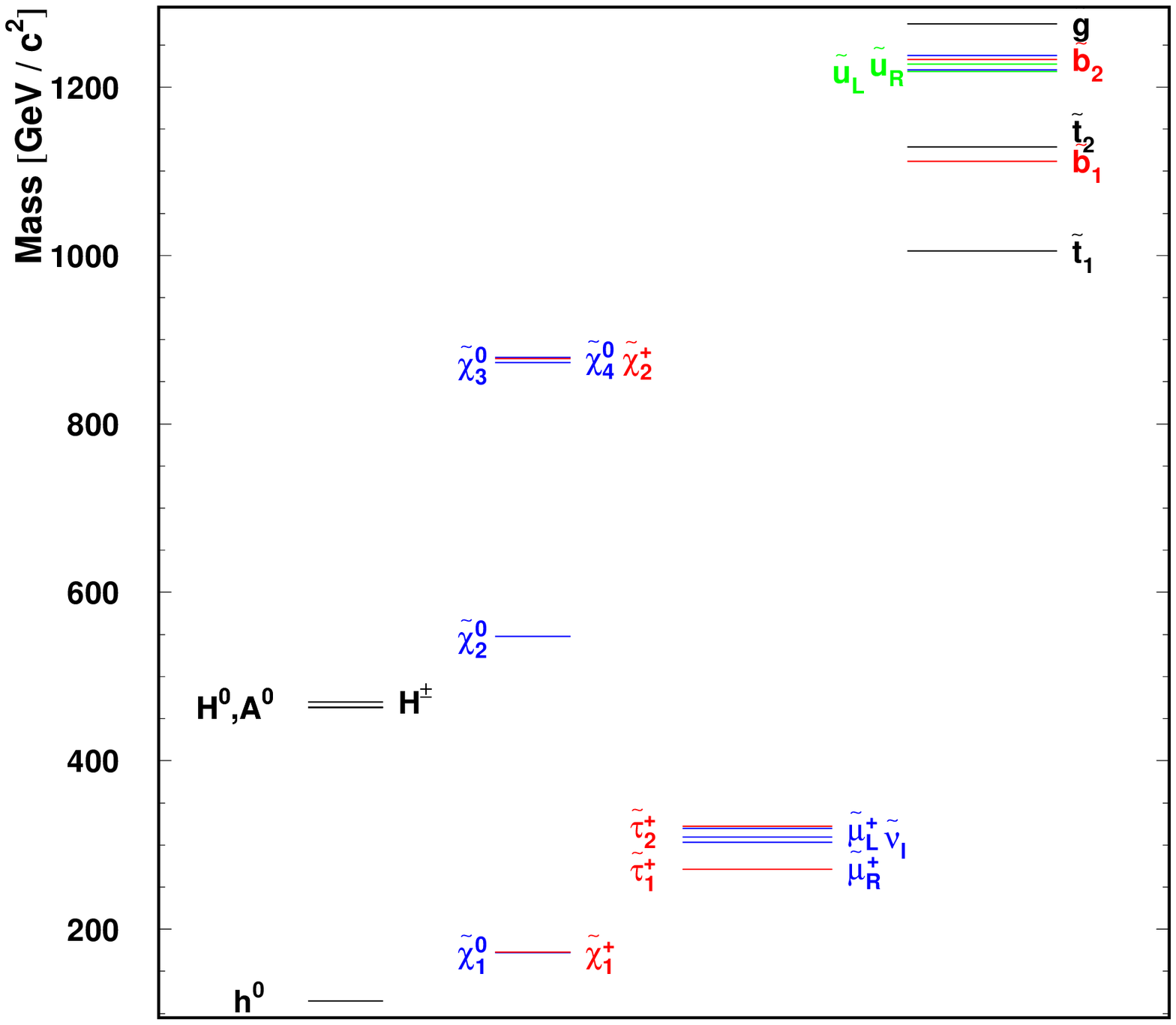,width=.9\textwidth}
  \caption{SPS 9 mass spectrum of \susygen}
\end{figure}
\newpage
\normalsize
\noindent
{\bf \susygen\  parameters} 
\vskip -1cm
\begin{verbatim}
 Susygen inputs:
 --------------
 M2      =   175.530   mu       =   869.900
 M1      =   550.600   M3       =  1232.270
 TANB    =    10.000

 Sparticle masses: (input from Isajet)
 ----------------

 SUPR     1227.  SUPL      1218.
 SDNR     1238.  SDNL      1221.
 SELR      303.  SELL       320. 
 SNU       310.
 STP1     1005.  STP2      1129. cosmix =  0.262
 SBT1     1112.  SBT2      1233. cosmix =  0.996
 STA1      271.  STA2       323. cosmix=   0.493
 SGLU     1275.

 Gaugino masses:
 --------------

 M1 =    550.600 M2 =    175.530 M3 =   1232.270

 NEUTRALINO m, CP, ph/zi/ha/hb 1 =  172.5  1. -0.475 -0.874  0.038  0.094
 NEUTRALINO m, CP, ph/zi/ha/hb 2 =  547.6  1.  0.879 -0.466  0.067  0.079
 NEUTRALINO m, CP, ph/zi/ha/hb 3 =  872.9 -1.  0.006  0.052 -0.634  0.772
 NEUTRALINO m, CP, ph/zi/ha/hb 4 =  878.9  1.  0.048 -0.126 -0.769 -0.624

 CHARGINO MASSES    =   172.486   877.866
 CHARGINO ETA       =    -1.000     1.000

 U matrix      WINO      HIGGSINO      V matrix      WINO     HIGGSINO 
 W1SS+        -0.991     0.136         W1SS-         0.999   -0.040
 W2SS+         0.136     0.991         W2SS-         0.040    0.999

 Higgses masses: 
 --------------
 Light CP-even Higgs =   114.830
 Heavy CP-even Higgs =   912.560
       CP-odd  Higgs =   911.740
       Charged Higgs =   915.830
       sin(a-b)      =     0.102
       cos(a-b)      =     0.995
\end{verbatim}
\end{small}
\clearpage

\subsection{Spectrum \& parameters of  PYTHIA 6.2/00}
\begin{figure}[h] \centering
  \epsfig{file=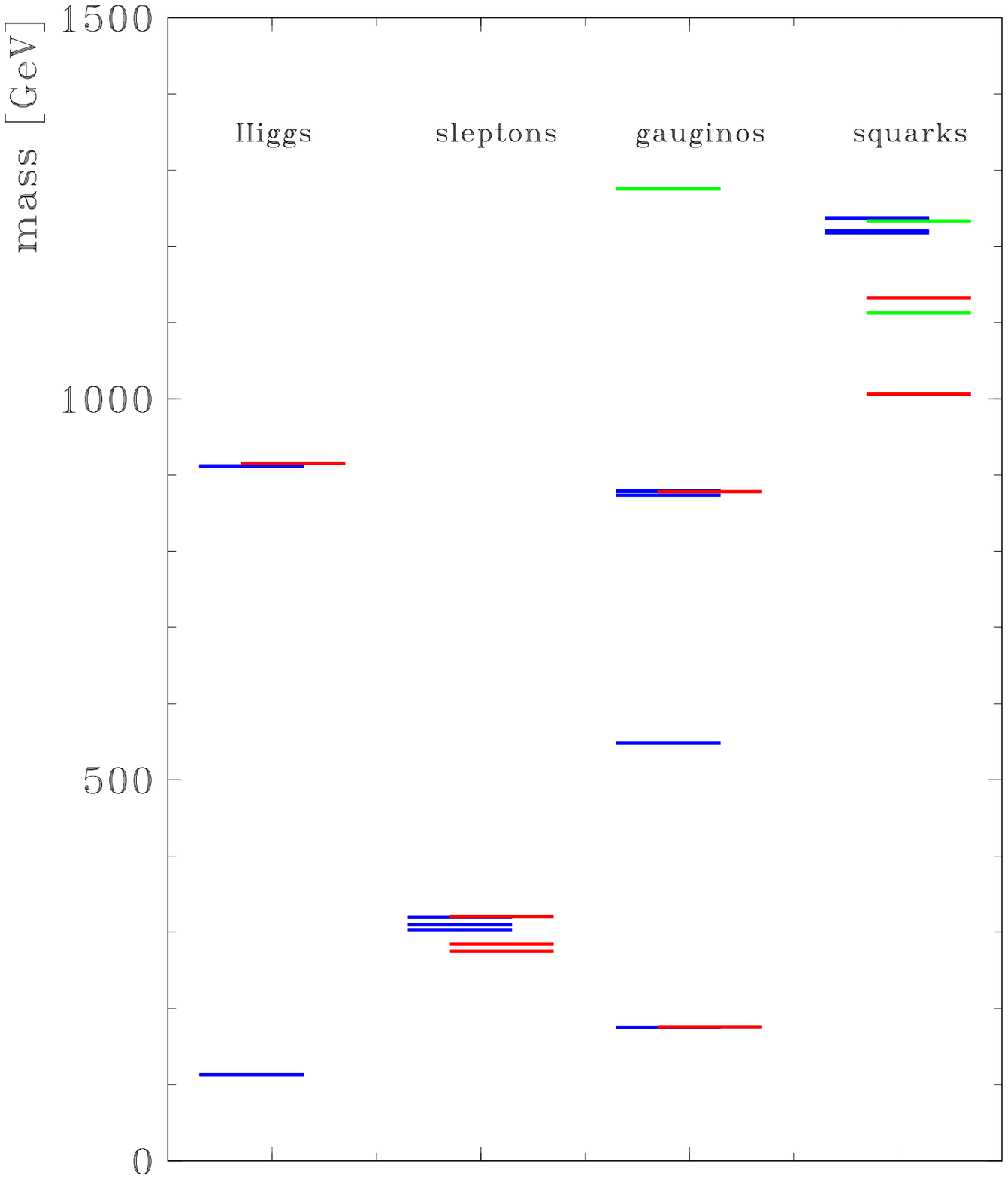,width=.9\textwidth}
  \caption{SPS 9 mass spectrum of \pythia}
\end{figure}
\newpage
\normalsize
\noindent
{\bf \pythia\  parameters} 
\begin{small}
\begin{verbatim}
   MSSM input parameters
   ---------------------
   M_1         RMSS( 1) =   550.6    
   M_2         RMSS( 2) =   178.5    
   M_3         RMSS( 3) =   1200.    
   tan_beta    RMSS( 5) =   10.00    
   mu          RMSS( 4) =   870.0    

   sparticle masses & widths
   -------------------------
   M_se_R   303.0 ( 0.000)   M_se_L   319.7 ( 1.949)   M_sne_L  309.6 ( 1.807)
   M_sm_R   303.0 ( 0.000)   M_sm_L   319.7 ( 1.949)   M_snm_L  309.6 ( 1.807)
   M_st_1   275.4 ( 0.619)   M_st_2   320.4 ( 0.959)   M_snt_L  284.5 ( 1.381)

   M_ch0_1  175.4 ( 0.000)   M_ch0_2  547.6 ( 6.016)   M_ch0_3  873.0 ( 6.869)
   M_ch0_4  879.0 ( 7.023)
   M_ch+_1  175.7 ( 0.000)   M_ch+_2  878.0 ( 7.060)

   M_h0     113.3 ( 0.004)   M_H0     911.4 ( 2.071)   M_A0     911.1 ( 2.112)
   M_H+     915.3 ( 2.236)

   M_g~    1275.4 ( 5.103)
   M_uL    1217.8 (15.156)   M_uR    1236.5 ( 1.836)   M_dL    1220.4 (15.051)
   M_dR    1237.2 ( 0.460)
   M_b1    1112.5 (15.069)   M_b2    1233.2 ( 0.654)   M_t1    1005.5 ( 3.225)
   M_t2    1132.0 (15.159)

   parameter settings IMSS, RMSS
   -----------------------------
   IMSS( 1) = 1   IMSS( 4) = 1   IMSS( 7) = 0   IMSS(10) = 0
   IMSS( 2) = 0   IMSS( 5) = 0   IMSS( 8) = 0   IMSS(11) = 0
   IMSS( 3) = 0   IMSS( 6) = 0   IMSS( 9) = 0   IMSS(12) = 0

   RMSS( 1) =   550.6       RMSS( 9) =   1237.       RMSS(17) =   1162.    
   RMSS( 2) =   178.5       RMSS(10) =   1112.       RMSS(18) = -0.1020    
   RMSS( 3) =   1200.       RMSS(11) =   1232.       RMSS(19) =   912.0    
   RMSS( 4) =   870.0       RMSS(12) =   1003.       RMSS(20) =  0.4100E-01
   RMSS( 5) =   10.00       RMSS(13) =   295.4       RMSS(21) =   1.000    
   RMSS( 6) =   319.7       RMSS(14) =   295.2       RMSS(22) =   1237.    
   RMSS( 7) =   303.0       RMSS(15) =   216.4       RMSS(23) =  0.1000E+05
   RMSS( 8) =   1219.       RMSS(16) =  -350.3       RMSS(24) =  0.1000E+05
\end{verbatim}
\end{small}

\clearpage

\subsection{Decay modes}

 \begin{table}[h] \centering
   \begin{tabular}{|l|c c c|c|c c c|}
   \hdick & & & \\[-2.ex]
   particle & $m_I$ & $m_S$ & $m_P$ & \ decay \ & ${\cal B}_I$ & ${\cal B}_S$ & ${\cal B}_P$ 
   \\[.5ex] \hdick
   $\ser    $ &  303.0 & 303.0  &  303.0 & $\nt_1    e^-               $ & 0.103 & 1.000 & 1.000 \\
              &        &        &        & $\stau_1  e^-      \tau^+   $ & 0.240 &       &       \\
              &        &        &        & $\staup_1 e^-      \tau^-   $ & 0.657 &       &       \\
   \hline
   $\sel    $ &  319.7 & 319.7  &  319.7 & $\nt_1    e^-               $ & 0.336 & 0.334 & 0.335 \\
              &        &        &        & $\cm_1    \nu_e             $ & 0.664 & 0.667 & 0.665 \\
   \hline
   $\sne    $ &  309.7 & 309.7  &  309.6 & $\nt_1    \nu_e             $ & 0.332 & 0.333 & 0.333 \\
              &        &        &        & $\cp_1    e^-               $ & 0.668 & 0.667 & 0.667 \\
   \hline
   $\smur   $ &  303.0 & 303.0  &  303.0 & $\nt_1    \mu^-             $ & 0.103 & 0.886 & 0.888 \\
              &        &        &        & $\cm_1    \nu_\mu           $ &       & 0.114 & 0.112 \\
              &        &        &        & $\stau_1  \mu^-    \tau^+   $ & 0.240 &       &       \\
              &        &        &        & $\staup_1 \mu^-    \tau^-   $ & 0.657 &       &       \\
   \hline
   $\smul   $ &  319.7 & 319.7  & 319.7  & $\nt_1    \mu^-             $ & 0.336 & 0.334 & 0.335 \\
              &        &        &        & $\cm_1    \nu_\mu           $ & 0.664 & 0.666 & 0.665 \\
   \hline
   $\snm    $ &  309.7 & 309.7  & 309.6  & $\nt_1    \nu_\mu           $ & 0.332 & 0.333 & 0.333 \\
              &        &        &        & $\cp_1    \mu^-             $ & 0.668 & 0.667 & 0.667 \\
   \hline
   $\stau_1 $ &  271.3 & 271.3  & 275.4  & $\nt_1    \tau^-            $ & 0.336 & 0.334 & 0.335 \\
              &        &        &        & $\cm_1    \nu_\tau          $ & 0.664 & 0.666 & 0.665 \\
   \hline
   $\stau_2 $ &  322.5 & 322.5  & 320.4  & $\nt_1    \tau^-            $ & 0.336 & 0.334 & 0.335 \\
              &        &        &        & $\cm_1    \nu_\tau          $ & 0.664 & 0.666 & 0.665 \\
   \hline
   $\snt    $ &  300.7 & 300.7  & 284.5  & $\nt_1    \nu_\tau          $ & 0.332 & 0.333 & 0.333 \\
              &        &        &        & $\cp_1    \tau^-            $ & 0.668 & 0.667 & 0.667 \\
   \hline
   \end{tabular}
   \caption{Slepton masses (GeV) and significant branching ratios ($>3\%$) 
            from \isajet~(I), \susygen~(S) and \pythia~(P)}
 \end{table}
  
 \begin{table}[h] \centering
   \begin{tabular}{|l|c c c|c|c c c|}
   \hdick & & & \\[-2.ex]
   particle & $m_I$ & $m_S$ & $m_P$ & \ decay \ & ${\cal B}_I$ & ${\cal B}_S$ & ${\cal B}_P$ 
   \\[.5ex] \hdick
   $\nt_1   $ &  175.5 &  172.5 & 175.4  & $                           $ & 1.000 & 1.000 & 1.000 \\
   \hline
   $\nt_2   $ &  549.0 & 547.6  & 547.6  & $\ser     e^+               $ & 0.110 & 0.109 & 0.110 \\
              &        &        &        & $\serp    e^-               $ & 0.110 & 0.109 & 0.110 \\
              &        &        &        & $\smur    \mu^+             $ & 0.110 & 0.109 & 0.110 \\
              &        &        &        & $\smurp   \mu^-             $ & 0.110 & 0.109 & 0.110 \\
              &        &        &        & $\stau_1  \tau^+            $ & 0.108 & 0.107 & 0.082 \\
              &        &        &        & $\staup_1 \tau^-            $ & 0.108 & 0.107 & 0.082 \\
              &        &        &        & $\stau_2  \tau^+            $ & 0.042 & 0.042 & 0.062 \\
              &        &        &        & $\staup_2 \tau^-            $ & 0.042 & 0.042 & 0.062 \\
   \hline
   $\nt_3   $ &  874.4 & 872.9  & 873.0  & $\cp_1    W^-               $ & 0.304 & 0.292 & 0.302 \\
              &        &        &        & $\cm_1    W^+               $ & 0.304 & 0.292 & 0.302 \\
              &        &        &        & $\nt_1    Z^0               $ & 0.181 & 0.221 & 0.230 \\
              &        &        &        & $\nt_2    Z^0               $ & 0.057 & 0.067 & 0.069 \\
              &        &        &        & $\nt_1    h^0               $ & 0.115 & 0.097 & 0.067 \\
   \hline
   $\nt_4   $ &  876.0 & 878.9  & 879.0  & $\cp_1    W^-               $ & 0.296 & 0.322 & 0.291 \\
              &        &        &        & $\cm_1    W^+               $ & 0.296 & 0.322 & 0.291 \\
              &        &        &        & $\nt_1    Z^0               $ & 0.118 & 0.075 & 0.068 \\
              &        &        &        & $\nt_1    h^0               $ & 0.171 & 0.158 & 0.219 \\
              &        &        &        & $\nt_2    h^0               $ & 0.054 & 0.049 & 0.064 \\
   \hline
   \end{tabular}
   \caption{Neutralino masses (GeV) and significant branching ratios ($>3\%$) 
            from \isajet~(I), \susygen~(S) and \pythia~(P)}
 \end{table}
  
 \begin{table}[h] \centering
   \begin{tabular}{|l|c c c|c|c c c|}
   \hdick & & & \\[-2.ex]
   particle & $m_I$ & $m_S$ & $m_P$ & \ decay \ & ${\cal B}_I$ & ${\cal B}_S$ & ${\cal B}_P$ 
   \\[.5ex] \hdick
   $\cp_1   $ &  175.7 & 172.5  & 175.7  & $\nt_1    \pi^+             $ & 0.960 &       &       \\
              &        &        &        & $\nt_1    \nu_e e^+         $ &       & 1.000 & 0.500 \\
              &        &        &        & $\nt_1    \nu_\mu \mu^+     $ &       &       & 0.500 \\
   \hline
   $\cp_2   $ &  877.2 & 877.9  & 878.0  & $\nt_1    W^+               $ & 0.300 & 0.416 & 0.298 \\
              &        &        &        & $\nt_2    W^+               $ & 0.066 & 0.112 & 0.079 \\
              &        &        &        & $\cp_1    Z^0               $ & 0.298 & 0.421 & 0.293 \\
              &        &        &        & $\cp_1    h^0               $ & 0.286 &       & 0.283 \\
   \hline
   \end{tabular}
   \caption{Chargino masses (GeV) and significant branching ratios ($>3\%$) 
            from \isajet~(I), \susygen~(S) and \pythia~(P)}
 \end{table}
  
 \begin{table}[h] \centering
   \begin{tabular}{|l|c c c|c|c c c|}
   \hdick & & & \\[-2.ex]
   particle & $m_I$ & $m_S$ & $m_P$ & \ decay \ & ${\cal B}_I$ & ${\cal B}_S$ & ${\cal B}_P$ 
   \\[.5ex] \hdick
   $h^0     $ &  114.8 & 114.8  & 113.3  & $\tau^-   \tau^+            $ & 0.050 & 0.071 & 0.065 \\
              &        &        &        & $b        \bar b            $ & 0.827 & 0.824 & 0.783 \\
              &        &        &        & $c        \bar c            $ & 0.040 &       & 0.048 \\
              &        &        &        & $g        g                 $ & 0.030 & 0.051 & 0.038 \\
              &        &        &        & $W^+      W^-               $ &       & 0.032 & 0.054 \\
   \hline
   $H^0     $ &  912.6 &  912.6 & 911.4  & $\tau^-   \tau^+            $ & 0.055 & 0.089 & 0.092 \\
              &        &        &        & $b        \bar b            $ & 0.680 & 0.808 & 0.726 \\
              &        &        &        & $t        \bar t            $ & 0.133 & 0.057 & 0.178 \\
              &        &        &        & $\cp_1    \cm_1             $ & 0.030 &       &       \\
   \hline
   $A^0     $ &  911.7 & 911.7  & 911.1  & $\tau^-   \tau^+            $ & 0.054 & 0.081 & 0.090 \\
              &        &        &        & $b        \bar b            $ & 0.670 & 0.735 & 0.713 \\
              &        &        &        & $t        \bar t            $ & 0.143 & 0.147 & 0.195 \\
              &        &        &        & $\stau_1  \staup_1          $ & 0.030 &       &       \\
              &        &        &        & $\cp_1 \cm_1                $ &       & 0.033 &       \\
   \hline
   $H^+     $ &  915.8 & 915.8  & 915.3  & $\nu_\tau \tau^+            $ & 0.064 & 0.097 & 0.086 \\
              &        &        &        & $t        \bar b            $ & 0.826 & 0.873 & 0.913 \\
              &        &        &        & $\staup_1 \snt              $ & 0.066 &       &       \\
   \hline
   \end{tabular}
   \caption{Higgs masses (GeV) and significant branching ratios ($>3\%$) 
            from \isajet~(I), \susygen~(S) and \pythia~(P)}
 \end{table}
  
 \begin{table}[h] \centering
   \begin{tabular}{|l|c c c|c|c c c|}
   \hdick & & & \\[-2.ex]
   particle & $m_I$ & $m_S$ & $m_P$ & \ decay \ & ${\cal B}_I$ & ${\cal B}_S$ & ${\cal B}_P$ 
   \\[.5ex] \hdick
   $\st_1   $ & 1005.2 &  1005.2& 1005.5 & $\nt_1    t                 $ & 0.095 & 0.116 & 0.129 \\
              &        &        &        & $\nt_2    t                 $ & 0.341 & 0.291 & 0.302 \\
              &        &        &        & $\cp_1    b                 $ & 0.205 & 0.259 & 0.280 \\
              &        &        &        & $\cp_2    b                 $ & 0.359 & 0.334 & 0.289 \\
   \hline
   $\st_2   $ & 1128.8 & 1128.8 & 1132.0 & $\cp_1    b                 $ & 0.558 & 0.442 & 0.545 \\
              &        &        &        & $\cp_2    b                 $ &       & 0.184 &       \\
              &        &        &        & $\nt_1    t                 $ & 0.262 & 0.198 & 0.256 \\
              &        &        &        & $\nt_3    t                 $ & 0.088 & 0.082 & 0.054 \\
              &        &        &        & $\nt_4    t                 $ & 0.042 & 0.087 & 0.103 \\
   \hline
   $\sb_1   $ & 1112.1 & 1112.1 & 1112.5 & $\nt_1    b                 $ & 0.301 & 0.333 & 0.297 \\
              &        &        &        & $\cm_1    t                 $ & 0.565 & 0.624 & 0.556 \\
              &        &        &        & $\cm_2    t                 $ & 0.114 &       & 0.133 \\
   \hline
   $\sb_2   $ & 1232.9 & 1232.9 & 1233.2 & $\nt_1    b                 $ &       &       & 0.063 \\
              &        &        &        & $\nt_2    b                 $ & 0.434 & 0.303 & 0.690 \\
              &        &        &        & $\nt_3    b                 $ & 0.097 & 0.182 &       \\
              &        &        &        & $\nt_4    b                 $ & 0.098 & 0.181 &       \\
              &        &        &        & $\cm_1    t                 $ &       &       & 0.119 \\
              &        &        &        & $\cm_2    t                 $ & 0.256 & 0.324 & 0.061 \\
              &        &        &        & $\tilde{t}_1 W^-            $ &       &       & 0.037 \\
   \hline
   \end{tabular}
   \caption{Light squark masses (GeV) and significant branching ratios ($>3\%$) 
            from \isajet~(I), \susygen~(S) and \pythia~(P)}
 \end{table}

\end{document}